\documentclass{article}
\PassOptionsToPackage{svgnames}{xcolor}
\usepackage{lmodern}
\usepackage{geometry}
\usepackage{anyfontsize}
\usepackage{microtype} 
\usepackage{amsmath,amssymb,amsopn,amsthm,mathrsfs,mathtools}
\usepackage{graphicx}
\usepackage{bm}
\usepackage{appendix}
\usepackage{enumitem}
\usepackage[affil-it]{authblk}
\usepackage{tikzit}
\usepackage{tcolorbox}
\usepackage{cancel}
\usepackage{comment}
\usepackage{cite}
\usepackage{soul}

\usepackage{ulem}
\usepackage{tabu}
\usepackage{multirow}
\usepackage{caption}
\usepackage{subcaption}
\usepackage{setspace}

\newtheorem{assumption}{Property}
\newtheorem{linassumption}{Property}[subsubsection]

\newtheorem{thm}{Theorem}[section]
\newtheorem{cor}[thm]{Corollary}

\theoremstyle{definition}
\newtheorem{defn}{Definition}[section]

\theoremstyle{remark}
\newtheorem{remark}{Remark}[section]

\tcbset{boxrule=0.5pt,
        arc=2mm,
        top=1mm,bottom=1mm,right=1mm,left=1mm}

\newcommand{\M}{\mathcal{M}}
\newcommand{\PI}{\Pi_{\mathcal{M}}}

\newcommand{\bracke}[1]{\ensuremath{\left\langle #1 \right\rangle}}

\newcommand{\tepsilon}{\bm{\tilde{\varepsilon}}}
\newcommand{\tepsilonu}[1]{\bm{\tilde{\varepsilon}}_{#1}(\bm{u})}

\newcommand{\divut}{\text{div}\bm{\tilde{u}}}
\newcommand{\uD}{\bm{u}\!\cdot\!\nabla}

\newcommand{\vdx}[1]{v_{#1}\partial_{x_{#1}}}
\newcommand{\dx}[1]{\partial_{x_{#1}}}

\newcommand{\trho}{\tilde{\rho}}

\newcommand{\tu}{\tilde{u}}

\newcommand{\ttheta}{\tilde{\theta}}

\newcommand{\Id}{\text{Id}}
\newcommand{\St}{\text{St}}

\newcommand{\parent}[1]{\left( #1\right)}

\newcommand{\sqbrac}[1]{\left[#1 \right]}
\newcommand{\abrac}[1]{\left\langle #1 \right\rangle}

\newcommand{\Lmu}{\mathcal{L}_{\mu} }
\newcommand{\iLmu}{\mathcal{L}_{\mu}^{-1}}

\newcommand{\LM}{\mathcal{L}_{\mathcal{M}} }
\newcommand{\iLM}{\mathcal{L}_{\mathcal{M}}^{-1}}

\newcommand{\iLm}{\mathcal{L}_{M}^{-1}}
\newcommand{\Lm}{\mathcal{L}_{M}}

\newcommand{\KrPhi}{\bm{\Phi}}
\newcommand{\odir}{\bm{\hat\eta}}

\newcommand{\gauss}{\mathtt{E}}
\newcommand{\LG}{\mathbb{L}_{\gauss}}
\newcommand{\iLG}{\mathbb{L}^{-1}_{\gauss}}
\newcommand{\Ebrac}[1]{\left\langle\!\left\langle #1 \right\rangle\! \right\rangle_{\!\gauss}}

\makeatletter
\newif\iftag@here

\newcommand*{\taghere}[1][0pt]
{\ifmeasuring@\else
  \global\tag@heretrue
  \tikz[remember picture,overlay]{\coordinate (taghere) at (0pt,#1);}%
\fi}

\def\place@tag{%
    \iftagsleft@
      \kern-\tagshift@
      \iftag@here
        \global\tag@herefalse
        \tikz[remember picture,overlay]%
          {\path (taghere) -| node[anchor=base]{\rlap{\boxz@}} (0pt,0pt);}%
      \else
        \if1\shift@tag\row@\relax
            \rlap{\vbox{%
                \normalbaselines
                \boxz@
                \vbox to\lineht@{}%
                \raise@tag
            }}%
        \else
            \rlap{\boxz@}%
        \fi
        \kern\displaywidth@
      \fi
    \else
      \kern-\tagshift@
      \iftag@here
        \global\tag@herefalse
        \tikz[remember picture,overlay]%
          {\path  (taghere) -|  node[anchor=bse]{\llap{\boxz@}} (0pt,0pt);}%
      \else
        \if1\shift@tag\row@\relax
            \llap{\vtop{%
                \raise@tag
                \normalbaselines
                \setbox\@ne\null
                \dp\@ne\lineht@
                \box\@ne
                \boxz@
            }}%
        \else \llap{\boxz@}%
        \fi
      \fi
    \fi
}
\makeatother
\author{FA Baidoo$^1$, IM Gamba$^1$, TJR Hughes$^1$, MRA Abdelmalik$^2$} %
\affil{$^1$Oden Institute for Computational Sciences and Engineering, University of Texas at Austin\\
$^2$Department of Mechanical Engineering, Eindhoven University of Technology}

\title{Extensions to the Navier-Stokes-Fourier Equations for Rarefied Transport: Variational Multiscale Moment Methods for the Boltzmann Equation}

\date{}

\begin{document}
\maketitle

\begin{abstract}
    We derive a fourth order entropy stable extension of the Navier-Stokes-Fourier equations into the transition regime of rarefied gases. We do this through a novel reformulation of the closure of conservation equations derived from the Boltzmann equation that subsumes existing methods such as the Chapman-Enskog expansion. We apply the linearized version of this extension to the stationary heat problem and the Poiseuille channel and compare our analytical solutions to asymptotic and numerical solutions of the linearized Boltzmann equation. In both model problems, our solutions compare remarkably well in the transition regime. For some macroscopic variables, this agreement even extends far beyond the transition regime.
\end{abstract}

\section{Introduction}
Gases in the transition regime have a mean free path that is large enough relative to macroscopic reference length scales to render continuum models like the Navier-Stokes-Fourier equations invalid while still being small enough that statistical methods such as Direct Simulation Monte Carlo are computationally expensive. This regime finds applications that span many orders of magnitude from 
the flows around objects in the thin atmosphere of near-earth orbit to flows within microchannels of microelectromechanical systems.

Perhaps the most storied macroscopic model for gases in this regime is the Burnett equations. These third-order moment equations are derived from the Chapman-Enskog expansion \cite{chapman1990mathematical} of the Boltzmann equation, where they arise as a correction to the Navier-Stokes-Fourier equations. Unfortunately, the Burnett equations prove to be a far from ideal extension into the transition regime. For example, Bobylev \cite{bobylev1982chapman, bobylev2006} showed that the Burnett equations are unstable with respect to short wavelength perturbations, while Comeaux et al. \cite{Comeaux1995} showed that the Burnett equations violate the second law of thermodynamics at large enough Knudsen numbers. The Burnett equations have also been shown to produce unphysical stationary solutions \cite{Struchtrup20053}.

Many modifications have been made to the Burnett equations to overcome these shortcomings. To name a few, Zhong \cite{Zhong1993} adds new terms to improve stability creating the augmented Burnett equations while Jin and Slemrod \cite{Jin2001} propose relaxing the deviatoric pressure and heat flux with rate equations.  Bobylev \cite{bobylev2006, Bobylev2018} advocates for a transformation of the vector of hydrodynamic variables in order to obtain generalized Burnett equations which do not suffer from the instabilities of the original. {\color{black} If we look beyond modifications to the Burnett equations and also consider alternatives to it, we can include moment methods that build on the work of Grad\cite{Grad1963} such as the regularized 13 (and 26) moment equations of Struchtrup and Torrilhon\cite{Struchtrup2003} which are still an area of active theoretical \cite{Struchtrup20052, Gu2009,Kauf2010, Cai2024, Hu2020  } and numerical research \cite{rana2013,Gu2007} as well as the work of Levermore \cite{levermore1996,Levermore1998gaussian}. There is also work on alternative entropy stable extensions to the Navier-Stokes-Fourier equations in progress \cite{LevermoreUnpublished}.
}

In this paper, we will use the framework of the Variational Multiscale (VMS) method to derive an alternative to the Burnett equations by modifying the process by which we obtain constitutive relations for the deviatoric stress and heat flux in the conservation equations obtained from the Boltzmann equation. The VMS method \cite{hughes1995, Hughes1998, hughes2007} was originally created as a framework to derive stable finite element schemes for highly advective partial differential equations and later for closures for turbulence modeling \cite{Bazilevs2007}. If we think of the  conservation equations (or moment equations more generally)  as a weak form in microscopic velocity of the Boltzmann equation, then it is not too surprising that an idea from the finite element method can be brought to bear on the problem of finding closures.
The methodology is conceptually simple. Assuming that the distribution is a perturbation of a Maxwellian, we separate the Boltzmann equation into a finite-dimensional coarse scale equation from which the macroscopic equations will arise and an infinite dimensional fine-scale equation which encodes the macroscopic variables we seek closures for. Within this framework, deriving a closure amounts to substituting an approximation of the solution of the fine-scale equation into the coarse-scale equation. The framework is general enough that it can be used to describe the Chapman-Enskog expansion whilst opening the door to considering new closures. For our particular alternative to the Burnett equation, we take advantage of the fact that the coarse-scale equation also gives an equation for the macroscopic entropy of the system of conservation equations. We create an approximation of the fine-scale equation that ensures that the conservation equations derived are entropy stable. In our alternative to the Burnett equations, the momentum and energy equations are fourth-order extensions to the Navier-Stokes-Fourier equations.

\textcolor{black}{The main contributions of this work are as follows:
\begin{itemize}
\item We propose a variational framework for the derivation of fluid dynamic closure relations for the conservation equations in both the linear and nonlinear settings. Our framework subsumes traditional methods like the Chapman-Enskog expansion while enabling the derivation of entropy-stable closures beyond the Burnett level. Such entropy-stable closures address known instabilities and thermodynamic violations in higher-order moment models.
\item We derive a fourth-order entropy-stable extension to the Navier-Stokes-Fourier equations in both the linear and nonlinear settings. To demonstrate the efficacy of such a closure, we focus on the linear setting to derive analytical solutions of our extended Navier-Stokes-Fourier equations to benchmark problems (stationary heat transfer and Poiseuille flow), and compare such analytical solutions to those of asymptotic and numerical approximations of solutions of the linearized Boltzmann equation.
\item We derive the fluid transport coefficients of a subsystem of the nonlinear entropy stable extension to the Navier-Stokes-Fourier equations, and show the entropy dissipation inequality associated with it. 
\end{itemize}
}

The paper is organized as follows. In Section 2, we introduce the Boltzmann equation and the conservation equations that arise from it. We discuss the collision operator (with a special emphasis on the linearized collision operator) and the properties that make the analysis that follows possible. The material presented in this section is well known (see for example, \cite{cercignani1988,cercignani1994, golse2005,saint-raymond2009, levermore1996,bardos:1991vf}) but is presented here in order to keep the work self-contained. 
In Section 3, we elaborate on the VMS moment method for deriving closures. We illustrate the process with the Chapman-Enskog expansion, showing that the Euler and Navier-Stokes-Fourier equations are entropy stable whilst the Burnett equations and higher are not. We then introduce the methodology to generate an entropy stable alternative to the Burnett equations. In Section 4, we apply the linearized version of the extension to the stationary heat problem and the Poiseuille channel problem and compare the analytical solutions generated to asymptotic and numerical solutions to the Boltzmann equation found in \cite{SoneBook,Ohwada1989, Ohwada1989StatHeat}. Supplementary calculations can be found in the Appendices at the end.

\section{The Boltzmann Equation}
Consider the distribution  function $F=F(t,\bm x,\bm v)$ that at time $t\in\mathbb{R}_{\geq0}$ and spatial position $\bm x=(x_1, \dots,x_D)\in\Omega \subset\mathbb{R}^D$ gives the density of particles of a monatomic fluid with phase velocity $\bm v=(v_1,\dots,v_D)\in\mathbb{R}^D$, where $D$ is the number of dimensions. The evolution of $F$ is governed by the Boltzmann equation
\begin{align}\label{eq:be}
\text{St}\,\partial_t F(t,\bm x,\bm v) 
+ 
v_i\partial_{x_i} F(t, \bm x, \bm v)
=
\frac{1}{\epsilon}\,\mathcal{C}(F)(t,\bm x,\bm v) \quad \text{ where } (t,\bm x, \bm v) \in \mathbb R_+\times \mathbb R^D \times \mathbb R^D,
\end{align}
where in the above equation and throughout the text, we employ the Cartesian tensor notation \cite{CartesianTensorBook, Hashiguchi2020} with repeated indices implying summation up to dimension $D$. {\color{black} A brief introduction to this summer notation is given in Appendix \ref{sec:Tensor}} The collision operator $\mathcal{C}(F)$ acts globally on the velocity $\bm v$ dependence of $F$ but locally in time $t$ and position $\bm x$. The dimensionless number St is called the kinetic Strouhal number, defined as the ratio of a reference macroscopic length scale of the gas to the product of the \textit{thermal speed of sound} of the gas (defined later) and a reference time scale. The dimensionless number $\epsilon$ is the Knudsen number, defined as the ratio of the mean free path of the gas to a reference macroscopic length scale. 

\subsection{Macroscopic Observables}
{Macroscopic fluid variables are obtained from $F(t, \bm x, \bm v)$ by taking its moments in the velocity variable:}
\begin{subequations}\label{eq:macroparam}
\begin{align}
\int_{\mathbb R^D} F(t, \bm x, \bm v) \, d\bm v&=\rho \label{eq:mass}\\ 
\int_{\mathbb R^D} v_i\, F(t, \bm x, \bm v) \, d\bm v &= \rho\, u_i \label{eq:momentum}\\ ,
\int_{\mathbb R^D} v_iv_j\, F(t, \bm x, \bm v) \, d\bm v&= \rho u_i u_j+P_{ij} \label{eq:presstensor}\\
\int_{\mathbb R^D} v_i v_j v_k \,F(t, \bm x, \bm v) \, d\bm v &=  \rho u_i u_j u_k + u_i P_{jk} + u_j P_{ik} + u_k P_{ij} + Q_{ijk} \label{eq:heattensoreqn}
\end{align}
\end{subequations}
where $\rho=\rho(t,\bm x)$, $u_i=u_i(t,\bm x)$ are the macroscopic mass and bulk velocity vector, respectively. 
In addition, the macroscopic { stress} tensor
$P_{ij}=P_{ij}(t,\bm x)$ and the heat flux tensor $ Q_{ijk}=Q_{ijk}(t,\bm x)$  are derived from the second and third moments of $F$ with respect to random deviations { of the particle velocities} , that is 
\begin{align}
   P_{ij}(t, \bm x) &= \int_{\mathbb R^D} (v_i- u_i)(v_j - u_j)\, F(t, \bm x, \bm v)\, d\bm v \label{eq:fullstress}\\
   Q_{ijk}(t, \bm x) &= \int_{\mathbb R^D} (v_i- u_i)(v_j - u_j)(v_k - u_k)\, F(t, \bm x, \bm v)\, d\bm v\label{eq:heattensor}
\end{align}
The stress tensor is decomposed into an isotropic pressure $p(t,\bm x) = \frac{1}{D}\text{Tr}(P_{ij})$ and the accompanying deviatoric stress tensor 
\begin{equation}\sigma_{ij}(t, \bm x) =  p\,\delta_{ij}-P_{ij}\end{equation} 
With the internal energy  per degree of freedom given by 
\begin{align}\label{eq:intenergy}
  \rho\theta(t, \bm x) = \frac{1}{D}\int_{\mathbb R^D} |\bm v - \bm u|^2 \, F \,d\bm v
\end{align}
we thus have that $p = \rho \theta$.  For an ideal gas, this means that $\theta(t, \bm x)$ is the re-scaled temperature i.e. $\theta = \frac{k_BT}{m}$, where $k_B= 1.38 \cdot 10^{-23} \text{JK}^{-1}$ is the Boltzmann constant and $m$ is the mass of a gas particle. As such, given a reference temperature $\theta_0$, the thermal speed of sound of the gas is given by $\sqrt{\gamma\theta_0}$ where $\gamma$ is the adiabatic exponent. Furthermore, we may infer relations for the total energy and total energy flux by taking the traces of (\ref{eq:presstensor}) and (\ref{eq:heattensor}), respectively:
\begin{subequations}
\begin{align}
\frac{1}{2}\int_{\mathbb R^D} |{\bm v}|^2 F\, d\bm v &= \frac{1}{2}\rho |\bm u|^2 + \frac{D}{2}\rho\theta, \label{eq:energy}\\
\frac{1}{2}\int_{\mathbb R^D} |\bm v|^2 v_i F \, d\bm v &= \frac{1}{2}\rho|\bm u|^2 u_i + \frac{(D+2)}{2} \rho\theta u _i- \sigma_{ij} u_j + q_i, \label{eq:heat} 
\end{align}
\end{subequations}
where the vector $q_i=Q_{ijj}/2$ denotes the heat flux vector.

For our purposes, it is also necessary to consider the situation where our distribution $F$ is close to a stationary equilibrium state. In particular,
\begin{align*}
   \rho(t, \bm x) &= \rho_0 + \epsilon\,\tilde{\rho}(t, \bm x)\\
   \bm{u}(t, \bm x) &= \epsilon\, \tilde{\bm u}(t, \bm x) \\
   \theta(t, \bm x)  &= \theta_0 + \epsilon\,\tilde{\theta}(t, \bm x) 
\end{align*}
where $\rho_0$ and $\theta_0$ are a constant background density and temperature.
  In this situation, we consider the linearization of the macroscopic parameters (\ref{eq:macroparam}) with respect to $\epsilon$:
\begin{subequations}\label{eq:linmacroparam}
\begin{align}
\int_{\mathbb R^D} F(t, \bm x, \bm v) \, d\bm v&=\rho_0 + \epsilon\, \tilde{\rho} \label{eq:linmass}\\ 
\int_{\mathbb R^D} v_i\, F(t, \bm x, \bm v) \, d\bm v &= \epsilon\, \rho_0 \tilde{u}_i \label{eq:linmomentum}\\ 
\int_{\mathbb R^D} v_iv_j\, F(t, \bm x, \bm v) \, d\bm v&= \rho_0\theta_0\, \delta_{ij} + \epsilon\left( \rho_0\,\tilde{\theta}+ \tilde{\rho}\, \theta_0 \right)\delta_{ij} - \epsilon\tilde{\sigma}_{ij} \label{eq:linpresstensor}\\
\int_{\mathbb R^D} |\bm{v}|^2 v_i \,F(t, \bm x, \bm v) \, d\bm v &= \epsilon\,(D+2) \,\rho_0 \, \theta_{0}\,\tilde{u}_i  + 2\,  \epsilon\tilde{q}_i \label{eq:linheattensor}
\end{align}
\end{subequations}
where $\epsilon\tilde\sigma_{ij}$ and $\epsilon\tilde q_i$ represent the linearizations of $\sigma_{ij}$ and $q_i$ respectively.

\subsection{The Collision Operator}
The collision operator $\mathcal C(F)$ is defined on a set of functions $\mathscr{D}_{\bm v}(\mathcal{C})\subset L^1_2(\mathbb{R}_{\bm v}^D) \cap L\log L(\mathbb{R}_v^D)$ 
where  $L_2^1(\mathbb{R}_{\bm v}^D) := \{ F : \int_{\mathbb{R}^D} (1 + |\bm v|^2) F(t, \bm x, \bm v)\, d\bm v < \infty \}$ and $L\log L(\mathbb{R}_{\bm v}^D):= \{ F : \int_{\mathbb{R}^D}F \log(F) \, d\bm v < \infty\} $. Given any vector $\bm U\in\mathbb{R}^D$ and orthonormal matrix $\bm O\in \mathbb{R}^{D\times D}$, translations and rotations in the $\bm v$ variable are defined by $(\mathcal{T}_{\bm U}F)(v) := F(\bm v -\bm U)$ and $(\mathcal{T}_{\bm O} F)(\bm v)=F({\bm O}^T\bm v)$. It will be assumed that for all $F \in \mathscr{D}_{\bm v}(\mathcal{C})$, we also have that  $\mathcal{T}_{\bm U}F$ and $\mathcal{T}_{\bm O}F$ are contained in $\mathscr{D}_{\bm v}(\mathcal{C})$.

For our purposes, we will require that the collision operator satisfies certain properties we shall outline in the intervening subsections. These properties are well-known (see, for instance \cite{cercignani1988}) but are presented here to ensure coherence in the text.

\begin{assumption}
We require that the collision operator commutes with translations and rotations, that is \begin{equation}\label{eq:GalilInvar}
   \mathcal{T}_{\bm U}\mathcal{C}(F) =\mathcal{C}(\mathcal{T}_{\bm U}F),\quad\text{and}\quad
\mathcal{T}_{{\bm O}}\mathcal{C}(F)=\mathcal{C}(\mathcal{T}_{\bm O}F). 
\end{equation}

\end{assumption}
The invariances \eqref{eq:GalilInvar} are consistent with the Hamiltonian dynamics of the advective material derivative in the left hand side of the Boltzmann equation (\ref{eq:be}), meaning whenever $F(t,\bm x,\bm v)$ solves this  equation, then so do $F(t,\bm x - {\bm U}t,{\bm v}-{\bm U})$ and  $F(t,{\bm O}^T{\bm x}\,,{\bm O}^T{\bm v})$.

\subsubsection{Collision Invariance}
A scalar-valued function $\psi(\bm v)$ is a collision invariant of $\mathcal{C}$ if
\begin{align}\label{eq:defcollinvar}
\int_{\mathbb R^D} \psi\,\mathcal{C}(F) \, d\bm v = 0  \quad \forall F\in \mathscr{D}_{\bm v}(\mathcal{C}),
\end{align}
 The relation in (\ref{eq:defcollinvar}) associates a scalar conservation law to (\ref{eq:be}) with each collision invariant:
\begin{equation}
\label{eq:conslaw}
\text{St }\partial_t\int_{\mathbb R^D} \psi F\, d\bm v
+
\partial_{x_i} \int_{\mathbb R^D} v_i \psi F \, d\bm v 
= 
0
\end{equation}
\begin{assumption} \label{assume:collinvar}
The space of collision invariants of $\mathcal{C}$ is $\mathscr{I}:=\mathrm{span}\{1,v_1,\ldots,v_D, |\bm v|^2\}$. In other words,
\begin{equation}\label{eq:collinvar}
\int_{\mathbb R^D}
{}\psi(\bm v)\,
\mathcal{C}(F)
\, d\bm v
=
0
\ \quad
\forall{}F\in\mathscr{D}_{\bm v}(\mathcal{C})
\
\Leftrightarrow
\ \quad
\psi(\bm v)\in\mathscr{I}.
\end{equation}
\end{assumption}
Property \ref{assume:collinvar} implies that solutions of the Boltzmann equation (\ref{eq:be}) obey the conservation laws for compressible gas flow systems \footnote{The reader can recall that  the classical adiabatic exponent $\gamma=\gamma(D)$ depends on the space dimension $D$.  For monoatomic gases this relation is $\gamma = 1 + 2/D$, recovering the familiar version of these conservation equations in terms of the adiabatic number. 
}\small
\begin{subequations}
\label{eq:macroconslaws}
\begin{align}
\text{St }\partial_t\rho + \partial_{x_i} (\rho u_i)&=0; \label{eq:masslaw}\\
\text{St }\partial_t(\rho u_j) + \partial_{x_i} (\rho {u}_i{u}_j+ \rho \theta\delta_{ij} - \sigma_{ij})&=0;\label{eq:momentumlaw}\\
\text{St }\partial_t\left(\frac{1}{2}\rho|\bm u|^2+\frac{D}{2}\rho \theta\right) + \partial_{x_i} \left(\left(\frac{1}{2}\rho|\bm u|^2 u_i + \frac{D}{2} \rho\theta \right)u_i+ \rho\theta u_i- \sigma_{ij} u_j + q_i\right)&=0\label{eq:energylaw};
\end{align}
\end{subequations}
\normalsize
i.e. solutions of (\ref{eq:be}) conserve mass according to (\ref{eq:masslaw}), momentum according to (\ref{eq:momentumlaw}), and energy according to (\ref{eq:energylaw}). Note that the system of conservation laws (\ref{eq:macroconslaws}) is not closed since there are $(D^2+5D+2)/2$  variables and only $D+2$ relations. The closure of (\ref{eq:macroconslaws}) requires a constitutive modeling assumption that characterizes, both $\sigma_{ij}$ and $q_i$, in terms of the proposed unknowns $\rho$, $u_i$ and $\theta$. Using equations (\ref{eq:fullstress}), (\ref{eq:heattensor}) and (\ref{eq:heat}), one finds that
\begin{subequations}
\begin{align}
    \sigma_{ij} &= -\int_{\mathbb R^D} \left( (v_i-u_i) (v_j - u_j)-\frac{1}{D}|{\bm v}-\bm{u}|^2\delta_{ij}\right) F\, d\bm v, \label{eq:sigmaclose}\\
    q_{i} &= \frac{1}{2}\int_{\mathbb R^D}  \left(|\bm v - \bm{u}|^2 (v_i - u_i) - (D+2)\theta (v_i-u_i)\right)  F\, d\bm v, \label{eq:qclose}
\end{align}
\end{subequations}
Thus closures for (\ref{eq:macroconslaws}) can be found by approximating $F$ as a function that is solely parametrized by $\rho$, $u_i$ and $\theta$ along with their derivatives.

{
The above dynamic holds true when we linearize about a constant equilibrium state. We obtain linearized conservation equations:
\begin{subequations}
\label{eq:linmacroconslaws}
\begin{align}
\text{St }\partial_t\tilde{\rho} + \rho_0\partial_{x_i}  \tilde{u}_i&=0; \label{eq:linmasslaw}\\
\text{St }\rho_0\,\partial_t \tilde{u}_j + \partial_{x_i} (\tilde{\rho}\, \theta_0\, \delta_{ij} + \rho_0\, \tilde{\theta}\,\delta_{ij} - \tilde{\sigma}_{ij})&=0;\label{eq:linmomentumlaw}\\
\frac{D}{2}\text{St }\rho_0\,\partial_t \tilde{\theta} + \partial_{x_i} \left(\rho_0\theta_0\, \tilde{u}_i + \tilde{q}_i\right)&=0\label{eq:linenergylaw};
\end{align}
\end{subequations}
where
\begin{subequations}
\begin{align}
    \epsilon\,\tilde{\sigma}_{ij} = -\int_{\mathbb R^D} \left( v_i v_j-\frac{1}{D}|{\bm v}|^2\delta_{ij}\right) F\, d\bm v, \label{eq:linsigmaclose}\\
    \epsilon\,\tilde{q}_{i} = \frac{1}{2}\int_{\mathbb R^D}\left(|\bm v|^2 v_i - (D+2)\theta_0 v_i\right)  F\, d\bm v, \label{eq:linqclose}
\end{align}
\end{subequations}
and closures for (\ref{eq:linmacroconslaws}) can be found by approximating $F$ as a function that is solely parametrized by $\tilde{\rho}$, $\tilde{u}_i$ and $\tilde{\theta}$ along with their derivatives.
}

\subsubsection{$\mathcal H-$theorem: Entropy Dissipation and Equilibria}
\begin{assumption}\label{assume:dissipation}
$\mathcal{C}$ satisfies the local dissipation relation 
\begin{equation}
\label{eq:dissipation}
\int_{\mathbb R^D} \ln\left({F}\right)\,\mathcal{C}(F) \, d\bm v \leq 0 \quad \forall F\in\mathscr{D}(\mathcal{C}).
\end{equation}
 
\end{assumption} 

Relation (\ref{eq:dissipation}) leads to a statement of Boltzmann's $\mathcal H-$theorem. This is accomplished by weighting the Boltzmann equation (\ref{eq:be}) with $\ln\left(F\right )$ and integrating on the entire velocity space in order to obtain the local entropy-dissipation law:
\begin{align}\label{eq:entdissip}
    \text{St }\partial_t \int_{\mathbb R^D} \Big(F\ln\left({F}\right)-F\Big) \, d\bm v 
 +
 \partial_{x_i} \int_{\mathbb R^D} v_i \Big(F\ln\left({F}\right)-F\Big) \, d\bm v 
 = 
 \frac{1}{\epsilon}\int_{\mathbb R^D} \ln\left({F}\right)  \mathcal{C}(F)\, d\bm v\leq 0
\end{align}
 It is also important that when the Boltzmann equation is at equilibrium ($\mathcal C(F) = 0$), the entropy equation (\ref{eq:entdissip}) also be at equilibrium. As such 
\begin{assumption} \label{assume:equilibria}  
\begin{equation}\label{eq:equilibria}
\int_{\mathbb R^D} \ln\left({F}\right)\,\mathcal{C}(F) \, d\bm v  = 0 
\implies 
\mathcal{C}(F) = 0
\end{equation}
\end{assumption}
Using (\ref{eq:collinvar}) we have that equality in (\ref{eq:dissipation}) holds if and only if $\ln(F)\in\mathscr{I}$ in accordance with (\ref{eq:collinvar}). 
As a result,
\begin{equation}
\label{eq:equilibrium}
\mathcal{C}(F) = 0
\quad\Leftrightarrow\quad
\int_{\mathbb R^D} \ln\left(F\right)\,\mathcal{C}(F)\, d\bm v  = 0
\quad\Leftrightarrow\quad
\ln\left(F\right)\in\mathscr{I}
\end{equation}
By virtue of (\ref{eq:collinvar}), the second equivalence indicates that the form of such local equilibria is given by 
\begin{equation}\label{eq:gaussian}
F = \,e^{a + b_iv_i + c|\bm v|^2}
\end{equation}
{ for some coefficients $a(t,\bm x)$, $b_i(t,\bm x)$ and $c(t,\bm x)$. Furthermore, integrability implies that $c<0$ }. 

\begin{remark}
Without loss of generality we may reparameterize (\ref{eq:gaussian}) as a Maxwellian distribution 
\begin{equation} \label{eq:maxwellian}
\mathcal{M}_{\rho,{\bm u},\theta}({\bm v}) :=
    \frac{\rho}{\left(2\pi\theta\right)^{D/2}}\exp\left(-\frac{|\bm v-\bm u|^2}{2\theta}\right)
\end{equation}
for some $(\rho, {\bm u}, \theta)\in\mathbb{R}_{+}\times\mathbb{R}^D\times\mathbb{R}_{+}$.
\end{remark}
For the rest of the text, we shall use $\mathcal M$ to refer to any arbitrary Maxwellian and only include the subscript when we want to emphasize the Maxwellian's parameterization by $\rho$, $\bm u$ and $\theta$. Furthermore, the mass, momentum and energy moments of a distribution $F$, given by Equations (\ref{eq:mass}), (\ref{eq:momentum}) and (\ref{eq:energy}) respectively, can be used to derive what is called the  self-consistent Maxwellian which we shall denote by  $\mathcal \mu(F) :=\mathcal M_{\rho_{\mu},\bm u_{\mu},\theta_{\mu}}$ where
\begin{equation}\label{eq:selfeq}
\begin{pmatrix}
\rho_{\mu}(F)
\\ 
\rho_{\mu}(F) \bm u_{\mu}(F)
\\
\rho_{\mu}(F) |\bm u_{\mu}(F)|^2 + D\rho_{\mu}(F)\theta_{\mu}(F)
\end{pmatrix}
:= 
\begin{pmatrix} 
\int_{\mathbb R^D} F \, d\bm v
\\ 
\int_{\mathbb R^D} \bm v F \, d\bm v
\\ 
\int_{\mathbb R^D}|\bm v|^2 F \, d\bm v
\end{pmatrix} 
=
\begin{pmatrix} 
\int_{\mathbb R^D} \mathcal \mu(F) \, d\bm v
\\ 
\int_{\mathbb R^D} \bm v\, \mu(F) \, d\bm v
\\ 
\int_{\mathbb R^D}|\bm v|^2\, \mu(F) \, d\bm v
\end{pmatrix}.
\end{equation}
In other words, $\mu(F)$ is the Maxwellian that has the same $\rho(t,\bm{x})$, $u_i(t, \bm{x})$ and $\theta(t, \bm{x})$ as $F$. Because $\rho(t,\bm{x})$, $u_i(t, \bm{x})$ and $\theta(t, \bm{x})$ completely determine a Maxwellian, we also have that $F=\mu(F)$ if and only if $F$ is a Maxwellian.

\subsection{The Linearized Collision Operator}
\label{sec:lincoll}
{ To derive macroscopic fluid equations from (\ref{eq:be}) near equilibrium, we assume that the distribution $F$ is a perturbation in Knudsen number from some Maxwellian}. 
\begin{equation}\label{eq:renorm}
F(t,\bm x,\bm v) = \mathcal{M}(t,\bm x,\bm v)\, \big(1 + \epsilon f(t, \bm{x}, \bm{v})\big).
\end{equation}
A corresponding Taylor series expansion of the collision operator about the Maxwellian $\mathcal{M}$ gives 
\begin{equation}\label{eq:taylorcoll}
\mathcal C (\mathcal{M}(1 + \epsilon f)) = 
\epsilon \left.\frac{d}{d\epsilon}\left[\mathcal{C}\big(\mathcal{M}(1 + \epsilon f)\big) \right]\right|_{\epsilon=0}+ \frac{\epsilon^2}{2}  \left.\frac{d^2}{d\epsilon^2}\left[\mathcal{C}\big(\mathcal{M}(1 + \epsilon f)\big) \right]\right|_{\epsilon=0} + O(\epsilon^3).
\end{equation}
Substituting (\ref{eq:renorm}) and (\ref{eq:taylorcoll}) into the Boltzmann equation (\ref{eq:be}) gives 
\begin{equation}\label{eq:Elinearbe}
(\text{St}\,\partial_t  + v_i \partial_{x_i})
[\mathcal{M}+\epsilon \mathcal{M} f]
 = \mathcal{M}\mathcal{L}_{\mathcal M}[f] +  \frac{\epsilon}{2}  \left.\frac{d^2}{d\epsilon^2}\left[\mathcal{C}\big(\mathcal{M}(1 + \epsilon f)\big) \right]\right|_{\epsilon=0} + O(\epsilon^2)
\end{equation}
where the linearized collision operator
\begin{equation}\label{eq:lincoll}
\mathcal{L}_{\mathcal{M}}[f] := \frac{1}{\mathcal M} \left.\frac{d}{d\epsilon}\left[\mathcal{C}\big(\mathcal{M}(1 + \epsilon f)\big) \right]\right|_{\epsilon=0}
\end{equation}
is of particular importance and thus requires further study.

\begin{remark}
    For the bilinear collision operators $\mathcal{C}(F) = \mathcal{Q}(F, F)$, the linearized collision operator takes the form
    $$\mathcal{L}_{\mathcal{M}}^{\mathcal{Q}}[f] = \frac{1}{\mathcal{M}}\,\Big(\mathcal{Q}(\mathcal{M}f, \mathcal{M}) +\mathcal{Q}(\mathcal{M}, \mathcal{M}f) \,\Big) $$
    For example, the Hard Spheres operator will be 
    $$\mathcal{L}_{\mathcal{M}}^{\text{HS}}[f]= \frac{\epsilon}{2\pi\rho_0\ell\sqrt{2}}\int_{\mathbb{R}^3\times\mathbb{S}^2} \mathcal{M}(\bm{v}_*)\Big(f(\bm{v}{'})+ f(\bm v{'}_*) -f(\bm v_*) -f(\bm v)  \Big)\big|\bm{\hat\eta}\cdot(\bm{v}_*-\bm{v})\big|\ d\bm{\hat\eta}\, d\bm v_* $$
    where 
    $\bm{v}' = \bm{v}+ \big(\bm{\hat\eta}\cdot(\bm{v}_*-\bm{v})\big)\bm{\hat{\eta}}$ and $\bm{v}'_* = \bm{v}_*- \big(\bm{\hat\eta}\cdot(\bm{v}_*-\bm{v})\big)\bm{\hat{\eta}}$, $\ell$  is the mean free path of the gas. {\color{black} Note that $\ell = \frac{m}{d_m^2\rho_0\pi\sqrt{2}}$ where $d_m$ is the diameter of the gas particle and $m$ is the mass.}    
    
With the Bhatnagar-Gross-Krook (BGK) operator ~\cite{bhatnagar:1954hc, SoneBook}\! 
, 
$ \mathcal{C}^{\mathrm{BGK}}(F) = \frac{\mu(F) - F}{\tau(F)}$ where $\tau(F)$ assumed to not depend on $\boldsymbol{v}$, the linearized collision operator will be
\begin{equation}\label{eq:linBGKintro}
    \mathcal{L}^{\mathrm{BGK}}_{\mathcal M}[f] 
= 
- \frac{1}{\tau(\mathcal{M})}\left(\mathrm{Id}-\Pi_{\mathcal M} \right)[f]
\end{equation}
where $\mathrm{Id}$ is the identity operator and given a Maxwellian $\mathcal{M}$ with conservation moments $(\varrho,\, \boldsymbol{w}, \vartheta)$ and for any function $g \in \mathscr{L}^2(\mathcal{M}d\boldsymbol{v}) $
\begin{multline}\label{eq:orthproj}
    \Pi_{\mathcal M} [g] := \frac{1}{\varrho}\int_{\mathbb R^D} g\mathcal M \,d\bm v 
    +
    \frac{\bm v-\bm w}{\varrho\vartheta}
    \cdot
    \int_{\mathbb R^D} 
    (\bm v- \bm w)g\mathcal M \, d\bm v
    \\ 
    + 
    \left( \frac{|\bm v- \bm w|^2}{2\vartheta} -\frac{D}{2} \right) 
    \frac{2}{D\varrho}
    \int_{\mathbb R^D} 
    \left( \frac{|\bm v- \bm w|^2}{2\vartheta} -\frac{D}{2} \right) g\mathcal M
    \,d\bm v,
\end{multline}
is an orthogonal projection from $\mathscr{L}^2(\mathcal{M}d\boldsymbol{v})$ onto $\mathscr{I}$.
A proof that the expression \eqref{eq:orthproj} is indeed an orthogonal projection can be found in Appendix \ref{appendix:projection} and the calculation to derive \eqref{eq:linBGKintro} can be found in Appendix \ref{sec:BGK}.

\end{remark}

\subsubsection{Properties of the Linearized Collision Operator}
For our purposes, we require that the linearized collision operator satisfy a set of properties we will use extensively. 
Given an arbitrary Maxwellian $\mathcal{M}_{\rho, \bm{u},\theta}$, we denote the space of functions that are square-integrable in $\bm{v}$ when weighted by $\mathcal{M}_{\rho,\bm{u}, \theta}$ by $\mathscr{L}^2(\mathcal{M}\,d\bm{v})$. We also denote by $\mathcal{O}_{\bm{O}}$ the composition of operators $ \mathcal{T}_{\bm{u}}\mathcal{T}_{\bm O}\mathcal{T}_{\bm{u}}^{-1}$. The linearized collision operators $\mathcal{L}_{\mathcal{M}}$ we deal with are such that:
\begin{linassumption}\label{assume:selfadj}
$ \mathcal{L}_{\mathcal{M}}:\mathscr{L}^2(\mathcal{M}\,d\bm{v})\longrightarrow\,\mathscr{L}^2(\mathcal{M}\,d\bm{v}) $ is a self-adjoint operator with respect to the weighted inner product of $\mathscr{L}^2(
\mathcal{M}\,d\bm{v})$.
\end{linassumption}

\begin{linassumption} \label{assume:linsymm} The linearized collision operator commutes with $\mathcal{O}_{\bm O}$ i.e. 
$\\ \mathcal{O}_{\bm O} \mathcal{L}_{\mathcal{M}} [f] =\mathcal{L}_{\mathcal{M}}[\mathcal{O}_{\bm O}  f]$.
This property is inherited from the collision operator $\mathcal{C}$ because if we apply \eqref{eq:taylorcoll} to both sides of \eqref{eq:GalilInvar} we get \begin{multline}\label{eq:linGalilinvar}
\epsilon\, \mathcal{O}_{\bm O}\left[\left.\frac{d}{d\epsilon}\mathcal{C}\big(\mathcal{M}(1 + \epsilon f)\big) \right|_{\epsilon=0}\right] + \frac{\epsilon^2}{2} \mathcal{O}_{\bm{O}}\left[ \left.\frac{d^2}{d\epsilon^2}\mathcal{C}\big(\mathcal{M}(1 + \epsilon f)\big) \right|_{\epsilon=0}\right] + O(\epsilon^3) \\
= \epsilon\left.\frac{d}{d\epsilon}\mathcal{C}\left(\mathcal{O}_{\bm{O}}\Big[\mathcal{M}(1 + \epsilon f)\Big]\right) \right|_{\epsilon=0} + \frac{\epsilon^2}{2}  \left.\frac{d^2}{d\epsilon^2}\mathcal{C}\big(\mathcal{O}_{\bm{O}}\left[\mathcal{M}(1 + \epsilon f)\right]\big) \right|_{\epsilon=0} + O(\epsilon^3),
\end{multline}
Since $\epsilon$ is arbitrary and $\mathcal{O}_{\bm{O}}\mathcal{M} = \mathcal{M}$, the property follows immediately. This property greatly simplifies the form in which the closures for the deviatoric stress and heat flux take.
\end{linassumption}
\begin{linassumption}$\mathcal{L}_{\mathcal{M}}$ is negative semi-definite on $\mathscr{L}^2(\mathcal{M}\,d\bm{v})$.
This property is inherited from the dissipation property of the collision operator (Equation \eqref{eq:dissipation}) because 
\begin{equation}\label{eq:lindissipation}
\int_{\mathbb R^D} f\mathcal{M}\mathcal{L}_{\mathcal{M}}[f]\,d\bm v 
= 
\frac{1}{2}\frac{d^2}{d\epsilon^2} \left[ 
\left.\int_{\mathbb R^D} 
\ln\left({\mathcal M +\epsilon \mathcal M f}\right) \mathcal{C}\left({\mathcal M + \epsilon \mathcal M f}\right)
\, d\bm v
\right|_{\epsilon=0}\right]\leq0 
\end{equation}
where the inequality in \eqref{eq:lindissipation} follows from the fact that we attain a local maximum at $\epsilon = 0$ whilst the equality in \eqref{eq:lindissipation} can be obtained by a Taylor series expansion of the middle term
\begin{align}\label{eq:secondvar}
\frac{d^2}{d\epsilon^2}
\left[\left. 
\int_{\mathbb R^D}  
\left(\overbrace{\ln\left({\mathcal M} \right)}^{\in \mathscr{I}} +\overbrace{\ln(1+\epsilon f)}^{\epsilon f + O(\epsilon^2)}\right)\, \overbrace{\mathcal{C}(\mathcal M +\epsilon \mathcal M f)}^{\text{Equation \eqref{eq:taylorcoll}}}
\,d\bm v
 \right]\right|_{\epsilon=0}
=  
\int_{\mathbb R^D}
2 f\,\left.\frac{d}{d\epsilon}\mathcal{C}\big(\mathcal{M}(1 + \epsilon f)\big) \right|_{\epsilon=0}
\, d\bm v
\end{align}
\end{linassumption}
\begin{linassumption}\label{assume:ker}
The kernel of $\mathcal{L}_{\mathcal M}$ is the space of collision invariants $\mathscr{I}$.\\
This motivates the decomposition $\mathscr{L}^2(\mathcal{M}\,d\bm{v}) = \mathscr{I} \oplus \mathscr{I}^{\perp_{\mathcal{M}}}$, where $\mathscr{I}^{\perp_{\mathcal{M}}}$ is the orthogonal complement to the space of collision products with respect to the $\mathscr{L}^2(\mathcal{M}\,d\bm{v})$ inner product.
\end{linassumption} 
\begin{linassumption}[Fredholm Alternative]\label{assume:fredholm}
The linearized collision operator with domain restricted to $\mathscr{I}^{\perp_{\mathcal{M}}}$ is invertible. That is, the equation $\mathcal{L}_{\mathcal{M}}[f] = g$ has a unique solution if and only if $f, \,g\in \mathscr{I}^{\perp_{\mathcal{M}}}$. We denote the inverse by $\mathcal{L}^{-1}_{\mathcal{M}}$ and note that it can be extended to the rest of $\mathscr{L}^2(\mathcal{M}\,d\bm{v})$ if we assert $\mathscr{I}$ as its kernel. Abusing notation, we will also denote the extension by $\mathcal{L}^{-1}_{\mathcal{M}}$.
Thus for any $g\in \mathscr{L}^2(\mathcal{M}\,d\bm{v}) $,
$$\mathcal{L}_{\mathcal M}\mathcal{L}^{-1}_{\mathcal M}[g] = ({\mathrm{Id}} - \PI)[g] $$
where $\Pi_{\mathcal{M}}$ is given by \eqref{eq:orthproj}.
\end{linassumption}

 Of the properties outlined above, the Fredholm alternative property is the most restrictive since it does not apply to collision operators such as the soft-sphere collision operators. It is however known to hold for hard potential and Maxwell molecule operators  \cite{Grad1963,saint-raymond2014,golse2005} as well as the BGK operator by virtue of it being a scaled orthogonal projection onto $\mathscr{I}^{\perp_{\mathcal{M}}}$. 

\section{Variational Multiscale Moment Closures}
The first step will be to apply the projection operator $\Pi_\mathcal{M}$ to the equation

\begin{equation}\label{eq:base}
\frac{1}{\mathcal{M}}\left(\text{St}\,\partial_t + v_i\partial_{x_i} \right)[\mathcal{M} + \epsilon\mathcal{M}f] = \mathcal{L}_{\mathcal{M}}[f]
\end{equation} 
in order to separate it into a coarse-scale equation made up of the components of (\ref{eq:base}) from the space of collision invariants $\mathscr{I}$ and a fine-scale equation made up of components of (\ref{eq:base}) from the orthogonal complement of $\mathscr{I}^{\perp_{\mathcal{M}}}$:
\begin{subequations}\label{eq:projeqns}
\begin{align}
    \text{St }\partial_t
\ln\left({\mathcal{M}}\right) + \Pi_{\mathcal{M}}\left[ v_i\partial_{x_i}\ln({\mathcal{M}}) + \frac{\epsilon}{\mathcal{M}}\left(\text{St }\partial_t + v_i\partial_{x_i}\right)[\mathcal{M}f]\right] &= 0\label{eq:coarseproj}\\
(\text{Id} - \Pi_{\mathcal{M}})\left[ v_i\partial_{x_i}\ln({\mathcal{M}}) + \frac{\epsilon}{\mathcal{M}}\left(\text{St }\partial_t + v_i\partial_{x_i}\right)[\mathcal{M}f]\right]  &= \mathcal{L}_\mathcal{M}[f]
\end{align}
\end{subequations}
When we test the first equation with an arbitrary element of $\mathscr{I}$, denoted by $\bar{m}$, in the $\mathscr{L^2}(\mathcal{M}d\bm{v})$ inner product, we obtain conservation equations of the form:
\begin{equation}\label{eq:conserweak}
    \left\langle \bar{m},\text{St }\partial_t \mathcal{M} \right\rangle + \left\langle \bar{m}, v_i\partial_{x_i}\mathcal{M} \right\rangle + \epsilon \left\langle \bar{m}, (\text{St }\partial_t+v_i\partial_{x_i})[\mathcal{M}f]\right\rangle = 0
\end{equation}
Where we now use angular brackets to denote integration in $\bm{v}$ over $\mathbb{R}^D$
\[
\langle h(\bm v),g(\bm v)\rangle := \int_{\mathbb R^D} h(\bm v)\,g(\bm v)\,d\bm v
\]
The second equation can be re-written as 
\begin{align}\label{eq:finescaleeqn}
  (\text{Id} - \Pi_{\mathcal{M}})[f] = \mathcal{L}_\mathcal{M}^{-1} (\text{Id} - \Pi_{\mathcal{M}})\Bigg[ v_i\partial_{x_i}\ln({\mathcal{M}}) + \frac{\epsilon}{\mathcal{M}}\left(\text{St }\partial_t + v_i\partial_{x_i}\right)[\mathcal{M}f]\Bigg]
\end{align}
 giving an equation for the term $ (\text{Id} - \Pi_{\mathcal{M}})[f] $. As will be elaborated upon in the coming sections, the basic idea for deriving variational multiscale moment closures is to substitute a suitable approximation of \eqref{eq:finescaleeqn} into the conservation equation \eqref{eq:conserweak}.


We work with two types of Maxwellian within the variational multiscale framework. The first is  constant background $M(\bm{v}) :=\mathcal{M}_{\rho_0, \bm{0}, \theta_0}$, which we will use to elaborate on the closure derivation process for the linearized Boltzmann equation. The second is the self-consistent Maxwellian $\mu(F)$ to the distribution $F$ which gives us closures for the conservation equations in their general form. For brevity, we shall denote this Maxwellian by $\mu$ from now on. In both cases, the equations \eqref{eq:projeqns} simplify greatly.

\begin{remark}
Note that \eqref{eq:base} leaves out the higher order terms like $\\  \left.\frac{d^2}{d\epsilon^2}\left[\mathcal{C}\big(\mathcal{M}(1 + \epsilon f)\big) \right]\right|_{\epsilon=0} $. This simplifies the analysis without fundamentally changing its form. In particular, the coarse-scale equation \eqref{eq:coarseproj} remains unchanged whilst the fine-scale equation incorporates the higher-order collision terms in a straightforward manner:\footnote{An expression very similar to \eqref{eq:completefinescale} can be found in Section 1.1.3 of \cite{saint-raymond2014}. }
\begin{multline}\label{eq:completefinescale}
  (\text{Id} - \Pi_{\mathcal{M}})[f] = \mathcal{L}_\mathcal{M}^{-1} (\text{Id} - \Pi_{\mathcal{M}})\Bigg[ v_i\partial_{x_i}\ln({\mathcal{M}}) + \frac{\epsilon}{\mathcal{M}}\left(\text{St }\partial_t + v_i\partial_{x_i}\right)[\mathcal{M}f]\\ -\frac{\epsilon}{2\mathcal{M}} \left.\frac{d^2}{d\epsilon^2}\left[\mathcal{C}\big(\mathcal{M}(1 + \epsilon f)\big) \right]\right|_{\epsilon=0} + O(\epsilon^2)\Bigg]
\end{multline}
Thus a more complete closure procedure will involve adding terms that incorporate these higher order collision terms to whatever approximation of \eqref{eq:finescaleeqn} we come up with.
\end{remark}

\subsection{Linear Theory: The Constant Background Maxwellian Formulation}
With $\mathcal{M} = M$, we obtain the linearized Boltzmann equation from \eqref{eq:base}
\begin{equation}\label{eq:linBoltz}
    \left(\text{St}\,\partial_t + v_i\partial_{x_i} \right)[f] = \frac{1}{\epsilon}\, \mathcal{L}_{{M}}[f]
\end{equation}
Noting that $ \left\langle \bar{m},  M\,\partial_t (\text{Id} - \Pi_M)[f]\right\rangle = 0$, the conservation equations (\ref{eq:conserweak}) take the form  
\begin{equation}\label{eq:linearconser}
     \left\langle \bar{m},  M\,(\text{St }\partial_t +v_i\partial_{x_i})\bar{f}\right\rangle + \left\langle \bar{m},  M\,v_i\partial_{x_i}\breve{f}\right\rangle = 0
\end{equation}
where $\bar{f} := \Pi_M[f]$ and  $\breve{f} := (\text{Id} - \Pi_M)[f]$ and the fine-scale equation \eqref{eq:finescaleeqn} simplifies to 
\begin{equation}\label{eq:linfinescale}
 \breve{f} = \epsilon\, \mathcal{L}_M^{-1} (\text{Id} - \Pi_{M})\left[v_i\partial_{x_i}\bar{f} + \left(\text{St }\partial_t + v_i\partial_{x_i}\right) \,\breve{f}\,\right]
\end{equation}
Recalling the expressions \eqref{eq:linsigmaclose} and \eqref{eq:linqclose} for the linearized deviatoric stress and heat flux and observing that $(\text{Id} - \Pi_M)[v_iv_j] = v_iv_j - \frac{1}{D}|\bm{v}|^2\delta_{ij}$ and $(\text{Id} - \Pi_M)[|\bm{v}|^2v_i] = |\bm{v}|^2v_i - (D+2)\theta_0 v_i\,$, we see that the expressions for the deviatoric stress and heat flux are completely determined by the $ \langle v_i\,\bar{m},  M\,\partial_{x_i}\breve{f}\rangle$ term in \eqref{eq:linearconser}. This means that the process of closing the system of conservation equations derived from the Boltzmann equation amounts to a substitution of some approximation of \eqref{eq:linfinescale} into \eqref{eq:linearconser}. Our goal in this text is to ensure that the fine scale approximation is \textit{entropy stable}.


With $\bar{m} = \bar{f}$ in \eqref{eq:linearconser}, we have that
\begin{align} \label{eq:linentstabeqn}
\frac{\text{St}}{2}\partial_t\left\langle \bar{f},  M\bar{f} \right\rangle\, +\frac{1}{2}\partial_{x_i}\left\langle v_i\bar{f},  M\bar{f}\right\rangle + \partial_{x_i}\left\langle v_i\bar{f},  M\,\breve{f}\right\rangle  - \left\langle v_i\partial_{x_i}\bar{f}, M\, \breve{f} \right\rangle=0
\end{align}
An entropy stable (or entropy dissipative) closure refers to a fine-scale approximation $\breve{f}$ such that the spatial integral of 
$\left\langle v_i\partial_{x_i}\bar{f}, M\, \breve{f} \right\rangle$ is non-positive. This is because under the assumption that any terms in divergence form can be ignored (for example in an infinite spatial domain), equation \eqref{eq:linentstabeqn} gives
\begin{equation}\label{eq:linentineq}
   \text{St } \frac{d}{dt}\int_{\Omega}\frac{1}{2}\left\langle \bar{f},  M\bar{f} \right\rangle\,d\bm{x} \,=\, \int_{\Omega}\left\langle v_i\partial_{x_i}\bar{f}, M\, \breve{f} \right\rangle\,d\bm{x}\le 0
\end{equation}
The non-negative term $\frac{1}{2}\left\langle \bar{f},  M\bar{f} \right\rangle$ defines the macroscopic entropy for the closed conservation equations derived from the approximation $\breve{f}$ and \eqref{eq:linentineq} says that this entropy is non-increasing in time.\footnote{To obtain the \textit{non-decreasing} entropy more commonly used in physics, we would use $-\frac{1}{2}\left\langle \bar{f},  M\bar{f} \right\rangle$ instead of $\frac{1}{2}\left\langle \bar{f},  M\bar{f} \right\rangle$ as our macroscopic entropy.} 

An important motivation for \eqref{eq:linentineq} comes from testing the linearized Boltzmann equation \eqref{eq:linBoltz} with $f$ in the $\mathscr{L}^2(M\,d\bm{v})$ inner-product and integrating in space (ignoring the divergence term) to get 
\begin{equation}\label{eq:linboltzineq}
     \text{St } \frac{d}{dt}\int_{\Omega}\frac{1}{2}\left\langle {f},  M{f} \right\rangle\,d\bm{x} \,=\, \frac{1}{\epsilon}\int_{\Omega}\left\langle {f}, M\, \mathcal{L}_M[f] \right\rangle\,d\bm{x}\le 0
\end{equation}
The idea is to view the conservation equations as a weak form (in $\bm{v}$) of the (linearized) Boltzmann equation. The process of closing these conservation equations then amounts to solving for a finite-dimensional Galerkin approximation $\bar{f}$ of the unknown $f$ with the entropy stability criterion in \eqref{eq:linentineq} serving to ensure that $\bar{f}$ satisfies a "discrete" version of the entropy inequality \eqref{eq:linboltzineq}  that $f$ satisfies.  
\begin{remark}
Using \eqref{eq:linmacroparam} and \eqref{eq:orthproj}, we can show that  $$\bar{f} = \frac{\tilde{\rho}}{\rho_0} + \frac{\bm{v}\cdot\bm{\tilde{u}}}{\theta_0} + \left(\frac{|\bm v  |^2}{2\theta_0}-\frac{D}{2}\right) \frac{\tilde\theta}{\theta_0}$$ 
As such, the macroscopic entropy is given by
$$\frac{1}{2}\bracke{\bar{f},\, M\,\bar{f}} = \frac{1}{2\rho_0}\,\tilde{\rho}^2 + \frac{\rho_0}{2\theta_0}\,|\bm{\tilde{u}}|^2 + \frac{D\rho_0}{4\theta_0^2}\,\tilde{\theta}^2$$
\end{remark}

\subsection{Self-consistent Maxwellian formulation}

With $\mathcal{M} = \mu$, we have that $f$ belongs in the orthogonal complement $\mathscr{I}^{\perp_\mu}$ because by the definition of $\mu$, 
$$ \langle\bar{m}, \mu \rangle = \langle\bar{m}, F \rangle =\langle \bar{m}, \mu+ \epsilon\mu f \rangle $$
and thus  $ \,\langle\bar{m}, \mu \,f \rangle = 0\,$. Furthermore, we can show that $\langle \bar{m}, \partial_t[\mu \,f]\rangle = 0$. 
Thus the conservation equation (\ref{eq:conserweak}) takes the form
\begin{equation}\label{eq:nonlinearconser}
    \left\langle \bar{m},\text{St }\partial_t \mu \right\rangle + \left\langle \bar{m}, v_i\partial_{x_i}\mu \right\rangle + \epsilon \left\langle \bar{m},  v_i\partial_{x_i}[\mu\,f]\right\rangle = 0
\end{equation}
and the fine-scale equation takes the form
\begin{equation}\label{eq:nonlinfinescale}
 f = \mathcal{L}_\mu^{-1} (\text{Id} - \Pi_{\mu})\left[ v_i\partial_{x_i}\ln({\mu}) + \frac{\epsilon}{\mu}\left(\text{St }\partial_t + v_i\partial_{x_i}\right)[\mu \,f]\right]
\end{equation}
As with the linear case, the closure for the deviatoric stress and heat flux is wholly determined by the approximation $f$ through the fine-scale equation. If we use $\bar{m} = \ln(\mu)$ in \eqref{eq:nonlinearconser}, the resulting equation can be written as 
\begin{equation}\label{eq:nonlinentstabeqn}
       \text{St }\partial_t\left\langle \left(\mu\ln\left( {\mu}\right) - \mu\right), 1\right\rangle + \partial_{x_i} \left\langle \left(\mu\ln\left({\mu}\right) - \mu\right), v_i\right\rangle + \epsilon\,\partial_{x_i}\left\langle v_i \ln\left({\mu}\right), \mu\,{f}\right\rangle - \epsilon\,\left\langle v_i \partial_{x_i}\ln\left({\mu}\right),\mu\, {f}\right\rangle = 0 
\end{equation}
Entropy stability in this context refers to when the approximation of $f$ is such that the term $\epsilon\,\left\langle v_i \partial_{x_i}\ln\left({\mu}\right),\mu\, {f}\right\rangle$ is non-positive. This is because under circumstances in which we can do away with terms in divergence form when we integrate in space, we get
\begin{equation}\label{eq:nonlinearineq}
    \text{St }\frac{d}{dt} \int_{\Omega} \left\langle \left(\mu\ln\left({\mu}\right) - \mu\right), 1\right\rangle \, d\bm{x}  = \epsilon\,\int_\Omega \left\langle v_i \partial_{x_i}\ln\left({\mu}\right),\mu\, {f}\right\rangle\,d\bm{x} \le 0
\end{equation}
where $\left\langle \left(\mu\ln\left({\mu}\right) - \mu\right), 1\right\rangle$ is now the macroscopic entropy for the resulting conservation equations. This inequality is the "discrete" analogue to the local entropy dissipation law \eqref{eq:entdissip} which when integrated in space under similar conditions gives
$$  \text{St }\frac{d}{dt} \int_{\Omega} \left\langle \left(F\ln\left({F}\right) - F\right), 1\right\rangle\,d\bm{x} = \frac{1}{\epsilon} \int_{\Omega} \bracke{\ln(F),\, \mathcal{C}(F)} \,d\bm{x} \le 0$$
We can also draw a direct comparison with the linear case by making the substitution $\bar{f} = \ln(\mu)$ and $\breve{f} = \epsilon f$ into \eqref{eq:nonlinearconser} and \eqref{eq:nonlinfinescale} to get
\begin{align*}
     \bracke{\bar{m},\, \mu\, (\partial_t +v_i\partial_{x_i})[\bar{f}\,]} + \bracke{\bar{m}, \mu\, v_i\partial_{x_i}\breve{f}} + \bracke{\bar{m},\, \mu\, \breve{f}\,v_i\partial_{x_i}\bar{f}} = 0\\
\breve{f} = \epsilon\, \mathcal{L}^{-1}_{\mu}(\text{Id} - \Pi_{\mu})\left[ v_i\partial_{x_i}\bar{f} + \left(\text{St }\partial_t + v_i\partial_{x_i}\right)\breve{f} + \breve{f}\,v_i\partial_{x_i}\bar{f}\right] 
\end{align*}
We see that the primary differences from the linear case are in the fact that the Maxwellian depends on the coarse-scale term (i.e. $\mu = e^{\bar{f}}\,$) as well as the additional non-linear term $\breve{f}v_i\partial_{x_i}\bar{f}$ in both the coarse and fine scale equations. We would then think of the coarse-scale term $\ln(\mu)$ as a finite-dimensional Galerkin approximation to $\ln(F)$. It is, however, often more convenient to work with $\mu$ and $f$ for the self-consistent formulation.

\begin{remark}
In terms of the macroscopic variables, the "discrete" entropy is 
$$\left\langle \left(\mu\ln\left( {\mu}\right) - \mu\right), \, 1\right\rangle = \rho \left(\ln\left(\frac{\rho}{(2\pi \theta)^{\frac{D}{2}} }  \right) - \frac{D+2}{2}\right) $$
\end{remark}

In the sections that follow, we shall use the framework developed here to describe the Chapman-Enskog expansion, showing how entropy stability exists for the Euler and Navier-Stokes-Fourier equations but not for the Burnett equations and beyond. We will then describe a way to induce entropy stability beyond the  Navier-Stokes-Fourier equations.

\subsection{Chapman-Enskog Closures}
The first step to deriving the classical Chapman-Enskog closure \cite{chapman1990mathematical} from equations (\ref{eq:nonlinfinescale}) and (\ref{eq:linfinescale}) is to assume that the orthogonal complement term can be written as a formal power series:
\begin{equation}\label{eq:CE_power}
    \breve{f}(t, \bm{x}, \bm{v}) = \sum_{n=0}^\infty \epsilon^n \breve{f}_n(t, \bm{x}, \bm{v})
\end{equation}
If we substitute the power series into (\ref{eq:nonlinfinescale}) and (\ref{eq:linfinescale}) and arrange the resulting sequence of equations in orders of $\epsilon$, we get respectively
\begin{subequations}\label{eq:ce_seq}
\begin{align}
\label{eq:lin_ce_seq}
    {\breve f_0 =0; \quad \breve f_1 = \mathcal L^{-1}_{M}\left[v_i\partial_{x_i}\bar f\right];}
    &\quad
    \breve f_{n+1} = \mathcal L^{-1}_{M}
        (\text{St }\partial_t + v_i \partial_{x_i}) 
         \big[\breve f_n\big]
         \\
\label{eq:nonlin_ce_seq}
     f_0 = \mathcal L^{-1}_{\mu}\left[v_i\partial_{x_i}\ln({\mu})\right];
    &\quad 
     f_{n+1} = \mathcal L^{-1}_{\mu}
    \left[
        \frac{1}{\mu}(\text{St }\partial_t + v_i \partial_{x_i}) 
        [\mu\,  f_n]
    \right];
\end{align}
\end{subequations}
Truncating the power series at different orders of $\epsilon$ leads to different closures for the conservation equations. For example the zeroth order truncation gives the Navier-Stokes-Fourier equations in the self-consistent Maxwellian formulation and the linearized Euler equations in the background Maxwellian formulation.
For our purposes, it is more advantageous to obtain the Chapman-Enskog closure through a fixed point iteration on (\ref{eq:nonlinfinescale}) and (\ref{eq:linfinescale}) that generates the partial sums of the power series \eqref{eq:CE_power}:
\begin{subequations}
\label{eq:ce_fixed_point}
\begin{align}
   \label{eq:lince_fixed_point} \breve f_{(n+1)} 
    &=  \epsilon\,\mathcal L^{-1}_M
    \left[
    (\text{St }\partial_t + v_i \partial_{x_i}) 
    [\breve f_{(n)} ]
    + 
    v_i\partial_{x_i} \bar f\,
    \right]; \qquad \breve{f}_{(0)} = 0\\
    \label{eq:nonlince_fixed_point} f_{(n+1)}
    &= \mathcal L^{-1}_{\mu}
    \left[\frac{\epsilon}{\mu}
   {(\text{St }\partial_t + v_i \partial_{x_i}) 
     [\mu\, f_{(n)}]} 
    + 
    v_i\partial_{x_i}\ln\left({\mu}\right)
    \right]; \qquad f_{(0)} = 0
\end{align}
\end{subequations}

with
\begin{align*}
   \breve{f}_{(n+1)}= \sum_{j = 1}^{n+1} \epsilon^j \breve{f}_j; \qquad
   {f}_{(n+1)} = \sum_{j = 0}^n \epsilon^j {f}_j 
\end{align*}

\subsubsection{Euler Equations}
The Euler equations arise from using $ {f}_{(0)} = \breve{f}_{(0)} = 0$ as our closure for the fine-scale term. Thus the conservation equations (\ref{eq:linearconser}) and (\ref{eq:nonlinearconser}) become 
\begin{align}\label{eq:eulerweak}
\left\langle \bar{m},\text{St }\partial_t \mu \right\rangle + \left\langle \bar{m}, v_i\partial_{x_i}\mu \right\rangle &= 0\\
  \left\langle \bar{m},  M\,\text{St }\partial_t \bar{f}\right\rangle  +\left\langle \bar{m},  M\,v_i\partial_{x_i}\bar{f}\right\rangle  &= 0
\end{align}
with $\bar{m} = 1,\, v_i \text{ and } \frac{|v^2|}{2}$ macroscopic conservation of mass, momentum and energy explicitly take the form of \eqref{eq:linmacroconslaws} and \eqref{eq:macroconslaws} with $\sigma_{ij} = \tilde{\sigma}_{ij} = 0$ and $q_i = \tilde{q_i} = 0$.
The corresponding macroscopic entropy equations are given by
\begin{align}\label{eq:eulerent}
\frac{d}{dt} \int_{\Omega} \frac{1}{2}\left\langle \bar{f},  M\, \bar{f}\right\rangle   &= 0\\
 \frac{d}{dt}\int_\Omega \left\langle 1, \left(\mu\ln\left({\mu}\right) - \mu\right)\right\rangle  &= 0
\end{align}
which means that these equations are entropy stable. 

\subsubsection{Navier-Stokes-Fourier equations}
The Navier-Stokes-Fourier equations arise from using the correction directly above that of the Euler equations: 
\begin{align*}
  \breve{f}_{(1)} = \epsilon\mathcal{L}_M^{-1}\left[v_i \partial_{x_i}\bar{f}\, \right]   \quad \text{and} \quad f_{(1)} = \mathcal{L}_\mu^{-1}\left[v_i \partial_{x_i} \ln\left({\mu} \right) \right]
\end{align*}
This results in conservation equations (\ref{eq:macroconslaws}) and (\ref{eq:linmacroconslaws}) with deviatoric stress tensor and heat flux given by a Newtonian stress-tensor satisfying the Stokes hypothesis and Fourier's law of heat conduction respectively:
\begin{align*}
    \tilde{\sigma}_{ij}^{(1)} &= \tilde{\omega}\,\left(\partial_{x_j} \tilde{u}_i + \partial_{x_i} \tilde{u}_j - \frac{2}{D}\partial_{x_k} \tilde{u}_k\,\delta_{ij}\right) \quad &\tilde{q}^{(1)}_i = -\tilde{\kappa}\, \partial_{x_i}\tilde{\theta}\nonumber\\\sigma_{ij}^{(1)} &= \omega\,\left(\partial_{x_j} u_i + \partial_{x_i} u_j - \frac{2}{D}\partial_{x_k} u_k\,\delta_{ij}\right) \quad &q^{(1)}_i = -\kappa\, \partial_{x_i}\theta\nonumber
\end{align*}
where the viscosity and heat conductivity\footnote{{\color{black} Strictly speaking, due to the use of the re-scaled temperature $\theta =\frac{k_B T}{m}$ in place of the absolute temperature $T$, the heat conductivity defined here differs from the usual heat conductivity by a factor $\frac{k_B}{m}$. }} are given by
{\color{black}
 \begin{align}
    \tilde{\omega} \,&= - \frac{\epsilon}{\theta_0 }\left\langle A_{12}^M, M\, \mathcal{L}_{M}^{-1}\left[A_{12}^M\right] \right\rangle &(\ge 0)\nonumber\\
    \omega(t, \bm{x}) &= - \frac{\epsilon}{\theta}\left\langle A_{12}^\mu, \mu\, \mathcal{L}_{\mu}^{-1}\left[A_{12}^\mu\right] \right\rangle\, &(\ge0)\label{eq:omegadef}\\ 
    \nonumber\tilde{\kappa}\, &= -\frac{\epsilon}{\theta_0^2}\left\langle B_1^M, M \,\mathcal{L}^{-1}_M \left[ B_1^M\right] \right\rangle \,&(\ge 0)\\
    \kappa(t, \bm{x}) &= -\frac{\epsilon}{\theta^2}\left\langle B_1^\mu, \mu \,\mathcal{L}^{-1}_\mu \left[ B_1^\mu\right] \right\rangle\,&(\ge 0)\label{eq:kappadef}  
\end{align}
The details for this derivation  and the definitions of the tensors $A_{ij}^\mathcal{M}$ and $B_{i}^\mathcal{M}$ are given in Appendices \ref{appendix:alphabet} and \ref{appendix:alphabet2}.
}

The entropy relations read 
\begin{align*}
    \text{St }\frac{d}{dt}\int_\Omega\frac{1}{2}\left\langle \bar{f},  M\bar{f} \right\rangle\,d\bm{x}  &= \epsilon\int_{\Omega} \left\langle v_i\partial_{x_i}\bar{f}, M\, \mathcal{L}^{-1}_M\left[ v_i\partial_{x_i} \bar{f}\right] \right\rangle\,d\bm{x} \le 0\\
        \text{St }\frac{d}{dt}\int_{\Omega}\left\langle \left(\ln\left( {\mu}\right) - 1\right), \mu\right\rangle \, d\bm{x} &=\epsilon\int_\Omega \,\left\langle v_i \partial_{x_i}\ln\left({\mu}\right),\mu\, \mathcal{L}^{-1}_\mu\left[ v_i \partial_{x_i}\ln\left({\mu} \right) \right]\right\rangle\,d\bm{x}\le 0  
\end{align*}
where we use the negative-definiteness of the linearized collision operator to obtain the inequalities above. We also observe that entropy stability holds at all Knudsen number $\epsilon$ regardless of whether or not the Navier-Stokes-Fourier equations remain a valid model of gas flow. We will see shortly that this is not the case for the Burnett equations.

\subsubsection{Burnett equations (Part 1)}
We have:
\begin{align}
    {f}_{(2^{B1})} &= {f}_{(1)} + \epsilon \mathcal{L}_\mu^{-1}\left[\frac{1}{\mu}(\text{St } \partial_t + v_i\partial_{x_i})\left[\mu\, \mathcal{L}_\mu^{-1}\left[ v_i\partial_{x_i}\ln\left({\mu} \right)\right]\right]\right]\\
    \breve{f}_{(2^{B1})} &= \breve{f}_{(1)} + \epsilon^2 \mathcal{L}_M^{-1}(\text{St } \partial_t + v_i\partial_{x_i}) \mathcal{L}_M^{-1}\left[ v_i\partial_{x_i}\bar{f} \right] \label{eq:linburnettfinescale}
\end{align}
The stress and heat flux for the self-consistent Maxwellian formulation contains many terms so we will only write out the stress and heat flux for the background Maxwellian formulation:
\begin{align}
    \tilde{\sigma}^{(2^{B1})}_{ij} &= \tilde{\sigma}^{(1)}_{ij} - \Xi\,\text{St}\,\partial_t\left[\partial_{x_j} \tilde{u}_i + \partial_{x_i} \tilde{u}_j - \frac{2}{D}\partial_{x_k} \tilde{u}_k\,\delta_{ij}\right] -2\Psi \left(\partial_{x_i}\partial_{x_j}\tilde{\theta} - \frac{1}{D}\partial_{x_k}\partial_{x_k}\tilde{\theta}\delta_{ij} \right)\nonumber\\
    \tilde{q}_i^{(2^{B1})} &= \tilde{q}^{(1)}_i + \Upsilon\, \text{St}\,\partial_t\partial_{x_i}\tilde{\theta} + {\theta_0\Psi}\left( \partial_{x_k}\partial_{x_k}\tilde{u}_i + \left(1 - \frac{2}{D}\right) \partial_{x_i}\partial_{x_k}\tilde{u}_k \right)\label{eq:burnettstressheat1}
\end{align}
where
{\color{black}
\begin{align}
    \Xi &= \frac{\epsilon^2}{\theta_0}\left\langle \mathcal L^{-1}_M[A^M_{12}], M \mathcal L^{-1}_M[A^M_{12}]  \right\rangle\label{eq:xidef}\\
    \Psi &= \frac{\epsilon^2}{\theta_0^2} \left\langle \mathcal L^{-1}_M[A^M_{12}], M D^M_{12}\right\rangle\label{eq:psidef}\\
     \Upsilon &= \frac{\epsilon^2}{\theta_0^2 }\left\langle \mathcal L^{-1}_M[B^M_{1}], M \mathcal L^{-1}_M[B^M_{1}]  \right\rangle\label{eq:upsdef}
\end{align}
The   definition of the tensor $D^M_{mn}$ and the derivation of these new terms are given in Appendices \ref{appendix:alphabet} and \ref{appendix:alphabet2}.
}

To check for entropy stability, it suffices to analyse $\left\langle v_i\partial_{x_i}\bar{f}, M \breve{f}_{(2^{B1})}\right\rangle$ or, more specifically, the $\\ \epsilon^2\left\langle v_i\partial_{x_i}\bar{f}, M \mathcal{L}_M^{-1}\left(\text{St }\partial_t + v_i\partial_{x_i}\right) \mathcal{L}_M^{-1}\left[ v_i \partial_{x_i}\bar{f}\right] \right\rangle$  term contained within it. We have that
\begin{multline*}
    \left\langle v_i\partial_{x_i}\bar{f}, M \mathcal{L}_M^{-1}\left(\text{St }\partial_t + v_i\partial_{x_i}\right) \mathcal{L}_M^{-1}\left[ v_i \partial_{x_i}\bar{f}\right] \right\rangle\\ = \text{St }\left\langle\mathcal{L}_M^{-1}\left[ v_i\partial_{x_i}\bar{f}\right], M \partial_t \mathcal{L}_M^{-1}\left[ v_i \partial_{x_i}\bar{f}\right] \right\rangle \\+ \left\langle\mathcal{L}_M^{-1}\left[ v_i\partial_{x_i}\bar{f}\right], M v_i\partial_{x_i} \mathcal{L}_M^{-1}\left[ v_i \partial_{x_i}\bar{f}\right] \right\rangle \\ =  \frac{\text{St}}{2}\partial_{t}\left\langle   M, \left( \mathcal{L}_M^{-1}\left[ v_i\partial_{x_i} \bar{f}\right]\right)^2 \right\rangle + \frac{1}{2} \partial_{x_i}\left\langle   v_i, M \left( \mathcal{L}_M^{-1}\left[ v_j\partial_{x_j} \bar{f}\right]\right)^2 \right\rangle 
\end{multline*}
We used the self-adjointness of the linearized operator in the first step and used the product rule for derivatives in the second step.

The time derivative term in the above calculation does not have a definite sign and as such for large Knudsen numbers (when the $\epsilon^2$ terms dominate) a loss of entropy stability is possible. 
Considering that this would only be a problem at large enough Knudsen numbers, it might be argued that it only serves as a hard cap on the regime of validity of the Burnett equations\cite{GARCIACOLIN2008}. We argue that this loss of entropy stability is an undesirable property for a macroscopic conservation equation that extends the Navier-Stokes-Fourier equations for two reasons. First of all, the Euler equations and the Navier-Stokes-Fourier equations remain entropy stable for all Knudsen numbers even beyond the regime of validity for these equations. This suggests a pattern we should strive to preserve when extending these equations. Secondly, even if we were to allow for the loss of entropy stability in derived extended hydrodynamic equations, the above calculation shows that the loss of entropy stability is problem dependent because the sign of the potentially troublesome term $\frac{\text{St}}{2}\partial_{t}\left\langle   M, \left( \mathcal{L}_M^{-1}\left[ v_i\partial_{x_i} \bar{f}\right]\right)^2 \right\rangle$  depends on the conservation variables we would be using the conservation equations to solve for since they are contained in $\bar{f}$. Thus it becomes difficult to make an a priori determination of when entropy stability  is lost. This hampers the practicality of these equations for real world problems. The same issue is also present in the self-consistent Maxwellian formulation where the $O(\epsilon)$ term $\left\langle v_i \partial_{x_i}\ln\left({\mu}\right),\mu\, \mathcal{L}^{-1}_\mu\left[\frac{1}{\mu}\left(\text{St }\partial_t + v_i\partial_{x_i}\right)\left[\mu\, \mathcal{L}_\mu^{-1}\left[ v_i \partial_{x_i}\ln\left({\mu}\right)\right] \right]\right]\right\rangle$ can be rewritten as 

$$ \frac{\text{St}}{2}\left\langle   \frac{1}{\mu}\,,\, \partial_{t}\left(\mu\, \mathcal{L}_\mu^{-1}\left[ v_i\partial_{x_i} \ln\left({\mu}\right)\right]\right)^2 \right\rangle + \frac{1}{2} \left\langle   \frac{v_i}{\mu}\,,\, \partial_{x_i}\left(\mu\, \mathcal{L}_\mu^{-1}\left[ v_j\partial_{x_j} \ln\left({\mu}\right)\right]\right)^2 \right\rangle $$

\subsubsection{Burnett equations (Part 2)}
The Burnett equations are usually derived in such a way as to get rid of the time derivatives in the closure for the deviatoric stress and heat flux. This is accomplished by using the Euler equations. For (\ref{eq:burnettstressheat1}), this means that we make the following substitutions
\begin{align*}
\text{St }\partial_t \tilde{u}_j &= -\frac{1}{\rho_0}\partial_{x_j} [\rho_0 \tilde{\theta}+ \tilde{\rho} \theta_0]\nonumber\\
\text{St }\partial_t \tilde{\theta} &= -\frac{2}{D} \theta_0\,\partial_{x_i}  \tilde{u}_i\nonumber
\end{align*}
in order to obtain stress and heat flux given by 
\begin{align}
\tilde{\sigma}^{(2^{B2})}_{ij} &= \tilde{\sigma}^{(1)}_{ij} - 2(\Psi - \,\Xi) \left(\partial_{x_i}\partial_{x_j}\tilde{\theta} - \frac{1}{D}\partial_{x_k}\partial_{x_k}\tilde{\theta}\,\delta_{ij} \right)+\frac{2\,\Xi\theta_0}{\rho_0}\left(\partial_{x_i}\partial_{x_j}\tilde{\rho} - \frac{1}{D}\partial_{x_k}\partial_{x_k}\tilde{\rho}\, \delta_{ij} \right)\nonumber\\
    \tilde{q}_i^{(2^{B2})} &= \tilde{q}^{(1)}_i + \theta_0\left({\Psi}\left(1 - \frac{2}{D}\right)-\frac{2}{D}\Upsilon\right)\, \partial_{x_i}\partial_{x_k}\tilde{u}_k + {\theta_0\Psi} \partial_{x_k}\partial_{x_k}\tilde{u}_i \label{eq:burnettstressheat2}
\end{align}
On the level of the coarse and fine-scale equations, this can be accomplished by observing that equation (\ref{eq:linearconser}) can be written as
$$\text{St }\partial_t\bar{f} = - \Pi_M\left[ v_i\partial_{x_i}\bar{f} \right] + O(\epsilon)$$
where $O(\epsilon)$ represents any fine-scale corrections made. These correction terms will be at least an order of $\epsilon$ higher than the other terms in $\breve{f}_{(2)}$. As such to get the fine-scale term for the Burnett equations, we formally ignore these corrections when we make the substitution into (\ref{eq:linburnettfinescale}) to obtain
\begin{equation}\label{eq:convburnettfinescale}
\breve{f}_{(2^{B2})}=\breve{f}_{(1)}-\epsilon^2\mathcal{L}_M^{-2} v_j\partial_{x_j}\Pi_M\left[v_i \partial_{x_i}\bar{f} \right] +\epsilon^2\mathcal{L}_M^{-1}v_j\partial_{x_j}\mathcal{L}^{-1}_M\left[v_i \partial_{x_i}\bar{f} \right]
\end{equation}

The derivation of the above closure takes advantage of the fact that $\mathcal{L}_M^{-1}$ commutes with the time derivative.
This is not the case with the self-consistent formulation and as such we do not have a fine scale analogue to the substitution described in this section. 
That said, the fine-scale approximation \eqref{eq:convburnettfinescale} also does not provide unconditional entropy stability. We get
\begin{multline*}
    \bracke{v_i \partial_{x_i}\bar{f},\, M \breve{f}_{(2^{B2})}} = \underbrace{\bracke{v_i \partial_{x_i}\bar{f},\,M \breve{f}_{(1)}}}_{\text{Stokes-Fourier (good)}} + \frac{\epsilon^2}{2}\underbrace{\partial_{x_i}\bracke{v_i, \, M \left(\mathcal{L}^{-1}_M[v_j\partial_{x_j}\bar{f}] \right)^2 }}_{\text{Divergence term (ok)}} \\-{\epsilon^2}\underbrace{\bracke{\mathcal{L}^{-1}_M[v_i \partial_{x_i}\bar{f}],\,M\,\mathcal{L}^{-1}_M[v_j\partial_{x_j}\Pi_M[v_k \partial_{x_k}\bar{f}\,]]}}_{\text{Indeterminate sign (not good)}} 
\end{multline*}


\subsubsection{Super-Burnett equation}
We note that the next level of closure of the Chapman-Enskog expansion in the background Maxwellian formulation is given by 
\begin{equation*}
    \breve{f}_{(3^{B})} = \breve{f}_{(2^B)} + \epsilon^3\mathcal{L}^{-1}_M\big(\text{St }\partial_t + v_k \partial_{x_k}\big)\mathcal{L}^{-1}_M\big(\text{St }\partial_t + v_j \partial_{x_j} \big)\mathcal{L}^{-1}_M\big[v_i\partial_{x_i}\bar{f} \big]
\end{equation*}
and the new terms that do not end up on the boundary in the entropy equation (\ref{eq:linentstabeqn}) will come from the term\\ $\left\langle v_l\partial_{x_l} \bar{f} , M\,\mathcal{L}^{-1}_M\big(\text{St }\partial_t + v_k \partial_{x_k}\big)\mathcal{L}^{-1}_M\big(\text{St }\partial_t + v_j \partial_{x_j} \big)\mathcal{L}^{-1}_M\big[v_i\partial_{x_i}\bar{f} \big] \right\rangle$. Using self-adjointness and the product rule of derivatives this term can be rewritten to give
\begin{multline*}
  - \left\langle (\text{St }\partial_t + v_j\partial_{x_j})\breve{f}_{(1)},\,M\, \mathcal{L}^{-1}_M \left[ (\text{St }\partial_t + v_j\partial_{x_j})\breve{f}_{(1)}\right]\right\rangle\\ +\text{St }\partial_t\left\langle\breve{f}_{(1)}, M\,\mathcal{L}^{-1}_M(\text{St }\partial_t + v_j\partial_{x_j})\breve{f}_{(1)} \right\rangle \\+\partial_{x_k}\left\langle v_k\,\breve{f}_{(1)}, M\,\mathcal{L}^{-1}_M(\text{St }\partial_t + v_j\partial_{x_j})\breve{f}_{(1)} \right\rangle 
\end{multline*}
Due to the negative-definiteness of the linearized collision operator, we see that the first term above will be non-negative. As such it will have a detrimental effect on the entropy stability of the conservation equations that come from this closure. We are thus unable to deal with the problems of the Burnett equation by moving up to the closure above it.
\subsection{An Entropy Stable Fine-Scale Closure}
With the observations made of the entropy stability of the Burnett equation and Super-Burnett equations, we can now propose a fine-scale closure that does produce entropy stability. This closure is accomplished with two steps:

\begin{enumerate}
    \item We assume that the Strouhal number St is at least of the same order as the Knudsen number $\epsilon$. This means that the term $\text{St }\partial_t$ in the fine-scale equations (\ref{eq:nonlinfinescale}) and (\ref{eq:linfinescale}) will be at least an order higher in Knudsen number than the other terms and thus when we come up with a new closure term beyond the Navier-Stokes closure, we formally ignore this term.
    Thus our starting point for our alternative to the Burnett equations will be
    \begin{align*}
         {f} &= \mathcal{L}_\mu^{-1}\left[ v_i\partial_{x_i}\left[\ln({\mu})\right] + \frac{\epsilon}{\mu} v_i\partial_{x_i}\left[\mu \,{f}\right]\right] + O(\epsilon^2 f)\\
         \breve{f}&= \epsilon\, \mathcal{L}_M^{-1} v_i\partial_{x_i}\left[\bar{f} + \,\breve{f}\,\right] + O\big(\epsilon^2\breve{f}\,\big)
    \end{align*}
    
    \item The above equations can be trivially rewritten as:
    \begin{align*}
        {f} &= {f} + \epsilon\mathcal{L}_\mu^{-1} v_j\partial_{x_j}\left[{f} - \mathcal{L}_\mu^{-1}\left[ v_i\partial_{x_i}\left[\ln(\mu)\right] + \frac{\epsilon}{\mu} v_i\partial_{x_i}\left[\mu \,{f}\right]\right]\right]\\ 
        \breve{f}&= \breve{f}+\epsilon\mathcal{L}^{-1}_M v_j \partial_{x_j}\left[\breve{f}- \epsilon\, \mathcal{L}_M^{-1} v_i\partial_{x_i}\left[\bar{f} + \,\breve{f}\,\right]\right]
    \end{align*}
    Initializing with $f_{(1)}$ (resp. $\breve{f}_{(1)}$), we use the above form as the basis for a new recursive relation on the fine-scale:
     \begin{align}
        \label{eq:nonlinaltiter}{f}_{(k+1)} &= {f}_{(k)} +\epsilon \mathcal{L}_\mu^{-1}v_j\partial_{x_j}\left[{f}_{(k)} - \mathcal{L}_\mu^{-1}\left[ v_i\partial_{x_i}[\ln(\mu)] + \frac{\epsilon}{\mu} v_i\partial_{x_i}\left[\mu \,{f}_{(k)}\right]\right]\right]\\\label{eq:linaltiter}
        \breve{f}_{(k+1)} &= \breve{f}_{(k)} +\epsilon\mathcal{L}^{-1}_M v_j \partial_{x_j}\left[\breve{f}_{(k)} - \epsilon\, \mathcal{L}_M^{-1} v_i\partial_{x_i}\left[\bar{f} + \,\breve{f}_{(k)}\right]\right]
    \end{align}
\end{enumerate}
In particular, we get an alternative closure to the Burnett equations of the form
    \begin{align}
        \label{eq:nonlinaltburnclose}{f}_{(2^E)} &= {f}_{(1)} -\epsilon^2 \mathcal{L}_\mu^{-1}v_k\partial_{x_k}\left[ \mathcal{L}_\mu^{-1}\left[ \frac{1}{\mu} v_j\partial_{x_j}\left[\mu \,\mathcal{L}_\mu^{-1}v_i\partial_{x_i}\ln\left(\mu \right)\right]\right]\right]\\\label{eq:linaltburnclose}
        \breve{f}_{(2^E)} &= \breve{f}_{(1)} -\epsilon^3\mathcal{L}^{-1}_M v_k \partial_{x_k}\left[ \mathcal{L}_M^{-1} v_j\partial_{x_j}\left[ \mathcal{L}^{-1}_M\left[ v_i\partial_{x_i}\bar{f}\right]\right]\right]
    \end{align}  
{\color{black} For the constant background formulation, this leads to a deviatoric stress and heat flux given by:
\begin{align}
    \label{eq:altburnstress} \tilde{\sigma}^{(2^E)}_{ij} &= 2\,\tilde{\omega}\,\tepsilon_{ij}(\bm{\tu}) - \dx{k}\tilde{\mathfrak{S}}_{ijk} 
    \\
       \label{eq:altburnheat} \tilde{q}^{(2^E)}_i &= - \tilde{\kappa}\,\partial_{x_i}\tilde{\theta} + \dx{k}\tilde{\mathfrak{H}}_{ik}
    \end{align}

where 
\begin{align}
\label{eq:tepsilonDef} \tepsilon_{ij}(\bm{w}) &:= \frac{\partial_{x_j} w_i + \partial_{x_i} w_j}{2} - \frac{\delta_{ij}}{D}\ \partial_{x_k}\! w_k\\
     \label{eq:linESESpress} \tilde{\mathfrak{S}}_{ijk} &:= 2  \tilde{\alpha}_1 \left(\dx{k}\tepsilon_{ij}(\bm{\tu}) +\dx{i}\tepsilon_{jk}(\bm{\tu}) + \dx{j}\tepsilon_{ik}(\bm{\tu}) - \frac{2}{D}\delta_{ij}\,\dx{l}\tepsilon_{lk}(\bm{\tu})\right) \nonumber\\ &\hspace{1.cm}+ 2\tilde{\alpha}_2 \left(\delta_{ik}\dx{l}\tepsilon_{lj}(\bm{\tu}) +\delta_{jk}\dx{l}\tepsilon_{li}(\bm{\tu}) - \frac{2}{D} \delta_{ij}\,\dx{l}\tepsilon_{lk}(\bm{\tu}) \right)
    \\
    \tilde{\mathfrak{H}}_{ik} &:= \tilde{\Lambda}_0 \Delta\ttheta \, \delta_{ik} + \tilde{\Lambda}_1 \dx{i}\dx{k}\ttheta \label{eq:linESEHeatTensor}
  \end{align}

The process of deriving equations \eqref{eq:altburnstress} and \eqref{eq:altburnheat} and the definitions of the fluid parameters $\tilde{\alpha}_i$ and $\tilde{\Lambda}_i$ are elaborated upon in Appendices \ref{appendix:alphabet} and \ref{appendix:alphabet2}.
To prove entropy stability we need to show that the terms $\langle v_i\partial_{x_i}\bar{f}, M\, \breve{f}_{(2^E)} \rangle$  and $\langle v_i \partial_{x_i}\ln\left(\mu\right),\mu\, {f}_{(2^E)}\rangle$ yield non-positive terms and divergence terms. We use the self-adjointness of $\mathcal{L}^{-1}_{M}$ and the product rule of derivatives to show that
\begin{align*}
    \left\langle v_i\partial_{x_i}\bar{f}, M \left(-\epsilon^3\mathcal{L}^{-1}_M v_k \partial_{x_k}\left[ \mathcal{L}_M^{-1} v_j\partial_{x_j}\left[ \mathcal{L}^{-1}_M\left[ v_i\partial_{x_i}\bar{f}\right]\right]\right] \right)\right\rangle &= -\epsilon\left\langle \breve{f}_{(1)}, M\,  v_k \partial_{x_k}\left[ \mathcal{L}_M^{-1} v_j\partial_{x_j}\left[ \breve{f}_{(1)}\right]\right] \right\rangle\\ &= -\epsilon\,\partial_{x_k}\left\langle v_k\breve{f}_{(1)}, M\, \mathcal{L}_M^{-1} v_j\partial_{x_j}\left[ \breve{f}_{(1)}\right] \right\rangle\\&\quad +\underbrace{\epsilon\left\langle v_k\partial_{x_k}\breve{f}_{(1)}, M\, \mathcal{L}_M^{-1} \left[v_j\partial_{x_j} \breve{f}_{(1)}\right] \right\rangle}_{\le 0}
\end{align*}
In terms of macroscopic variables, the dissipative term $\epsilon\left\langle v_k\partial_{x_k}\breve{f}_{(1)}, M\, \mathcal{L}_M^{-1} \left[v_j\partial_{x_j} \breve{f}_{(1)}\right] \right\rangle$ is
\begin{multline*}
    -\frac{\dx{k}\dx{j}\tu_i}{\theta_0}\tilde{\mathfrak{S}}_{ijk} - \frac{\dx{k}\dx{j}\ttheta}{\theta_0^2}\tilde{\mathfrak{H}}_{jk} = -\frac{6\tilde{\alpha}_1}{\theta_0}\,\dx{(i}\tepsilon_{jk)}\!(\bm{\tu})\,\dx{(i}\tepsilon_{jk)}\!(\bm{\tu}) 
 - \frac{4\tilde{\alpha}_2}{\theta_0}|\text{div}\tepsilon(\bm{\tu})|^2 -\frac{\tilde{\Lambda}_0}{\theta_0^2}(\Delta\ttheta)^2\\ - \frac{\tilde{\Lambda}_1}{\theta_0^2} \,\nabla\nabla\ttheta\bm{:}\nabla\nabla\ttheta
\end{multline*}

where $$\dx{(k}\!\tepsilon_{ij)}\!(\bm{\tu}) := \frac{\dx{k}\!\tepsilon_{ij}(\bm{\tu})+\dx{i}\tepsilon_{jk}(\bm{\tu})+\dx{j}\tepsilon_{ik}(\bm{\tu})}{3}$$ is the symmetrization of $\dx{k}\!\tepsilon_{ij}(\bm{\tu})$.}

Dissipation for the non-linear case starts from the term $$ \left\langle v_i \partial_{x_i}\ln\left(\mu\right),\mu\, \left( -\epsilon^2 \mathcal{L}_\mu^{-1}v_k\partial_{x_k}\left[ \mathcal{L}_\mu^{-1}\left[ \frac{1}{\mu} v_j\partial_{x_j}\left[\mu \,\mathcal{L}_\mu^{-1}v_i\partial_{x_i}\ln\left(\mu \right)\right]\right]\right]\right)\right\rangle$$
which gives
\begin{align*}
    -\epsilon^2\left\langle {f}_{(1)}, \mu\,  v_k \partial_{x_k} \mathcal{L}_\mu^{-1}\left[\frac{1}{\mu} v_j\partial_{x_j}\left[\mu {f}_{(1)}\right]\right] \right\rangle &= -\epsilon^2\,\partial_{x_k}\left\langle v_k {f}_{(1)}, \mu\,   \mathcal{L}_\mu^{-1}\left[\frac{1}{\mu} v_j\partial_{x_j}\left[\mu {f}_{(1)}\right]\right] \right\rangle\\&\quad+\epsilon^2\underbrace{\left\langle \frac{1}{\mu}v_k \partial_{x_k}\left[\mu{f}_{(1)}\right], \mu\,   \mathcal{L}_\mu^{-1}\left[\frac{1}{\mu} v_j\partial_{x_j}\left[\mu {f}_{(1)}\right]\right] \right\rangle}_{\le 0}
\end{align*}
The iteration \eqref{eq:linaltiter} also leads to an entropy stable $\breve{f}_{(3^E)}\,$.
 Using $\breve{f}_{(1)} =  \epsilon \mathcal{L}^{-1}_M v_k\partial_{x_k}\bar{f}$, this approximation can be written 
 \begin{align*}
     \breve{f}_{(3^E)} &= \breve{f}_{(2^E)} + \epsilon \mathcal{L}^{-1}_M v_k\partial_{x_k}\left[\breve{f}_{(2^E)} - \epsilon\mathcal{L}^{-1}_M v_j\partial_{x_j}\left[ \bar{f} + \breve{f}_{(2^E)}\right] \right]\\
     &=  \breve{f}_{(2^E)} + \epsilon \mathcal{L}^{-1}_M v_k\partial_{x_k}\left[\breve{f}_{(2^E)} - \breve{f}_{(1)}\right] - \epsilon^2 \mathcal{L}^{-1}_M v_k\partial_{x_k}\mathcal{L}^{-1}_M v_j\partial_{x_j}\left[\breve{f}_{(2^E)} \right]\\
     &= \breve{f}_{(2^E)} - \epsilon^3 \mathcal{L}^{-1}_M v_k\partial_{x_k} \mathcal{L}^{-1}_M v_j\partial_{x_j} \mathcal{L}^{-1}_M v_i\partial_{x_i}\breve{f}_{(1)} - \epsilon^2 \mathcal{L}^{-1}_M v_k\partial_{x_k}\mathcal{L}^{-1}_M v_j\partial_{x_j}\left[\breve{f}_{(2^E)} \right]
 \end{align*}
Thus the term of interest for proving entropy stability takes the form
 \begin{multline*}
     \bracke{v_m \partial_{x_m}\bar{f},\, M\,\breve{f}_{(3^E)}}  = \bracke{v_m \partial_{x_m}\bar{f},\, M\,\breve{f}_{(2^E)}} -\epsilon^2 \bracke{\breve{f}_{(1)}, M\,v_k\partial_{x_k}\mathcal{L}^{-1}_M v_j\partial_{x_j}\mathcal{L}^{-1}_M v_i\partial_{x_i}\breve{f}_{(1)} } \\
     -\epsilon \bracke{\breve{f}_{(1)}, M\,v_k\partial_{x_k}\mathcal{L}^{-1}_M v_j\partial_{x_j}\breve{f}_{(2^E)}  }
 \end{multline*}
and we need only focus on just the last two terms from this expression.
 First
 \begin{multline*}
     -\epsilon^2 \bracke{\breve{f}_{(1)}, M\,v_k\partial_{x_k}\mathcal{L}^{-1}_M v_j\partial_{x_j}\mathcal{L}^{-1}_M v_i\partial_{x_i}\breve{f}_{(1)} } = \\ -\epsilon^2 \partial_k \bracke{v_k \breve{f}_{(1)}, M\, v_j\partial_{x_j}\mathcal{L}^{-1}_M v_i\partial_{x_i}\breve{f}_{(1)}} + \epsilon^2 \bracke{\mathcal{L}^{-1}_M v_k\partial_{x_k}\breve{f}_{(1)}, M\, v_j\partial_{x_j}\mathcal{L}^{-1}_M v_i\partial_{x_i}\breve{f}_{(1)}}=\\
     -\epsilon^2 \partial_k \bracke{v_k \breve{f}_{(1)}, M\, v_j\partial_{x_j}\mathcal{L}^{-1}_M v_i\partial_{x_i}\breve{f}_{(1)}} + \frac{\epsilon^2}{2}\partial_{x_j}\bracke{v_j,M \left(\mathcal{L}^{-1}_M v_k\partial_{x_k}\breve{f}_{(1)} \right)^2}
 \end{multline*}
 With the other term, applying the product rule to $\partial_{x_k}$ gives 
 \begin{align*}
   -\epsilon \bracke{\breve{f}_{(1)}, M\, v_k\partial_{x_k}\mathcal{L}^{-1}_M v_j\partial_{x_j}\breve{f}_{(2^E)}}  = -\partial_{x_k}\bracke{\dots} + \epsilon\bracke{\mathcal{L}^{-1}_M v_k\partial_{x_k}\breve{f}_{(1)}, M\, v_j\partial_{x_j}\breve{f}_{(2^E)}} 
 \end{align*}
 Noting that $\breve{f}_{(2^E)} = \breve{f}_{(1)} - \epsilon^2\mathcal{L}^{-1}_M v_n\partial_{x_n}\mathcal{L}^{-1}_M v_l\partial_{x_l}\breve{f}_{(1)}$, the second term in the above can be written as
 \begin{align*}
\epsilon\underbrace{\bracke{\mathcal{L}^{-1}_M v_k\partial_{x_k}\breve{f}_{(1)}, M\, v_j \partial_{x_j}\breve{f}_{(1)}}}_{\le 0} - \epsilon^3\bracke{\mathcal{L}^{-1}_M v_k\partial_{x_k}\breve{f}_{(1)},M\, v_j\partial_{x_j}\mathcal{L}^{-1}_M v_n\partial_{x_n}\mathcal{L}^{-1}_M v_l\partial_{x_l}\breve{f}_{(1)}}
 \end{align*}
 Applying the product rule to $\partial_{x_j}$ in the second term above gives
 \begin{align*}
      -\epsilon^3\,\partial_{x_j}\bracke{\dots} + \epsilon^3\underbrace{\bracke{v_j\partial_{x_j}\mathcal{L}^{-1}_M v_k\partial_{x_k}\breve{f}_{(1)},M\, \mathcal{L}^{-1}_M\left[ v_n\partial_{x_n}\mathcal{L}^{-1}_M v_l\partial_{x_l}\breve{f}_{(1)}\right]}}_{\le 0}
 \end{align*}
 In fact, up to about $\hat{f}_{(5)}$ the iteration \eqref{eq:linaltiter} leads to an entropy stable closure.
 
 Unfortunately, due to the dependence $\mu$ has on $\boldsymbol{x}$, the steps followed above cannot be used to prove entropy stability for the non-linear fine-scale approximation $f_{(3^E)}$ derived from \eqref{eq:nonlinaltiter}. Compared to its linear counterpart, we will have terms like
\begin{align*}
\bracke{\mathcal{L}_\mu^{-1}\left[\frac{1}{\mu}v_k\partial_{x_k}[\mu f_{(1)}]\right], \mu\,v_j\partial_{x_j}\left[ \mathcal{L}_\mu^{-1}\left[\frac{1}{\mu}v_i\partial_{x_i}[\mu f_{(1)}]\right]\right] }
    =\frac{\epsilon^2}{2} \bracke{v_j, \mu\, \partial_{x_j}\left( \mathcal{L}_\mu^{-1}\left[\frac{1}{\mu}v_k\partial_{x_k}[\mu f_{(1)}]\right]\right)^2}
\end{align*}
that cannot either be moved to the boundary or shown to be non-positive.


\subsection{Summary Of {\color{black} Constant Background} Fine-Scale Approximations and the Resulting  Closures}
We summarise the linearized closures derived in this section for $D=3$.
\begin{table}[h]
\centering
{\tabulinesep=1.2mm
\begin{tabu}{|m{7em}|l|}
\hline
 \multicolumn{2}{|c|}{\textbf{Chapman-Enskog Iteration} }\\
 \multicolumn{2}{|c|}{$\breve{f}_{(0)} = 0; \quad \breve f_{(n+1)} 
    =  \epsilon\,\mathcal L^{-1}_M
    \left[ \,\boldsymbol{v}\cdot\nabla \bar f+
    (\text{St }\partial_t + \boldsymbol{v}\cdot 
    \nabla) 
    [\breve f_{(n)} ]\,\right]$}\\
\hline
\multirow{2}{7em}{\textit{Stokes-Fourier}} &$ \boldsymbol{\tilde\sigma}^{(1)} =  \tilde{\omega}\,\Big( \nabla\boldsymbol{\tilde u} + (\nabla\boldsymbol{\tilde u})^T - \frac{2}{3}\text{div}\boldsymbol{\tilde u}\,\text{Id} \Big)=: 2\,\tilde{\omega}\,\tepsilon(\bm{\tilde{u}})$  \\
     & $\boldsymbol{\tilde{q}}^{(1)} = - \tilde{\kappa}\, \nabla \tilde{\theta}$ \\
     \hline
     \multirow{2}{7em}{\textit{Burnett I}} &$ \boldsymbol{\tilde\sigma}^{(2^{B1})} = \boldsymbol{\tilde\sigma}^{(1)} - 2\,\tilde{\Xi}\,\,\text{St}\,\partial_t\tepsilon(\bm{\tilde{u}}) - 2\Psi \left(\nabla\nabla - \frac{\text{Id}}{3}\Delta \right)[\,\tilde{\theta}\,]$  \\ & $\boldsymbol{\tilde{q}}^{(2^{B1})} = \boldsymbol{\tilde{q}}^{(1)} + \Upsilon\,\text{St} \partial_t\nabla\tilde{\theta} + {\theta_0 \Psi}\left(\Delta\boldsymbol{\tilde{u}} - \frac{1}{3}\nabla\text{div}\boldsymbol{\tilde{u}} \right)$\\
     \hline
     \multirow{2}{7em}{\textit{Burnett II}} 
     & $ \boldsymbol{\tilde\sigma}^{(2^{B2})} = \boldsymbol{\tilde\sigma}^{(1)} - 2\left(\nabla\nabla - \frac{\text{Id}}{3}\Delta \right) \left[(\Psi - \Xi)\tilde{\theta} - \frac{\theta_0\Xi}{\rho_0}\tilde{\rho} \right]$ \\
     & $\boldsymbol{\tilde{q}}^{(2^{B2})} = \boldsymbol{\tilde{q}}^{(1)} +{\theta_0\Psi}\Delta\boldsymbol{\tilde{u}} + \theta_0 \left(\frac{\Psi}{3} - \frac{2\Upsilon }{3}\right)\nabla\text{div}\boldsymbol{\tilde{u}}$\\
     \hline
     \multicolumn{2}{|c|}{\textbf{New Iteration}}\\
\multicolumn{2}{|c|}{ ${\breve{f}_{(1)} = \epsilon\, \mathcal{L}_M^{-1}\boldsymbol{v}\cdot\nabla\bar{f}}; \quad \breve{f}_{(k+1)} = \breve{f}_{(k)} +\epsilon\,\mathcal{L}^{-1}_M \boldsymbol{v}\cdot\nabla\left[\breve{f}_{(k)} - \epsilon\, \mathcal{L}_M^{-1} \boldsymbol{v}\cdot\nabla\left[\bar{f} + \,\breve{f}_{(k)}\right]\right] $}\\
\hline
\multirow{2}{7em}{\textit{Entropy Stable Extension}}
& {\color{black} $\boldsymbol{\tilde\sigma}^{(2^E)} = \boldsymbol{\tilde\sigma}^{(1)} - \text{div}\,\tilde{\mathfrak{S}}$ $\qquad (\tilde{\mathfrak{S}}$ defined in \eqref{eq:linESESpress} ) }\\
& {\color{black} $\boldsymbol{\tilde{q}}^{(2^E)} = \boldsymbol{\tilde{q}}^{(1)} + \text{div}\tilde{\mathfrak{H}}\qquad\,\,$ ($\tilde{{\mathfrak{H}}}$ defined in \eqref{eq:linESEHeatTensor} ) }\\
\hline
\end{tabu}
\caption{ Summary of Fine-Scale Approximations and Resulting Closures}
}
\end{table}

{\color{black}
The conservation equations for the linearized entropy stable extension will be
\begin{align*}
    0 &=\St\partial_t\tilde{\rho} + \rho_0\, \text{div}\bm{\tilde{u}} \\ 
      0 &=  \rho_0\,\St\partial_t\bm{\tilde{u}}  + \nabla[\rho_0\tilde{\theta} + \tilde{\rho}\theta_0] - \text{div}\left[2\tilde{\omega}\,\bm{\varepsilon}(\bm{\tilde{u}} ) - \text{div}\,\tilde{\mathfrak{S}}\right]\\
      0 & = \frac{3}{2}\rho_0\,\St \partial_t\tilde{\theta} +\rho_0\theta_0\, \text{div}\bm{\tilde{u}}\,+\text{div}\left[ -\tilde\kappa \nabla\tilde{\theta} + \text{div}\tilde{\mathfrak{H}}\,\right]
\end{align*} 
and the linear coarse-scale entropy dissipation inequality will be 
\begin{multline}\label{eq:linESEentropyEqn}
\frac{\St}{2}\partial_t\sqbrac{\frac{\trho^2}{\rho_0} + \frac{\rho_0|\bm{\tu}|^2}{\theta_0} + \frac{3}{2}\frac{\rho_0\ttheta^2}{2\theta_0^2}} + \frac{1}{2}\dx{i}\sqbrac{\frac{\tu_i}{\theta_0} \big(\trho \theta_0 + \rho_0\ttheta\,\big)} 
  \\
  + \dx{i}\sqbrac{-\frac{\tu_k\,}{\theta_0}\,\tilde{\sigma}^{(2^E)}_{ik} + \frac{\ttheta\, }{\theta_0^2}\,\tilde{q}_i^{(2^E)}} 
 -\dx{k}\Bigg[ 
 \frac{\dx{i}{\tu_j}}{\theta_0}\ \tilde{\mathfrak{S}}_{ijk}+\frac{\dx{i}\ttheta}{\theta_0^2}\,\,\tilde{\mathfrak{H}}_{ik} \Bigg] =
 \\
 -\frac{2\tilde{\omega}}{\theta_0} \tepsilon(\bm{\tu})\!\bm{:}\!\tepsilon(\bm{\tu}) - \frac{\tilde{\kappa}}{\theta_0^2}\left|\nabla\ttheta\right|^2\,
 - \frac{6\tilde{\alpha}_1}{\theta_0}\,\dx{(i}\tepsilon_{jk)}(\bm{\tu})\dx{(i}\tepsilon_{jk)}(\bm{\tu}) 
 \\
 - \frac{4\tilde{\alpha}_2}{\theta_0}|\text{div}\tepsilon(\bm{\tu})|^2 -\frac{\tilde{\Lambda}_0}{\theta_0^2}\parent{\!\Delta \ttheta}^2-\frac{\tilde{\Lambda}_1}{\theta_0^2} \nabla\nabla\ttheta\!\bm{:}\!\nabla\nabla\ttheta 
  \le 0 
\end{multline}

\section{The non-linear Entropy Stable Extension and a consistent modification}
We briefly discuss the macroscopic form of the non-linear Entropy Stable Extension \eqref{eq:nonlinaltburnclose}. In particular, we shall look at a smaller subset of the full non-linear extension that remains entropy dissipative and is a direct non-linear generalization of the linear Entropy Stable Extension given in \eqref{eq:altburnstress} and \eqref{eq:altburnheat}.

The main challenge in the non-linear theory comes from dealing with the derivative of the inverse linearized operator $\dx{k}\iLmu[g]$. For bilinear collision operators, we can show the following: 
\begin{thm}[Theorem 10.1 of \cite{Baidoo2025}]\label{thm:InvLDerivative}
    Given a linearized collision operator of the form $\mathcal{L}_{\mathcal{M}}[g] = \frac{2}{\mathcal{M}}Q(\mathcal{M},\, \mathcal{M}g)$ for a Maxwellian $\mathcal{M}$ and function $g$ that are differentiable in $\bm{x}$, we have that 
    \begin{align}\label{eq:InvLDerivative}
    \partial_{x_i} \mathcal{L}_{\mathcal{M}}^{-1}\left[g \right] &= \mathcal{L}_{\mathcal{M}}^{-1}\left[\partial_{x_i}g \right]\, -\, \mathcal{L}_{\mathcal{M}}^{-1}\left[g \right]\, \partial_{x_i}\ln\mathcal{M}  + \mathcal{X}_{\mathcal{M}}[\dx{i}\ln\mathcal{M},\, g]
\end{align}
where the bilinear operator $\mathcal{X}_{\mathcal{M}}$ is
\begin{align}
    \mathcal{X}_{\mathcal{M}}[h, g] = \mathcal{L}^{-1}_{\mathcal{M}}\left[h\,(\textrm{Id}-\Pi_{\mathcal{M}})[g] - \frac{2}{\mathcal{M}}Q\left(\mathcal{M}h, \mathcal{M}\,\mathcal{L}_{\mathcal{M}}^{-1}[g]\right)\right]
\end{align}
\end{thm}
\begin{remark}
    Equation \eqref{eq:InvLDerivative} applies to any suitable Maxwellian $\mathcal{M}$. It thus also applies to the case of the global Maxwellian $\mathcal{M}=M$ where it takes the familiar form
    \begin{equation}\label{eq:linDerivCommute}
        \dx{i}\iLm[g] = \iLm[\dx{i}g]
    \end{equation}
\end{remark}
An immediate consequence of Theorem \ref{thm:InvLDerivative} is  the following
\begin{cor}
\begin{align}\label{eq:DerivativeSwitchIdentity}
    \frac{1}{\mathcal{M}} \partial_{x_k}\left[\mathcal{M}\, \mathcal{L}_{\mathcal{M}}^{-1}\left[g \right] \right] &= \partial_{x_k}\mathcal{L}_{\mathcal{M}}^{-1}\left[g \right]+ \mathcal{L}_{\mathcal{M}}^{-1}\left[g \right]\,\partial_{x_k}\!\ln\mathcal{M}\nonumber  \\
    &= \mathcal{L}_{\mathcal{M}}^{-1}\left[\partial_{x_k}g \right] + \mathcal{X}_{\mathcal{M}}[\dx{k}\!\ln\mathcal{M}, \, g]
\end{align}
\end{cor}

Thus, for example, with the above identity, the deviatoric stress contribution of the Entropy Stable Extension will involve computing 
\begin{align}\label{eqCalc:ESEstressTerm}
    \frac{\sigma_{ij}^{(2^E)} - \sigma_{ij}^{(1)}}{\epsilon^3} &=  \abrac{v_i v_j,\mu \iLmu\vdx{k}\sqbrac{\iLmu \left[\frac{1}{\mu}\vdx{l}\sqbrac{\mu\,f_{(1)}}\right]}}\notag
\\
&= \abrac{v_k \iLmu[v_i v_j], \mu\, \dx{k}\iLmu \left[\frac{1}{\mu}\vdx{l}\sqbrac{\mu\,f_{(1)}}\right]}\notag
    \\
    &= \dx{k}\abrac{v_k \iLmu[v_i v_j], \mu\, \iLmu \left[\frac{1}{\mu}\vdx{l}\sqbrac{\mu\,\iLmu[\vdx{n}\!\ln\mu]}\right]}\notag
    \\ &\hspace{0.5cm}
    -\abrac{ \frac{1}{\mu}\vdx{k}\sqbrac{\mu\,\iLmu[v_i v_j]}, \mu\, \iLmu \left[\frac{1}{\mu}\vdx{l}\sqbrac{\mu\,\iLmu[\vdx{n}\!\ln\mu]}\right]}\notag
    \\
    &= \dx{k}\abrac{v_k \iLmu[v_i v_j], \mu \iLmu\sqbrac{v_l \iLmu[\vdx{n}\dx{l}\!\ln\mu] }   } \notag
    \\&\hspace{0.5cm}
    +\dx{k}\abrac{v_k \iLmu[v_i v_j], \mu \iLmu\sqbrac{ v_l\, \mathcal{X}_{\mu}[\dx{l}\!\ln\mu, \vdx{n}\ln\!\mu]}   }\notag
    \\&\hspace{0.5cm}
    -\abrac{ \frac{1}{\mu}\vdx{k}\sqbrac{\mu\,\iLmu[v_i v_j]}, \mu \iLmu\sqbrac{v_l \iLmu[\vdx{n}\dx{l}\!\ln\mu]}  }\notag
    \\&\hspace{0.5cm}
    -\abrac{ \frac{1}{\mu}\vdx{k}\sqbrac{\mu\,\iLmu[v_i v_j]}, \mu \iLmu\sqbrac{v_l \mathcal{X}_{\mu}[\dx{l}\!\ln\mu, \vdx{n}\ln\!\mu]}  }
\end{align}

 At the macroscopic level, the contributions due to the bilinear operator $\mathcal{X}_{\mathcal{M}}$ are products of the first and second derivatives of the conservation moments only. For example, the second term on the right-hand side of \eqref{eqCalc:ESEstressTerm}, will have terms such as 
 \begin{align*}
     &2\dx{k}\!\Bigg[\mathtt{a}_1\, \tepsilonu{ij}\dx{k}\theta 
    + \mathtt{a}_2\, \dx{n}\theta\! \left(\delta_{ik}\tepsilonu{nj} + \delta_{jk}\tepsilonu{ni} - \frac{2\delta_{ij}}{D}\tepsilonu{kn} \right) 
    \\&\hspace{1cm}
    +\mathtt{a}_3\! \parent{\tepsilonu{ik}\dx{j}\theta +\tepsilonu{kj} \dx{i}\theta-\frac{2\delta_{ij}}{D}\tepsilonu{kl}\dx{l}\theta }+\mathtt{A}\, u_k \uD[\theta]\tepsilonu{ij}
    \\ &\hspace{1.cm}
     + \mathtt{P}\,\uD[\theta]\parent{\frac{\delta_{ik}\dx{j}\theta+\delta_{jk}\dx{i}\theta}{2} - \frac{\delta_{ij}}{D}\dx{k}\theta}+\mathtt{D}\,u_k\parent{\dx{i}\theta\dx{j}\theta - \frac{\delta_{ij}}{D}|\nabla\theta|^2}
    &
    \Bigg]
\end{align*}
where the fluid parameters $\mathtt{a}_i$, $\mathtt{A}$, $\mathtt{P}$ and $\mathtt{D}$ depend on $\iLmu$ and $\mathcal{X}_{\mathcal{M}}$. Put another way, the highest order derivative contributions to the Entropy Stable Extension come from the integrals that do not involve $\mathcal{X}_{\mathcal{M}}$. 

To keep the number of terms presented to a manageable level, we shall, from now on, resort to excluding the contributions due to the bilinear operator $\mathcal{X}_{\mathcal{M}}$. This can be thought of as assuming
 \begin{equation}\label{eq:nonlinDerivCommute}
     \frac{1}{\mu}\dx{k}\!\big[\mu\iLmu[g]\big] = \iLmu[\dx{k}g]
 \end{equation}
 instead of Equation \eqref{eq:DerivativeSwitchIdentity}. In this regard, it is useful to compare \eqref{eq:nonlinDerivCommute} to \eqref{eq:linDerivCommute} and note the following:
 \begin{enumerate}[label=(\alph*)]
     \item The closure that results from this assumption remains entropy dissipative. This is because of the following change to the resulting coarse-scale dissipation contribution
 \begin{align*}
    &\epsilon^3\!\abrac{\frac{1}{\mu}\vdx{k}\!\big[\mu \iLmu[\vdx{j}\ln\mu] \big], \mu\, \iLmu\!\!\sqbrac{\frac{1}{\mu}\vdx{l}\!\sqbrac{\mu\, \iLmu[v_n\dx{n}\ln\mu]}\!}\!} \ \le 0\\
    &\vspace{0.5cm} \hspace{5cm} { \bm{\downarrow}}\\
    &\hspace{1cm}\epsilon^3\abrac{v_k \iLmu[\vdx{i}\dx{k}\!\ln\mu] , \mu\, \iLmu\!\sqbrac{v_l\, \iLmu[v_n\dx{n}\dx{l}\!\ln\mu]}}\ \le 0
\end{align*}
\item Equation \eqref{eq:nonlinDerivCommute} is a direct linearization of Equation \eqref{eq:linDerivCommute} and thus the linear Entropy Stable Extension closure \eqref{eq:altburnstress} and \eqref{eq:altburnheat} are a direct linearization of the non-linear closure that results from \eqref{eq:nonlinDerivCommute}. 
 \end{enumerate}
  
The macroscopic closures that arise from this simplification to the non-linear Entropy Stable Extension are: 
\begin{align}
    &\sigma_{ij}^{(2^{SE})} = 2 \omega\, \tepsilonu{ij} - \dx{k}\!\mathfrak{S}_{ijk}\label{eq:SESEdevStress}\\
    &q_i^{(2^{SE})} = - \kappa\, \dx{i}\theta +\dx{j} \mathfrak{H}_{ij}+ \dx{j}\!u_k\,\mathfrak{S}_{ijk}\label{eq:SESEheatFlux}
\end{align}
where $\omega$ and $\tepsilonu{ij}$ are given by Equations \eqref{eq:omegadef} and \eqref{eq:tepsilonDef} respectively and
\begin{align}
\mathfrak{S}_{ijk} &=  2 {\alpha}_1\! \left(\dx{k}\tepsilon_{ij}(\bm{u}) +  \dx{i}\tepsilon_{jk}(\bm{u})+  \dx{j}\tepsilon_{ik}(\bm{u})  -\frac{2}{D}\delta_{ij}\dx{l}\tepsilon_{lk}(\bm{\tu}) \right)\nonumber\\ \nonumber&\hspace{.75cm}
      +2\,{\alpha}_2 \left(\delta_{ik}\dx{l}\tepsilon_{lj}(\bm{u}) +\delta_{jk}\dx{l}\tepsilon_{li}(\bm{u}) - \frac{2}{D}\delta_{ij}\dx{l}\tepsilon_{lk}(\bm{u})\right)
      \\\nonumber&\hspace{1cm}
      +2 \mathcal{A}\, u_k\, \uD[\tepsilon_{ij}(\bm{u})]+ 2\mathcal{P}\, \bm{u}\!\cdot\!\nabla\!\!\left[\frac{\delta_{ik}\partial_{x_j}\theta +\delta_{jk}\partial_{x_i}\theta}{2} - \frac{1}{D}\delta_{ij}\partial_{x_k}\theta  \right] 
      \\&\hspace{1.5cm}
      + 2\mathcal{D}\, u_k\, \left(\partial_{x_i}\partial_{x_j} - \frac{\delta_{ij}}{D}\Delta\right)\!\![\theta]\label{eq:SESEdevSpress}\\\nonumber \\
    \mathfrak{H}_{ij} &= \Lambda_0\, \Delta\theta \delta_{ij} + \Lambda_1 \dx{i}\dx{j}\theta + \mathcal{B}\,u_j\uD[\dx{i}\theta]+ 2\mathcal{D}\,\theta\, \uD[\tepsilonu{ij}] \nonumber\\& \hspace{1cm} + 2\mathcal{P}\, \theta \, u_j \dx{n}\tepsilonu{ni} \label{eq:SESEheatTensor}
\end{align}
with fluid parameters $\alpha_i(t, \bm{x})$, $\Lambda_i(t,\bm{x})$, $\mathcal{A}(t, \bm{x})$, $\mathcal{B}(t,\bm{x})$, $\mathcal{P}(t, \bm{x})$ and $\mathcal{D}(t, \bm{x})$ that depend on the inverse linearized operator $\iLmu$. Notice that $\mathfrak{S}_{ijk}$ and $\mathfrak{H}_{ik}$ are the non-linear versions of $\tilde{\mathfrak{S}}_{ijk}$ and $\tilde{\mathfrak{H}}_{ik}$ defined earlier in \eqref{eq:linESESpress} and \eqref{eq:linESEHeatTensor} respectively. Unlike their linear counterparts, $\mathfrak{S}_{ijk}$ and $\mathfrak{H}_{ik}$ explicitly couple the bulk velocity $\bm{u}$ to the temperature $\theta$ through the terms with fluid parameters $\mathcal{D}$ and $\mathcal{P}$. The values under Hard Spheres and BGK for the fluid parameters are given in Table \ref{tab:nonLinearFluidParameters} and the procedure for their computation is described in Appendix \ref{appendix:HardSphereCalc}.

\begin{table}[ht]
    \centering
    {\tabulinesep=2mm
    \begin{tabu}{|m{6cm}|  m{6cm}| m{6cm} | }
    \hline
    \multicolumn{1}{|c|}{\textbf{Fluid Parameters}} & \multicolumn{1}{c|}{\textbf{BGK}} & \multicolumn{1}{c|}{\textbf{Hard Spheres}}
    \\
     \hline
        \multicolumn{1}{|c|}{$\tau$} &\multicolumn{1}{c|}{$-$} & \multicolumn{1}{c|}{$L  \sqrt{{\pi}/ ({8 \,\theta(t, \bm{x})})}$}  
    \\
     \hline
    \multicolumn{1}{|c|}{$ (\epsilon \tau \rho_0 \theta)^{-1}\, \omega$} &\multicolumn{1}{c|}{1} & \multicolumn{1}{c|}{1.270042 } 
    \\
    \hline
        \multicolumn{1}{|c|}{$\frac{2}{5}(\epsilon\tau \rho_0 \theta)^{-1}\,\kappa$} &\multicolumn{1}{c|}{1} & \multicolumn{1}{c|}{1.922284} 
    \\
     \hline
     \multicolumn{1}{|c|}{$\left(\epsilon^3\tau^3 \,{\rho_0^3 \theta}/{\rho^2}\right)^{-1}  \mathcal{A}$ } & \multicolumn{1}{c|}{1} & \multicolumn{1}{c|}{2.134497}
     \\ 
      \hline
     \multicolumn{1}{|c|}{$\frac{2}{5}\left(\epsilon^3\tau^3 \,{\rho_0^3 \theta}/{\rho^2}\right)^{-1}  \mathcal{B}$ } & \multicolumn{1}{c|}{1} & \multicolumn{1}{c|}{7.440228}
     \\
      \hline
     \multicolumn{1}{|c|}{$\left(\epsilon^3\tau^3 \,{\rho_0^3 \theta}/{\rho^2}\right)^{-1}  \mathcal{D}$ } & \multicolumn{1}{c|}{1} & \multicolumn{1}{c|}{1.946232}
     \\
      \hline
     \multicolumn{1}{|c|}{$\left(\epsilon^3\tau^3 \,{\rho_0^3 \theta}/{\rho^2}\right)^{-1} \mathcal{P}$ } & \multicolumn{1}{c|}{1} & \multicolumn{1}{c|}{3.856049}
     \\ 
      \hline
     \multicolumn{1}{|c|}{$\left(\epsilon^3\tau^3 \,{\rho_0^3 \theta^2}/{\rho^2}\right)^{-1} \alpha_1$ } & \multicolumn{1}{c|}{1} & \multicolumn{1}{c|}{1.274657}
     \\ 
       \hline
     \multicolumn{1}{|c|}{$\left(\epsilon^3\tau^3 \,{\rho_0^3 \theta^2}/{\rho^2}\right)^{-1} \alpha_2$ } & \multicolumn{1}{c|}{0} & \multicolumn{1}{c|}{0.2895342}
     \\ 
       \hline
     \multicolumn{1}{|c|}{$\left(\epsilon^3\tau^3 \,{\rho_0^3 \theta^2}/{\rho^2}\right)^{-1} \Lambda_0$ } & \multicolumn{1}{c|}{$\frac{1}{3}$} & \multicolumn{1}{c|}{7.534324}
     \\ 
       \hline
     \multicolumn{1}{|c|}{$\left(\epsilon^3\tau^3 \,{\rho_0^3 \theta^2}/\rho^2\right)^{-1} \Lambda_1$ } & \multicolumn{1}{c|}{9 } & \multicolumn{1}{c|}{29.93582}
     \\ 
     \hline
    \end{tabu}
    }
    \caption{{\color{black} Non-dimensionalized non-linear fluid parameters for BGK and Hard Spheres collision operators. The constant $L$ in the definition of the Hard Spheres $\tau$ is the macroscopic length scale for the system under consideration i.e. $\epsilon = \frac{\ell
    }{L}$.}}
    \label{tab:nonLinearFluidParameters}
\end{table}

The conservation equations that result from using the simplified Entropy Stable Extension closures \eqref{eq:SESEdevStress} and \eqref{eq:SESEheatFlux} are:

\begin{align}
\text{St }\partial_t\rho + \partial_{x_i} [\rho u_i]&=0 \label{eq:SESEmasslaw}\\
\text{St }\partial_t[\rho u_j] + \partial_{x_i} [\rho {u}_i{u}_j+ \rho \theta\delta_{ij} - 2\omega\, \tepsilon_{ij}(\bm{u}) + \partial_{x_k}\mathfrak{S}_{ijk}]&=0\label{eq:SESEmomentumlaw}\\
\text{St }\partial_t\left[\frac{1}{2}\rho|\bm u|^2+\frac{D}{2}\rho \theta\right] + \partial_{x_i} \Bigg[\left(\frac{1}{2}\rho|\bm u|^2 + \frac{D}{2} \rho\theta \right)u_i+ \rho\theta u_i - 2\omega\,\tepsilon_{ij}(\bm{u})\, u_j  \hspace{.5cm}\nonumber& \\ -\kappa\, \partial_{x_i}\theta+\, \partial_{x_k}\!\left[\mathfrak{S}_{ijk}u_j + \mathfrak{H}_{ik} \right]\Bigg]&=0\label{eq:SESEenergylaw}
\end{align}

This system of equations possesses an entropy dissipation inequality of the form
\begin{align*}
    &\St\, \partial_t \bar{H} + \dx{i}\big[\bar{H}u_i \big] - \dx{i}\left[\frac{q_i^{(2^{SE})}}{\theta} \right]   -\, \dx{k}\!\sqbrac{\frac{ \dx{i}u_j}{\theta}\mathfrak{S}_{ijk} + \frac{ \dx{i}\theta}{\theta^2}\mathfrak{H}_{ik}} = \mathfrak{D} \le 0
\end{align*}
where the entropy $\bar{H}$ and the entropy dissipation rate $\mathfrak{D}$ are given by
\begin{align*}
  &\bar{H} := \abrac{\ln\mu - 1, \mu} = \rho\, \left( \ln\!\left(\!\frac{\rho}{(2\pi\theta)^{\frac{D}{2}}}\! \right)  -\frac{D+2}{2}\right)\\
  &\mathfrak{D} := \epsilon\!\abrac{\vdx{i}\!\ln\mu, \mu\iLmu[\vdx{k}\!\ln\mu]} + \epsilon^3\! \abrac{v_k \iLmu[\vdx{i}\dx{k}\!\ln\mu] , \mu\, \iLmu\!\sqbrac{v_l\, \iLmu[v_n\dx{n}\dx{l}\!\ln\mu]}}\\
  &\hspace{0.26cm} = -\frac{2\omega}{\theta}\,\tepsilonu{}\bm{:}\tepsilonu{} - \frac{\kappa}{\theta^2}\,|\nabla\theta|^2 - \frac{\dx{k}\dx{i}u_j}{\theta} \mathfrak{S}_{ijk} - \frac{\dx{k}\dx{i}\theta}{\theta^2}\mathfrak{H}_{ik}\\
  &\hspace{0.26cm}=  -\frac{2\omega}{\theta}\,\tepsilonu{}\bm{:}\tepsilonu{} - \frac{\kappa}{\theta^2}\,|\nabla\theta|^2  -\frac{6\alpha_1}{\theta}\dx{(k}\tepsilonu{ij)}\,\dx{(k}\tepsilonu{ij)} - \frac{4\alpha_2}{\theta}|\text{div}\tepsilonu{}|^2 
  \\&\hspace{0.75cm}
 -\frac{\Lambda_0}{\theta^2}(\Delta\theta)^2 - \frac{\Lambda_1}{\theta^2} \nabla\nabla\theta\!\bm{:}\!\nabla\nabla\theta
 - \frac{2\mathcal{A}}{\theta}\ \uD[\tepsilonu{ij}]\uD[\tepsilonu{ij}]
 \\&\hspace{0.75cm}
 - \frac{\mathcal{B}}{\theta^2}\ \uD[\dx{i}\theta]\ \uD[\dx{i}\theta]
  - \frac{4\mathcal{P}}{\theta} \dx{k}\tepsilonu{kj} \ \uD[\dx{j}\theta] 
  \\&\hspace{0.75cm}
  -\ \frac{4\mathcal{D}}{\theta}\ \uD[\tepsilonu{ij}]\ \dx{i}\dx{j}\theta  
\end{align*}
The macroscopic form given to the entropy dissipation $\mathfrak{D}$ is likely not the most natural representation as it does not explicitly show that the dissipation rate is non-positive even though the fine-scale representation it is derived from is non-positive. In particular, there is likely to be a way to combine the $\mathcal{A}$, $\mathcal{B}$, $\mathcal{D}$ and $\mathcal{P}$ terms to achieve such a natural, explicitly non-positive representation.

}

\section{Model Problems {\color{black} for Linear Equations} }
\begin{figure}[h]
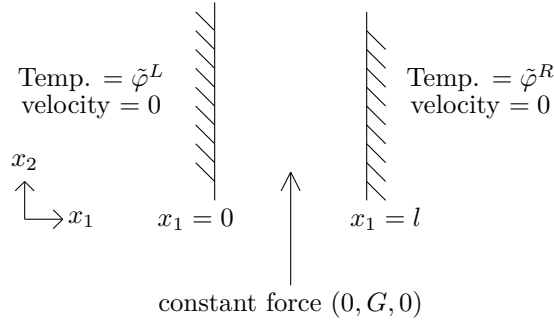

    \centering
    \ctikzfig{fig2}
    \caption{\small The setup consists of two parallel plates, assumed to be infinite in two dimensions, that are a distance $l$ apart. A constant force $G$ parallel to the surface of the plate is applied to the gas within the channel. }
    \label{fig:problemSetup}
\end{figure}

We now test the  {\color{black} linear} Entropy Stable Extension on the parallel plates problem illustrated in Figure \ref{fig:problemSetup}.
The advantage of this setup is that the macroscopic parameters of the system depend of just the coordinate perpendicular to the walls of the channel $x_1$. By imposing the condition $\,\partial_{x_2}[\tilde{\rho}\theta_0 + \rho_0\tilde{\theta}] = -G\,$ on the entire domain, the equations simplify to:
\begin{align}\label{eq:1Daltburn}
    \partial_{x_i}\tilde{u}_i &= 0  &-\tilde{\omega}\, \partial_{x_1}^2\tilde{u}_2 + \tilde{\Phi}\,\partial_{x_1}^4\tilde{u}_2 &= G \nonumber\\
     \rho_0\,\partial_{x_1}\tilde{\theta} + \theta_0\,\partial_{x_1}\tilde{\rho} &= 0
    &-\tilde{\kappa}\, \partial_{x_1}^2\tilde{\theta} + \tilde{\Lambda}\,\partial_{x_1}^4\tilde{\theta} &= 0
\end{align}
where {\color{black}$\tilde{\Phi} = 2\tilde{\alpha}_1+\tilde{\alpha}_2$ and $\tilde{\Lambda}= \tilde{\Lambda}_0 + \tilde{\Lambda}_1$}. 
For our purposes, we only focus on the two equations on the right. 
The corresponding equations for the two versions of the Burnett equations discussed, as well as the Navier-Stokes-Fourier equations, are 
 \begin{align}\label{eq:1Dburnett}
    -\tilde{\omega}\,\partial_{x_1}^2\tilde{u}_2 = G; \qquad
    -\tilde{\kappa}\, \partial_{x_1}^2\tilde{\theta} = 0
\end{align}
All variables and parameters are non-dimensionalized and the dependence on the Knudsen number extracted by introducing the following variables:
\begin{align}
    x &= \frac{x_1}{l} \quad &u &= \frac{\tilde{u}_2}{\sqrt{\theta_0}}\quad &\theta &= \frac{\tilde{\theta}}{\theta_0}\nonumber\\
    B &= \epsilon\frac{Gl^2}{\tilde{\omega}\sqrt{\theta_0}} \quad &k_u &= \epsilon l\,\sqrt{\frac{\tilde{\omega}}{\tilde{\Phi}}}\quad &k_\theta &=\epsilon l\,\sqrt{\frac{\tilde{\kappa}}{\tilde{\Lambda}}}\nonumber
\end{align}
Substituting these variables into \eqref{eq:1Daltburn} gives the following ODEs
\begin{align}
    -\frac{d^2 u}{dx^2} \,+\, \frac{\epsilon^2}{k_u^2}\,\frac{d^4u}{dx^4} = \frac{B}{\epsilon}\label{eq:nondim1Deqnvel}\\\label{eq:nondim1Deqntemp}
     -\frac{d^2 \theta}{dx^2} \,+\, \frac{\epsilon^2}{k_\theta^2}\,\frac{d^4\theta}{dx^4} = 0
\end{align}
We will use this setup to study two problems. The stationary heat transfer problem will focus exclusively on \eqref{eq:nondim1Deqntemp} and the Poiseuille channel problem will focus exclusively on \eqref{eq:nondim1Deqnvel}.

The constants in \eqref{eq:1Daltburn} take the form:
\begin{align*}
    \tilde{\omega} &= \gamma_1\,\epsilon\tau \rho_0\theta_0 \qquad &\tilde{\kappa} &= \frac{5}{2}\gamma_2\,\epsilon\tau\rho_0\theta_0\\ \tilde{\Phi}&=\Gamma_1\,\,\epsilon^3\tau^3\rho_0\theta_0^2 \qquad &\tilde{\Lambda} &= \Gamma_2\,\,\epsilon^3\tau^3\rho_0\theta_0^2
\end{align*}
where $\gamma_1$, $\gamma_2$, $\Gamma_1$ and $\Gamma_2$ are parameters that depend on the choice of collision operator and $\tau = l \sqrt{\frac{\pi}{8\theta_0}}$ for what follows.\footnote{This value of $\tau$ can be derived from equation (18b) and (21) of \cite{Aoki2017}. The mean collision frequency of a hard sphere gas at the equilibrium state described by the global Maxwellian $M$ is $\frac{1}{\epsilon \tau}$.} These can be calculated using the method described in Appendix \ref{appendix:HardSphereCalc}. We list their values for the Hard Spheres and BGK operators in Table \ref{tab:FluidParameters}. 

From this point on unless otherwise indicated, we shall work exclusively with the Hard Spheres parameters when concrete solutions need to be computed. 

\begin{table}[h]
   \begin{center}
{\tabulinesep=1.2mm
\begin{tabu}{|m{5.5em}|m{5.5em}|m{5.5em}|m{5.5em}|m{5.5em}|}
\hline
 \multicolumn{5}{|c|}{\textbf{Calculated Fluid Parameters for BGK and Hard spheres} }\\
\hline
 & \multicolumn{1}{c|}{$\gamma_1$} & \multicolumn{1}{c|}{$\gamma_2$} & \multicolumn{1}{c|}{$\Gamma_1$} & \multicolumn{1}{c|}{$\Gamma_2$}
\\
\hline
\multicolumn{1}{|c|}{\textbf{BGK}} & \multicolumn{1}{c|}{1} & \multicolumn{1}{c|}{1} & \multicolumn{1}{c|}{2} & \multicolumn{1}{c|}{$\frac{28}{3}$}\\
\hline
\multicolumn{1}{|c|}{\textbf{Hard Spheres}} & \multicolumn{1}{c|}{1.270042} & \multicolumn{1}{c|}{1.922284} & \multicolumn{1}{c|}{2.838848} & \multicolumn{1}{c|}{$37.47015$}\\
\hline 
\end{tabu}
}
\end{center}
    \caption{ }
    \label{tab:FluidParameters}
\end{table}
{\color{black}
We end by briefly noting that for the parallel plates problem in Figure \ref{fig:problemSetup}, non-linear Simplified Entropy Stable Extension equations \eqref{eq:SESEmasslaw}-\eqref{eq:SESEenergylaw} will be 
\begin{align*}
    \dx{1}[\rho u_1] &= 0\\
    \dx{1}\!\sqbrac{\rho u_1^2 + \rho \theta - \frac{4\omega}{3}\dx{1}u_1  +\dx{1}\mathfrak{S}_{111}} &= 0\\
    \dx{1}\!\sqbrac{\rho u_1 u_2 - \omega\, \dx{1}u_2 + \dx{1}\mathfrak{S}_{121}} &=0\\
    \dx{1}\!\sqbrac{\frac{1}{2}\big( |\bm{u}|^2 + 5\theta \big) \rho u_1 - \frac{4\omega}{3}u_1\dx{1}u_1 - \omega\, u_2\,\dx{1}u_2  - \kappa \dx{1}\theta + \dx{1}\!\sqbrac{\mathfrak{S}_{111}u_1 + \mathfrak{S}_{121}u_2 + \mathfrak{H}_{11}}} &= 0
\end{align*}
where 
\begin{align*}
    \mathfrak{S}_{111} &= \frac{4}{9}(7\alpha_1 + 2 \alpha_2) \dx{1}u_1 + \frac{4}{3}\mathcal{A} \,u_1^2\dx{1}u_1 + \frac{4}{3}(\mathcal{P} + \mathcal{D})\,u_1 \,\dx{1}^2\theta\\
    \mathfrak{S}_{121} &= (2\alpha_1 + \alpha_2)\,\dx{1}^2u_2 + \mathcal{A}\,u_1^2\,  \dx{1}^2u_2\\
    \mathfrak{H}_{11} &= (\Lambda_0+\Lambda_1)\,\dx{1}^2\theta +\mathcal{B}\,u_1^2\dx{1}^2\theta + \frac{2}{3}(\mathcal{D}+\mathcal{P})\,\theta\,u_1\,\dx{1}^2u_1
\end{align*}
with the fluid parameters $\alpha_i$, $\Lambda_i$, $\mathcal{A}$, etc given by Table \ref{tab:nonLinearFluidParameters}. It is straightforward to check that the linear equations \eqref{eq:1Daltburn} can be derived from the above non-linear equations by substituting in $\rho = \rho_0 + \epsilon \trho$, $\bm{u} = \epsilon\bm{\tu}$ and $\theta = \theta_0 + \epsilon \ttheta$ and keeping only the linear terms that result from the substitution.
}

\subsection{Boundary conditions}
In general, the derivation of boundary conditions for a given macroscopic closure requires its own specialized analysis. For example, Aoki et al. \cite{Aoki2017} and Coron \cite{Coron1989} employ an asymptotic boundary layer analysis on the Boltzmann equation to derive slip boundary conditions for the Navier-Stokes-Fourier equations while Gu and Emerson \cite{Gu2007} use a Hermite polynomial expansion of the distribution function supplemented by correction factors obtained from DSMC modeling of planar Couette flow to derive a complete set of boundary conditions for the regularized 13 moment equations \cite{Struchtrup2003}. 

In order to obtain boundary conditions for the entropy stable extension, we first determine complementary pairs of essential and natural boundary conditions for the two main equations of \eqref{eq:1Daltburn} by deriving a suitable weak form for them. Using test functions $w,\, \vartheta \in H^2([0,l])$ and integrating by parts {\color{black}(once on the Stokes-Fourier terms and twice for the Entropy Stable Extension terms)}, we get 
\begin{align*}
   \int_0^l w\, G\,dx_1 &= \tilde{\omega}\!\int_0^l \partial_{x_1}\! w\ \partial_{x_1}\!u_2 \,dx_1 + \tilde\Phi\!\int_0^l \partial_{x_1}^2\! w\, \partial_{x_1}^2\!u_2 \,dx_1 + \left(\!-w \ n_1\tilde{\sigma}_{12}^{(2^E)} \, - \, \partial_{x_1}\!w\  n_1\tilde{\Phi} \partial_{x_1}^2\!\tilde{u}_2\!\right)\Bigg|_{x=0}^{x=l}\\
    0 &= \tilde{\kappa}\!\int_0^l \partial_{x_1}\! \vartheta\ \partial_{x_1}\!\tilde\theta \,dx_1 + \tilde\Lambda\!\int_0^l \partial_{x_1}^2\! \vartheta\, \partial_{x_1}^2\!\tilde\theta \,dx_1 + \left(\vartheta \ n_1\tilde{q}_1^{(2^E)} \, - \, \partial_{x_1}\!\vartheta\  n_1\,\tilde{\Lambda}\, \partial_{x_1}^2\!\tilde{\theta}\right)\Bigg|_{x=0}^{x=l}
\end{align*}\normalsize
 where $n_1$ is the outward pointing normal at the boundary (so $n_1(0) = -1$ and $n_1(1) = 1$ ). We observe two pairs of complementary essential and natural boundary conditions inherent to the equations: the first pair indicates that we either specify the temperature $\tilde\theta$ (resp. $\tilde u_2$) at the boundary or specify the heat flux $n_1\tilde q_1^{(2^E)}$ (resp. $n_1\tilde \sigma^{(2^E)}_{12} $) at the boundary and the second pair indicates that we  either specify $n_1\partial_{x_1}\tilde{\theta}$  (resp.$n_1\partial_{x_1}\tilde u_2$) or specify the second derivative term $\tilde\Lambda\partial^2_{x_1}\tilde\theta$ (resp. $\tilde\Phi_1\partial_{x_1}^2\tilde u_2$). {\color{black} A more general version of this analysis is given at the beginning of Chapter 7 of \cite{Baidoo2025}.} 
 
 This analysis only tells us where to impose boundary conditions but not what the imposed boundary conditions should be. As a starting point for this latter goal, we replace the linearized Boltzmann equation's distribution function with our closure in the kinetic boundary condition. For the diffuse reflection boundary condition, which will be our focus in this paper, this amounts to computing:
 
\footnotesize
\begin{align}
 \int_{\{v_1n_1\ge0\}}\!  \left( \bar{f} + \breve{f}_{(2^E)}\!\right)M\,v_1 n_1 d\bm v + \int_{\{v_1n_1\le0\}} \! \left(\! \frac{\rho_b}{\rho_0} + \frac{\tilde{\bm{u}}_b\cdot\bm{v}}{\theta_0}+ \frac{\tilde{\varphi}_b}{2 \theta_0^2}(|\bm v|^2-D\theta_0)\!\right)\!M v_1 n_1 d\bm v &= 0  \label{eq:ZeroMassFlux}\\
 \int_{\{v_1n_1\le0\}}\!\bar{m}\!\left( \bar{f} + \breve{f}_{(2^E)}\!\right)\!M v_1 n_1  d\bm v - \int_{\{v_1n_1\le0\}}\!\bar{m}\!  \left( \frac{\rho_b}{\rho_0} + \frac{\tilde{\bm{u}}_b\cdot\bm{v}}{\theta_0}+ \frac{\tilde{\varphi}_b}{2 \theta_0^2}(|\bm v|^2-D\theta_0)\right)\!M v_1 n_1  d\bm v&=0\label{eq:MomentAccom}
\end{align}
\normalsize
 where $\bar{m} \in \{1, v_2, |\bm{v}|^2 \}$, $\tilde{\varphi}_b$ and $\tilde{\bm{u}}_b$ are the temperature and velocity of the boundary and $\rho_b$ is the mass density at the boundary required to ensure that the zero mass flux condition \eqref{eq:ZeroMassFlux} is satisfied. At $x=0$, the computation of \eqref{eq:ZeroMassFlux} and \eqref{eq:MomentAccom} yields
\begin{align}
    \tilde{u}_1 &= 0\label{eq:noflux0}\\
     \frac{1}{\rho_0}\big(\rho(0) - \rho_b^L\big) + \frac{1}{2}\,\big(\theta(0) - \varphi^L\big) &= 0\label{eq:halfmass0}\\
      \sqrt{\frac{2}{\pi}}\big(u(0)-u_b(0)\big) - \frac{\tilde{\omega}}{\epsilon l\rho_0\sqrt{\theta_0} }\left(\epsilon\,\frac{du}{dx}(0)\right)  + \frac{\tilde{\Phi}}{\epsilon^3 l^3\rho_0\sqrt{\theta_0} }\left(\epsilon^3\frac{d^3u}{dx^3}(0)\right)&= 0 \label{eq:halfTangentialMomentum0}\\
       \sqrt{\frac{2}{\pi}}\left(\frac{2(\rho(0) - \rho_b^L)}{\rho_0} +3\big(\theta(0) - \varphi^L\big)\right) -\frac{\tilde{\kappa}}{\epsilon l\rho_0\sqrt{\theta_0}}\,\left(\epsilon\frac{d\theta}{dx}(0)\right)\nonumber\\+\,\frac{\tilde{\Lambda}}{\epsilon^3l^3\rho_0\sqrt{\theta_0}}\,\left(\epsilon^3\frac{d^3\theta}{dx^3}(0)\right) &=0\label{eq:halfenergy0}
\end{align}
\normalsize
{\color{black} At $x=1$, the signs on the derivative terms switch.}
From these equations, we obtain natural boundary conditions of the form:  
\begin{align}
       - \sigma_{12}^{(2^E)}(x^b)\, n_1(x^b):=-\left(\!\epsilon\underline{\omega}\,\frac{du}{dx}(x^b)-\epsilon^3\underline{\Phi}\frac{d^3u}{dx^3}(x^b)\!\right)n_1(x^b)  &=  \sqrt{\frac{2}{\pi}}\,\Big(u(x^b) - u_b(x^b)\Big)\label{eq:halfTangentialMomentum}\\
       \,q^{(2^E)}_1(x^b)\,n_1(x^b):=\left(\!-\epsilon\underline{\kappa}\,\frac{d\theta}{dx}(x^b) +\epsilon^3 \underline{\Lambda}\frac{d^3\theta}{dx^3}(x^b)\!\right)n_1(x^b)\,    &=  2\sqrt{\frac{2}{\pi}}\,\Big(\theta(x^b) - \varphi_b\Big)\label{eq:halfenergy}
\end{align}\normalsize
where $x^b \in \{0,1\},\,$ $\underline{\omega}=\frac{\tilde{\omega}}{\epsilon l \rho_0\sqrt{\theta_0}},\,$ $\underline{\kappa}=\frac{\tilde{\kappa}}{\epsilon l \rho_0\sqrt{\theta_0}},\,$ $\underline{\Phi}=\frac{\tilde{\Phi}_1}{\epsilon^3 l^3 \rho_0\sqrt{\theta_0}}$ and  $\underline{\Lambda}=\frac{\tilde{\Lambda}}{\epsilon^3 l^3 \rho_0\sqrt{\theta_0}}$ . {\color{black} Observe that these appear to be a generalization of the Maxwell slip \cite{Maxwell1879, Le2012} and Smoluchowski jump \cite{SmoluchowskivonSmolan1898, Le2012} boundary conditions for the 1D channel.}

We would still need boundary conditions for $\underline{\Phi}\frac{d^2 u}{dx^2}$ and $\underline{\Lambda}\frac{d^2\theta}{dx^2}$ in order to completely determine a solution. A prescription for each will be made in the model problems that follow.


\subsection{Stationary Heat Transfer Problem}
The general solution to Equation \eqref{eq:nondim1Deqntemp} is
\begin{equation}\label{eq:stationaryheatsolution}
  \theta(x) = c_1 \exp\left(\frac{k_\theta x}{\epsilon}\right)\, +\, c_2 \exp\left(-\frac{k_\theta x}{\epsilon}\right) \,+\,c_3 x \,+\, c_4
\end{equation}
and the heat flux across the channel will be 
\begin{align}\label{eq:StatHeat_HeatFlux}
    q^{(2^E)}_1 :=  -\epsilon \underline\kappa\frac{d\theta}{dx} +\epsilon^3 \underline\Lambda \frac{d^3\theta}{dx^3} = -\epsilon\underline\kappa\left(\frac{d\theta}{dx} -\frac{\epsilon^2}{k_\theta^2}  \frac{d^3\theta}{dx^3} \right)  = -\,\epsilon \underline\kappa\,c_3. 
\end{align}
In order solve the $c_i$ uniquely, we use \eqref{eq:halfenergy} in addition to the following boundary condition:
\begin{equation}\label{eq:statHeatXtraBC}
\epsilon^2\underline{\Lambda}\frac{d^2\theta}{dx^2}(x^b) = \epsilon\, \underline{\zeta}\, \frac{d\theta}{dx}(x^b)n_1(x^b) + \underline{\chi}\, \big(\theta(x^b) - \varphi_b\big)
\end{equation}
where $\underline\zeta$ and $\underline\chi$ are as yet undetermined constants. As a result,
\begin{align}\label{eq:parametersoln}
   c_1 &= -c_2\, e^{-\frac{k_\theta}{\epsilon} }  = - \frac{\epsilon}{2c_0}\big(4\underline\zeta - \underline\chi\, \underline{\kappa}\sqrt{2\pi} \big)(\varphi^R - \varphi^L)\nonumber\\
   c_3 &= \frac{2}{c_0}\Big(k_\theta\, \underline\zeta\, \big(1+ e^{\frac{k_\theta}{\epsilon} }  \big) + \underline{\kappa}\, \big(1- e^{\frac{k_\theta}{\epsilon} }  \big) \Big)(\varphi^R-\varphi^L)\\
   c_4 &= \frac{1}{2c_0}\Bigg(\epsilon(\varphi^R+\varphi^L)\left(k_\theta\underline\zeta\underline\kappa\sqrt{2\pi}\big(1+ e^{\frac{k_\theta}{\epsilon} }  \big)  + \big(4\underline\zeta+\underline\kappa(\underline\kappa-\underline\chi)\sqrt{2\pi} \big)\big(1- e^{\frac{k_\theta}{\epsilon} }  \big) \right)\nonumber\\
&\qquad\qquad\qquad\quad\qquad\qquad\qquad\qquad\qquad\,
+4\varphi^L\Big(k_\theta\underline\zeta\big(1+ e^{\frac{k_\theta}{\epsilon} }  \big) +\underline\kappa\big(1- e^{\frac{k_\theta}{\epsilon} }  \big)  \Big)\Bigg)\nonumber
\end{align}
where $c_0 = \underline\zeta k_\theta\,(2+\epsilon\underline\kappa\sqrt{2\pi})\big(1+ e^{\frac{k_\theta}{\epsilon} }  \big) + \big(\epsilon(4\underline\zeta+\underline\kappa(\underline\kappa-\underline\chi)\sqrt{2\pi}) + 2\underline\kappa\big)\big(1- e^{\frac{k_\theta}{\epsilon} }  \big)  $. 
We still need to determine good values for $\underline{\zeta}$ and $\underline{\chi}$. To this end, we use the data points in Table 4.1 of \cite{SoneBook} which were obtained by numerically solving the linearized Boltzmann equation with the hard spheres operator. With $\varphi^R=1$ and $\varphi^L=0$, this heat flux data, $q^{\text{Data}}$, and our non-dimensional heat flux \eqref{eq:StatHeat_HeatFlux} are related by  $q^{\text{Data}} = -q^{(2^E)}_1\sqrt{2}$ while the Knudsen number $k$ in \cite{SoneBook} and our Knudsen number  $\epsilon$ are related by $k = \frac{\sqrt{\pi}}{2}\epsilon$.  Leaving $\underline\zeta$ and $\underline\chi$ as free parameters, we use the \textit{fitnlm} function in MATLAB to find the values that best fit \eqref{eq:StatHeat_HeatFlux} to the rescaled linearized Boltzmann solution data $-\frac{1}{\sqrt{2}}q^{\text{Data}}$. This process yields values of $\underline\zeta = -0.96491$ and $\underline\chi = -0.94298$ with $R^2 = 1$ and root mean square error of $0.000671$ between the data and our function.

\begin{figure}[ht]
    \centering
    \includegraphics[width=.8\textwidth]{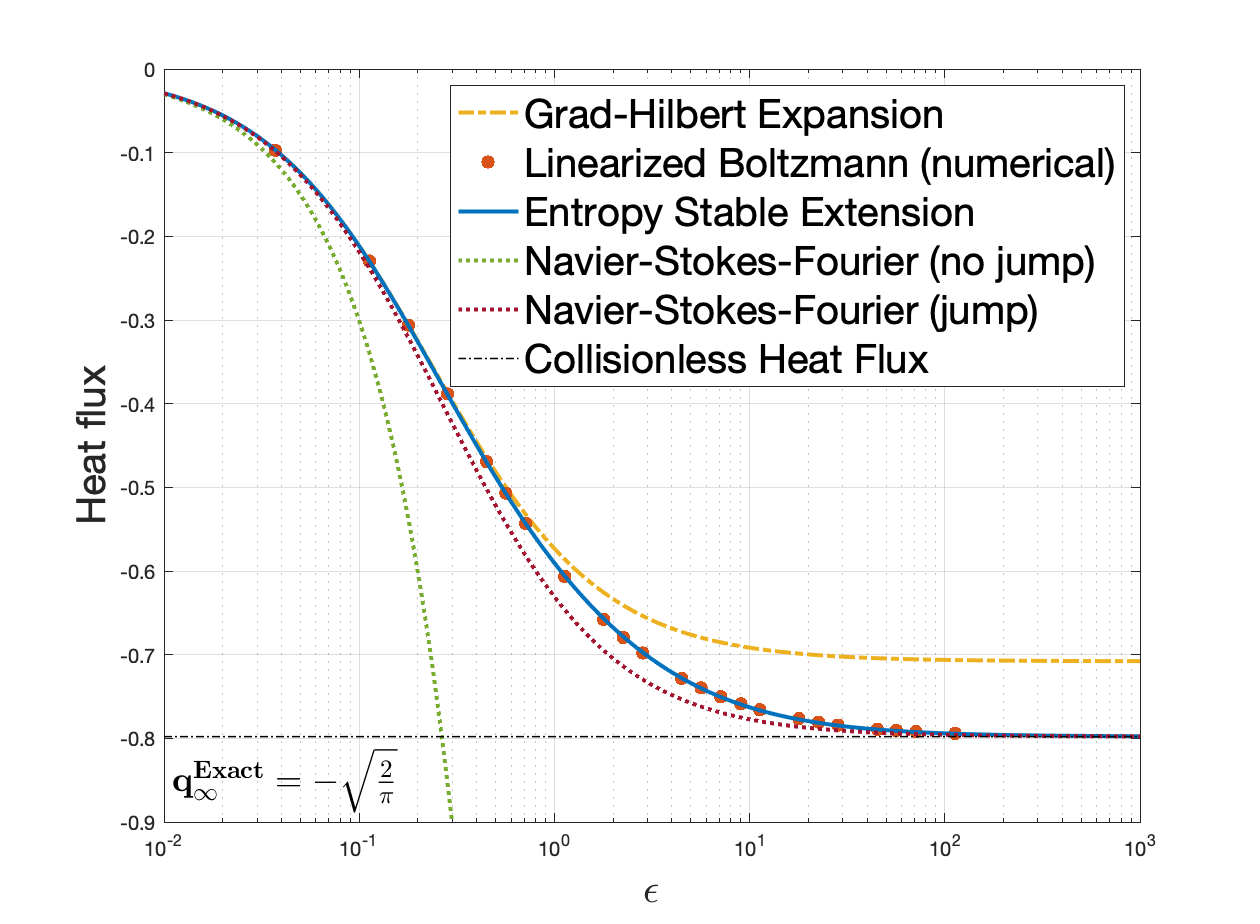}
    \caption{\small Heat flux as a function of the Knudsen number in the stationary heat problem. All the methods compared agree in the small $\epsilon$ regime but the Navier-Stokes-Fourier solution without a jump condition (green dotted line) deviates first as $\epsilon$ increases. The Navier-Stokes-Fourier solution with a jump condition (maroon dotted line) remains relatively close to the linearized Boltzmann solution and even converges to the correct collisionless limit heat flux. The Grad-Hilbert solution (orange dot-dashed line) does better than the Navier-Stokes-Fourier solution with a jump condition for $\epsilon\le1$ but converges to the wrong value in the collisionless limit. Finally, the solution due to our entropy stable extension agrees with the linearized Boltzmann solution remarkably well over the entire range of Knudsen numbers observed. }
    \label{fig:heatfluxcompare}
\end{figure}

In Figure \ref{fig:heatfluxcompare}, we plot Equation \eqref{eq:StatHeat_HeatFlux} with the obtained fitted values and $-\frac{1}{\sqrt{2}}q^{\text{Data}}$. We also plot the Navier-Stokes-Fourier heat flux solution with no temperature jump $q_1^{(1)} = -\epsilon \underline\kappa$, as well as a Navier-Stokes-Fourier heat flux solution with a temperature jump obtained by solving Equation \eqref{eq:1Dburnett} using the boundary condition \eqref{eq:halfenergy} without the third order derivative term $$q_1^{(1)} = -\frac{\epsilon \underline\kappa}{1+\epsilon\underline\kappa\sqrt{\frac{\pi}{2}}}$$
We also include the solution due to the Grad-Hilbert expansion given in Equation (4.18c) of \cite{SoneBook}. In our non-dimensionalization, this will be 
$$q^{\text{GH}} = - \frac{\epsilon \underline\kappa}{1+\epsilon\, d_1\sqrt{\pi}},$$ where $d_1 = 2.4001$ for hard spheres. Finally, we note that in the collisionless limit $\epsilon\rightarrow\infty$, the linearized Boltzmann equation can be solved exactly with heat flux calculated to be $q_{\infty}^{\text{Exact}} = - \sqrt{\frac{2}{\pi}}$. Due to the boundary condition \eqref{eq:halfenergy}, it turns out that the heat flux due to our entropy stable extension  will always converge to $q_{\infty}^{\text{Exact}}$ regardless of the value of $\underline\kappa$, $\underline\Lambda$, $\underline\zeta$ and $\underline\chi$ used. {\color{black} This is also true for the Stokes-Fourier solution with the jump boundary conditions.} 

\begin{figure}[ht]
    \centering
    \includegraphics[width=\textwidth]{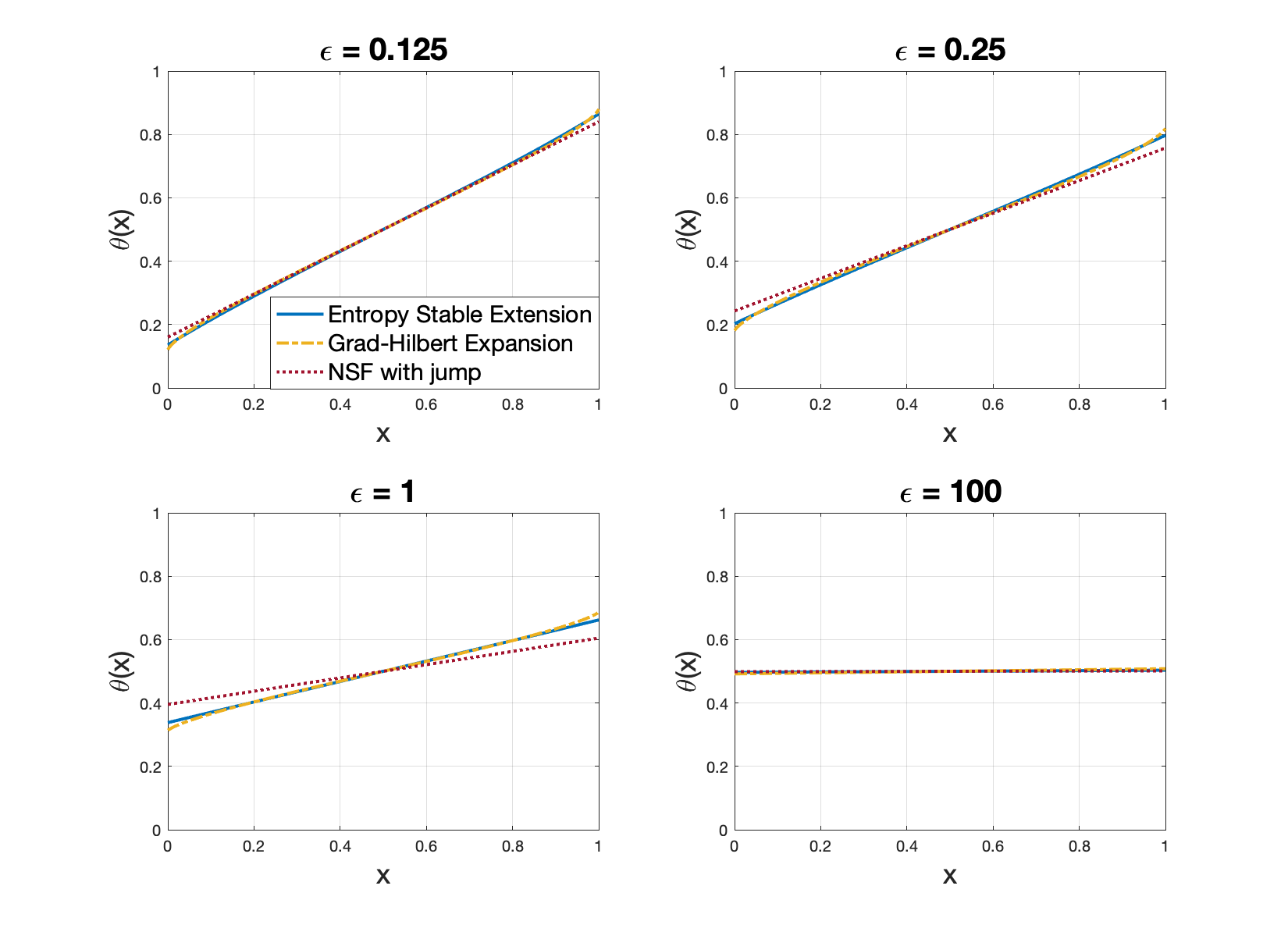}
    \caption{\small Comparison of the non-dimensional temperature distribution obtained via the entropy stable extension (blue line), the Grad-Hilbert expansion (orange dashed) and the Navier-Stokes-Fourier equations with the jump boundary conditions described in the text (red dotted). The three solutions agree strongly in the interior of the channel for smaller Knudsen numbers with a discrepancy at the walls that extends into the domain as the Knudsen number increases ($\epsilon = 0.125,\,0.25,\,\text{and } 1$). The solution due to the entropy stable extension hews more closely to the Grad-Hilbert temperature distribution than the Navier-Stokes-Fourier solution does. Because all three solutions converge to the correct collisionless limit temperature distribution, they again agree strongly for very large Knudsen numbers ($\epsilon=100$).}
    \label{fig:SoneStatHeatTempCompare}
\end{figure}

In Figure \ref{fig:SoneStatHeatTempCompare}, we plot the temperature distribution that results from using $\underline\zeta = -0.96491$ and $\underline\chi= -0.94298$ in Equation \eqref{eq:parametersoln} at $\epsilon = 0.125,\  0.25,\ 1.0 \text{ and } 100$. For comparison, we also plot the temperature distribution due to the Grad-Hilbert extension (Equation (4.18b) of \cite{SoneBook}\footnote{With $\varphi^R=1$ and $\varphi^L=0$, the non-dimensional temperature in \cite{SoneBook} $\frac{T_H-T_0}{T_1-T_0}$ equals $\theta$.}) and the Navier-Stokes-Fourier solution with jump conditions. 

In the collisionless limit, the temperature distribution predicted by the linearized Boltzmann equation is $\frac{\varphi^L+\varphi^R}{2} = \frac{1}{2}$ and we see that all three solutions plotted converge to this limit solution at large Knudsen numbers approach this limit. Thus all solutions are again in agreement at very large Knudsen numbers. {\color{black} It is interesting to note that the temperature solutions converge to the correct collisionless limit solution regardless of the value of $\underline{\kappa}$, $\underline{\Lambda}$, $\underline{\zeta}$ and $\underline{\chi}$ used.} 

\subsection{Poiseuille Channel Flow and the Knudsen Minimum}

The general solution to \eqref{eq:nondim1Deqnvel} is 
\begin{equation}\label{eq:flowprofile}
u(x) = -\frac{Bx^2}{2\epsilon} + d_3 x + d_4 + d_1 \exp\left( \frac{k_u x}{\epsilon}\right) + d_2 \exp\left( -\frac{k_u x}{\epsilon}\right)
\end{equation}
As in the case of the stationary heat problem, we need to specify a boundary condition on $\underline\Phi \frac{d^2u}{dx^2}(x^b)$ in addition to the boundary condition \eqref{eq:halfTangentialMomentum}, in order to fully determine the solution. However we cannot use a boundary condition like \eqref{eq:statHeatXtraBC} because Equation \eqref{eq:nondim1Deqnvel} shows that in the limit where $\epsilon$ goes to zero $\frac{d^2u}{dx^2} = O(\epsilon^{-1})$ and this would be inconsistent with a boundary condition that takes a form similar to \eqref{eq:statHeatXtraBC}. With this in mind, we instead opt for the following boundary conditions 
\begin{align*}
\left(-\epsilon\underline{\omega}\,\frac{du}{dx}(x^b)+\epsilon^3\underline{\Phi}\frac{d^3u}{dx^3}(x^b)\right)n_1(x^b)  = u(x^b) \sqrt{\frac{2}{\pi}}; \qquad \frac{d^2u}{dx^2}(x^b) =   \big(-1 + C(\epsilon)\big)\frac{B}{\epsilon};\qquad x^b\in\{0,\,1\} 
\end{align*}
The second boundary condition can be thought of as a perturbation of the  Navier-Stokes-Fourier solution's second derivative $\left(\text{i.e. }-\frac{B}{\epsilon}\right)$ by the function $C(\epsilon)$. The solution that results is
\begin{align}\label{eq:AltBPoiss}
  \frac{u(x)}{B} &= \frac{(x - x^2)}{2\epsilon} + \frac{\underline\omega}{2}\sqrt{\frac{\pi}{2}} + \frac{\epsilon C(\epsilon)}{k_u^2}\Bigg(\frac{\exp\left(\frac{k_ux}{\epsilon}\right)+\exp\left(\frac{k_u(1-x)}{\epsilon}\right)}{\exp\left(\frac{k_u}{\epsilon}\right)+1} - 1\Bigg)  
\end{align}
where the first two terms represent the Stokes-Fourier contribution to the solution whilst the term with $C(\epsilon)$ is the contribution of our entropy stable extension. Notice that the velocity slip at the boundary is constant for all Knudsen numbers because the $C(\epsilon)$ term vanishes at the boundaries. This means that the current set of boundary conditions produces a velocity slip at the boundary equal to that given by the Navier-Stokes-Fourier solution. This shortcoming will be dealt with in the next section, but first we explore Equation \eqref{eq:AltBPoiss} a bit further.

The perturbation $C(\epsilon)$ determines how the extension to the Stokes-Fourier solution behaves within the domain and thus needs to be carefully chosen. To that end, we calculate the mass flux within the channel. It has been experimentally observed  that the mass flux of the gas in the channel is observed to initially decrease with increasing Knudsen number until it reaches a particular Knudsen number, \textit{the Knudsen minimum}, where the mass flux then increases monotonically \cite{Knudsen1909}. This phenomenon cannot be replicated by the Navier-Stokes-Fourier equations which predict a monotonically decreasing mass flux. Equation \eqref{eq:AltBPoiss} gives the mass flux in the channel as 
\begin{align}\label{eq:altBurnMassFlux}
   \frac{Q(\epsilon)}{B} := \int_0^1 \frac{u(x)}{B}dx
   = \frac{1}{12\epsilon} + \frac{\underline\omega}{2}\sqrt{\frac{\pi}{2}}+ C(\epsilon)\underbrace{\frac{2\epsilon^2}{k_u^3}\left(  \tanh\left(\frac{k_u}{2\epsilon}\right) -\frac{k_u}{2\epsilon}\right) }_{:=\,D(\epsilon)}
\end{align}
and we immediately see that the $C(\epsilon)D(\epsilon)$ term will have to counteract the monotone decreasing behavior of the first two terms as $\epsilon$ increases in order to produce a Knudsen minimum. Because $D(\epsilon)$ is fixed, we can carefully choose $C(\epsilon)$ to select for a particular Knudsen minimum and growth behavior in the collisionless limit.

To illustrate this point, we note that Cercignani and Daneri  
show in \cite{Cercignani1963} that $\frac{Q(\epsilon)}{B} \sim  \frac{\gamma_1}{4}\ln(\epsilon)$ in the collisionless limit. We therefore choose a perturbation of the form $C(\epsilon) = - \epsilon\big(h_1 \ln\big(1 + \,\epsilon\big) + h_2\big)$, where $h_1$ and $h_2$ are constants. This will ensure that $C(0) = 0$ and $C(\epsilon)D(\epsilon) = O(\ln(\epsilon))$ in the collisionless limit. Roughly speaking, $h_1$ tunes the logarithmic growth at large $\epsilon$ whilst $h_2$ tunes the Knudsen minimum in the transition regime. This ansatz is chosen for its simplicity; it is by no means unique nor provably the best possible choice. To determine good values for $h_1$ and $h_2$, we use the data in Table V of the paper by Ohwada et al. \cite{Ohwada1989}, where this problem is solved numerically for the Hard Spheres linearized Boltzmann equation over a wide range of Knudsen numbers. 

\begin{figure}[ht]
    \centering
    \includegraphics[width=.8\textwidth]{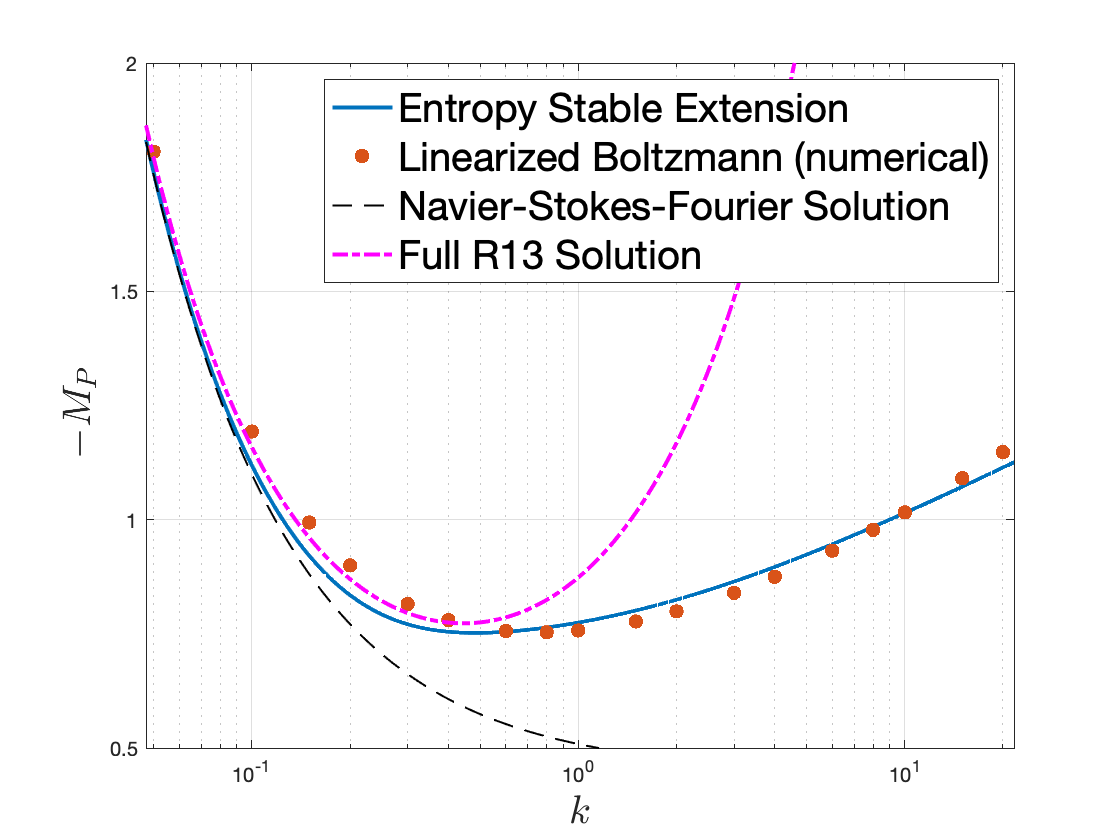}
    \caption{\small We use $C(\epsilon) = - \epsilon\,\big(2.1246\ln(1 + \epsilon ) + 2.3066\big)$ in \eqref{eq:altBurnMassFlux} and plot $\frac{2}{B\gamma_1\sqrt{\pi}}Q\big(\frac{2}{\sqrt{\pi}}k\big)$ against $k$ (blue line).  Also plotted are the linearized Boltzmann data points  due to Ohwada, Sone and Aoki \cite{Ohwada1989} (red dots), the Navier-Stokes-Fourier solution ($C(\epsilon)=0$) (black dash line) and the R13 moments solution from Struchtrup and Torrilhon \cite{Struchtrup2007} (magenta dot-dash). We observe a reasonable agreement between our function and the data points from Ohwada et al., quantified by  $R^2= 0.978$. The Knudsen minimum for the extension occurs around attained at $k = 0.48$. The Stokes solution for the mass flux (black dashed) does not exhibit a Knudsen minimum and decreases monotonically whilst R13 solution is really good for smaller values of $k$ but diverges from the Boltzmann solution as $k$ increases. } 
    \label{fig:sonefluxcompare}
\end{figure}

We first note that the relationship between the non-dimensional mass flux ${M}_P$ in \cite{Ohwada1989} and ours is $\,\frac{Q(\epsilon)}{B} = - \frac{\gamma_1\sqrt{\pi}}{2} M_P\,$ and that the conversion from our Knudsen parameter $\epsilon$ to their Knudsen parameter $k$ is given by $k = \frac{\sqrt{\pi}}{2}\,\epsilon$.
Using the \textit{fitnlm} function in MATLAB gives best fit parameters of $h_1 = 2.1246$ and  $h_2 = 2.3066$ with $R^2= 0.978$ between Equation \eqref{eq:altBurnMassFlux} and the data. In Figure \ref{fig:sonefluxcompare}, we compare the fitted solution $\frac{2}{\gamma_1\sqrt{\pi}}\frac{Q}{B}$ to the mass flux $-M_P$ in Table V, the Navier-Stokes-Fourier equation  solution (i.e. $C(\epsilon) = 0$) and
the regularized 13 moment (R13) equations solution calculated for Maxwell molecules by Struchtrup and Torrilhon in \cite{Struchtrup2007}. A summary of the conversions used to compare these solutions is given in Table \ref{tab:PoissConv}. 

\begin{table}[h]
   \begin{center}
{\tabulinesep=1.2mm
\begin{tabu}{|m{8.5em}|m{8.5em}|m{8.5em}|}
\hline
 \multicolumn{3}{|c|}{\textbf{Poiseuille Channel Conversions} }\\
\hline
{\textit{Linearized Boltzmann  \cite{Ohwada1989}}} & {\textit{Entropy Stable Extension}} & \textit{R13 moment equations \cite{Struchtrup2007}}
\\
\hline
\multicolumn{1}{|c|}{$k$} & \multicolumn{1}{c|}{$\frac{\sqrt{\pi}}{2}\epsilon$} & \multicolumn{1}{c|}{$\frac{4\sqrt{2}}{5}\text{Kn}$} \\
\hline
\multicolumn{1}{|c|}{$-u_P$} & \multicolumn{1}{c|}{$\frac{2}{\gamma_1\sqrt{\pi}}\frac{u}{B}$} & \multicolumn{1}{c|}{$\frac{4\gamma_1\sqrt{2}}{5F}v$}\\
\hline
 \multicolumn{1}{|c|}{$-M_P$} & \multicolumn{1}{c|}{$\frac{2}{\gamma_1\sqrt{\pi}} \frac{Q}{B}$} & \multicolumn{1}{c|}{$\frac{4\gamma_1\sqrt{2}}{5F} J$}
\\%
\hline
\end{tabu}
}
\end{center}
    \caption{\small The conversion scheme for a direct comparison of our entropy stable extension solutions to that of Ohwada et al. \cite{Ohwada1989} and Struchtrup and Torrilhon \cite{Struchtrup2007}. Here, $\gamma_1 = 1.270042$ }
    \label{tab:PoissConv}
\end{table}

Figure \ref{fig:sonefluxcompare} shows that the mass flux due to our entropy stable extension is in reasonable agreement with the linearized Boltzmann solution for the range of Knudsen numbers studied. The R13 solution is just as good as our solution for smaller values of $k$ but diverges wildly from linearized Boltzmann solution shortly after attaining a Knudsen minimum.

\begin{figure}[ht]
    \centering
    \includegraphics[width = \textwidth]{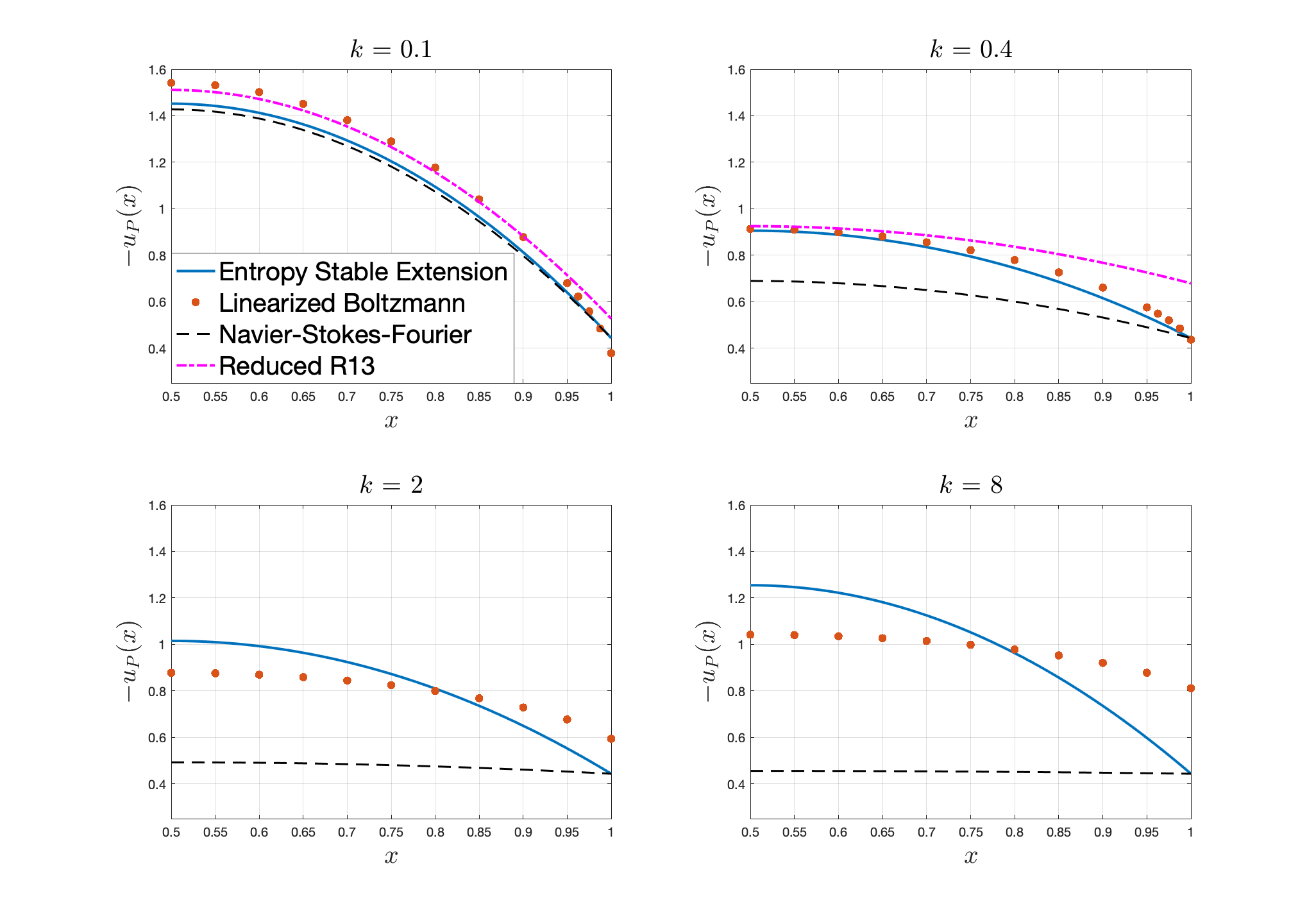}
    \caption{\small We use $C(\epsilon) = - \epsilon\,\big(2.1246\ln(1 + \epsilon ) + 2.3066\big)$ in \eqref{eq:AltBPoiss} to compare $\frac{2}{\gamma_1\sqrt{\pi}}\,u\big(x;\, \frac{2}{\sqrt{\pi}}k\big)$ (blue line)  to the data points for the velocity profile provided in \cite{Ohwada1989} (red dots), the reduced R13 moments solution in \cite{Struchtrup2008} (magenta dot-dash) and the Navier-Stokes-Fourier solution (black dash). w Notice that the Navier-Stokes-Fourier solution and the entropy stable extension always agree at $x=1$. The best results occur for low intermediate Knudsen numbers (represented by $k=0.4$). We omit the R13 solution at $k=2$ and $k=8$ because these are outside its range of validity.} 
    \label{fig:soneflowmix}
\end{figure}

Next, we use the fitted parameters in the velocity equation \eqref{eq:AltBPoiss} and compare the result to velocity data taken from Table I and II of \cite{Ohwada1989} in Figures \ref{fig:soneflowmix}. Because our solution provides no correction to the velocity slip, the velocity given by \eqref{eq:AltBPoiss} is a poor match to the Boltzmann solution. The best results are observed for low intermediate Knudsen numbers (eg. $k=0.4$ in Figure \eqref{fig:soneflowmix}) but the flow profile at smaller or bigger Knudsen regimes leaves something to be desired. For smaller Knudsen numbers, the inaccuracy could be remedied by a more judicious choice of $C(\epsilon)$. However at larger Knudsen numbers, no choice of $C(\epsilon)$ could remedy the inaccuracy that results due to a lack of a correction to the velocity slip in Equation \eqref{eq:AltBPoiss}. 

In the next section, we describe a set of boundary conditions that introduce a non-constant velocity slip for our solution, leading to a better solution overall.

\subsection{The Effects of an Added Slip Function}

In order to introduce a non-constant velocity slip to our solution, we replace $u_b(x^b) = 0$ with $u_b(x^b) = B\mathfrak{s}(\epsilon)$ in the boundary condition \eqref{eq:halfTangentialMomentum}, where $\mathfrak{s}(\epsilon)$ is a slip function we shall describe later. To keep things as simple as possible, our boundary conditions will now be:
\begin{align*}
\left(-\epsilon\underline{\omega}\,\frac{du}{dx}(x^b)+\epsilon^3\underline{\Phi}\frac{d^3u}{dx^3}(x^b)\right)n_1(x^b)  =  \sqrt{\frac{2}{\pi}}\Big(u(x^b) - B\mathfrak{s}(\epsilon)\Big); \qquad
    \epsilon^2\,\frac{d^2u}{dx^2}(x^b) &=   -\alpha \,B\mathfrak{s}(\epsilon);\qquad x^b\in\{0,\,1\} 
\end{align*}
where $\alpha$ is a yet to be determined constant. Notice that requiring that $\frac{d^2u}{dx^2}= O(\epsilon^{-1})$ as $\epsilon$ goes to zero implies that $\mathfrak{s}(\epsilon) = O(\epsilon)$ in this limit. The velocity and mass flux that result from these boundary conditions are 
\begin{align}
    \frac{u}{B} &= \frac{x-x^2}{2\epsilon} + \frac{\underline\omega}{2}\sqrt{\frac{\pi}{2}} + \mathfrak{s}(\epsilon) +\frac{\epsilon-\alpha\mathfrak{s}(\epsilon)}{k_u^2}\left(\frac{\exp\big(\frac{k_ux}{\epsilon} \big)+\exp\big(\frac{k_u(1-x)}{\epsilon} \big)}{\exp\big(\frac{k_u}{\epsilon} \big)+1} \,-\,1\right)\label{eq:velSlip}\\
    \frac{Q}{B}&= \frac{1}{12\epsilon}+\frac{\underline\omega}{2}\sqrt{\frac{\pi}{2}}+\mathfrak{s}(\epsilon)+\frac{2\epsilon\big(\epsilon -\alpha\mathfrak{s}(\epsilon)\big)}{k_u^3}\left(\tanh\left(\frac{k_u}{2\epsilon} \right) - \frac{k_u}{2\epsilon} \right)\label{eq:massFluxSlip}
\end{align}
We observe that $\frac{Q}{B}\sim \mathfrak{s}(\epsilon)$ in the collisionless limit which motivates the ansatz $\mathfrak{s}(\epsilon) = h_1 \ln(1+h_2\epsilon)$. This ensures that $\mathfrak s(\epsilon) = O(\epsilon)$ as $\epsilon$ goes to zero and $\mathfrak{s}(\epsilon) = O(\ln(\epsilon))$ for large $\epsilon$. There are now three parameters to determine, namely, $h_1$, $h_2$ and $\alpha$.

\begin{figure}[ht]
    \centering
    \includegraphics[width = .8\textwidth]{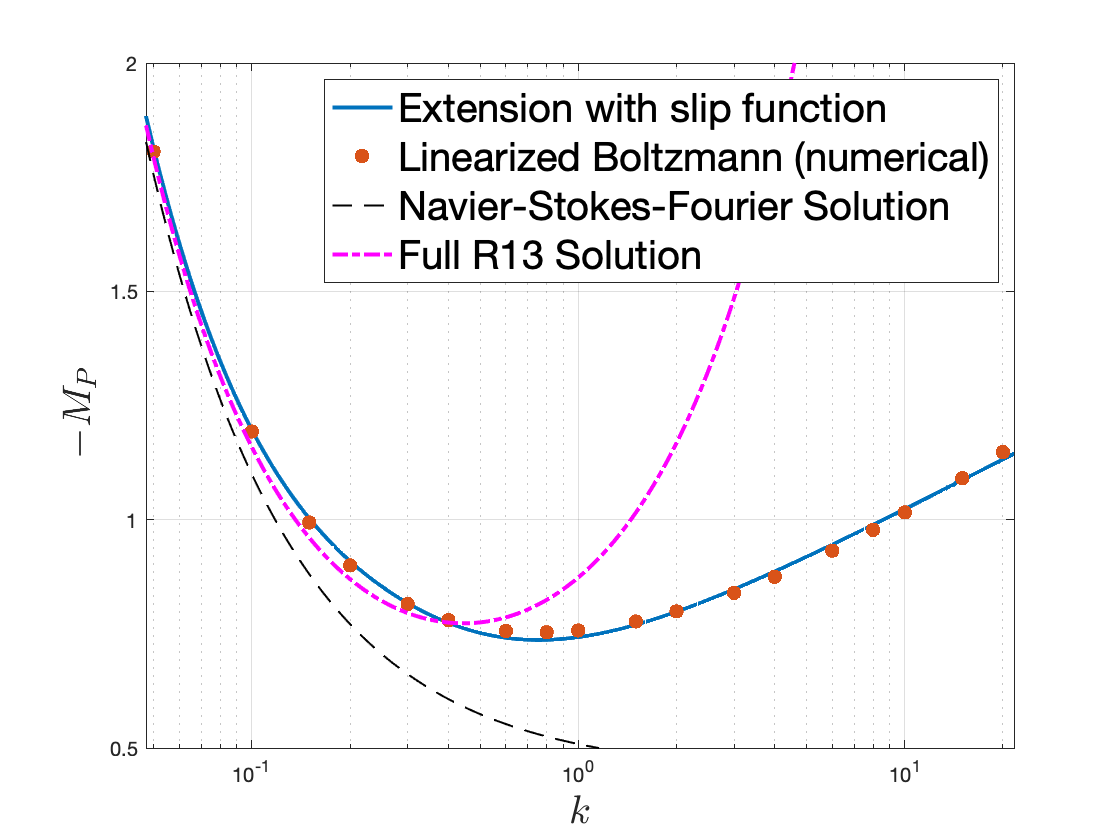}
    \caption{\small Plot of $\frac{2}{B\gamma_1\sqrt{\pi}}{Q}\left(\frac{2}{\sqrt{\pi}}k\right)$ from \eqref{eq:massFluxSlip} with $\alpha = 3.543$, $h_1 = 0.18199$, $h_2=3.0858$. Also plotted are the data points from Table V of \cite{Ohwada1989}, the R13 moment solution from \cite{Struchtrup2007} and the Navier-Stokes-Fourier solution. We observe an even stronger agreement to the data points than the solution given in Figure \ref{fig:sonefluxcompare} quantified by a higher $R^2$ of $0.998$. The Knudsen minimum for the entropy stable extension with an added slip function, located at $k=0.76$, is also in strong agreement with that of the Boltzmann solution from \cite{Ohwada1989}.}
    \label{fig:massfluxslip}
\end{figure}

As before, we determine good values for $\alpha$, $h_1$ and $h_2$ by comparing $\frac{2}{B\gamma_1\sqrt{\pi}}{Q}\left(\frac{2}{\sqrt{\pi}}k\right)$ to the $-M_p$ values in Table V of \cite{Ohwada1989} using the \textit{fitnlm} function in MATLAB. This process gives $\alpha = 3.543$, $h_1 = 0.18199$ and $h_2 = 3.0858$ with $R^2 = 0.998$.
In Figure \ref{fig:massfluxslip}, we compare our fitted solution to the data points of \cite{Ohwada1989}, the full R13 solution and the Navier-Stokes-Fourier solution. 

\begin{figure}[ht]
    \centering
    \includegraphics[width=.95\textwidth]{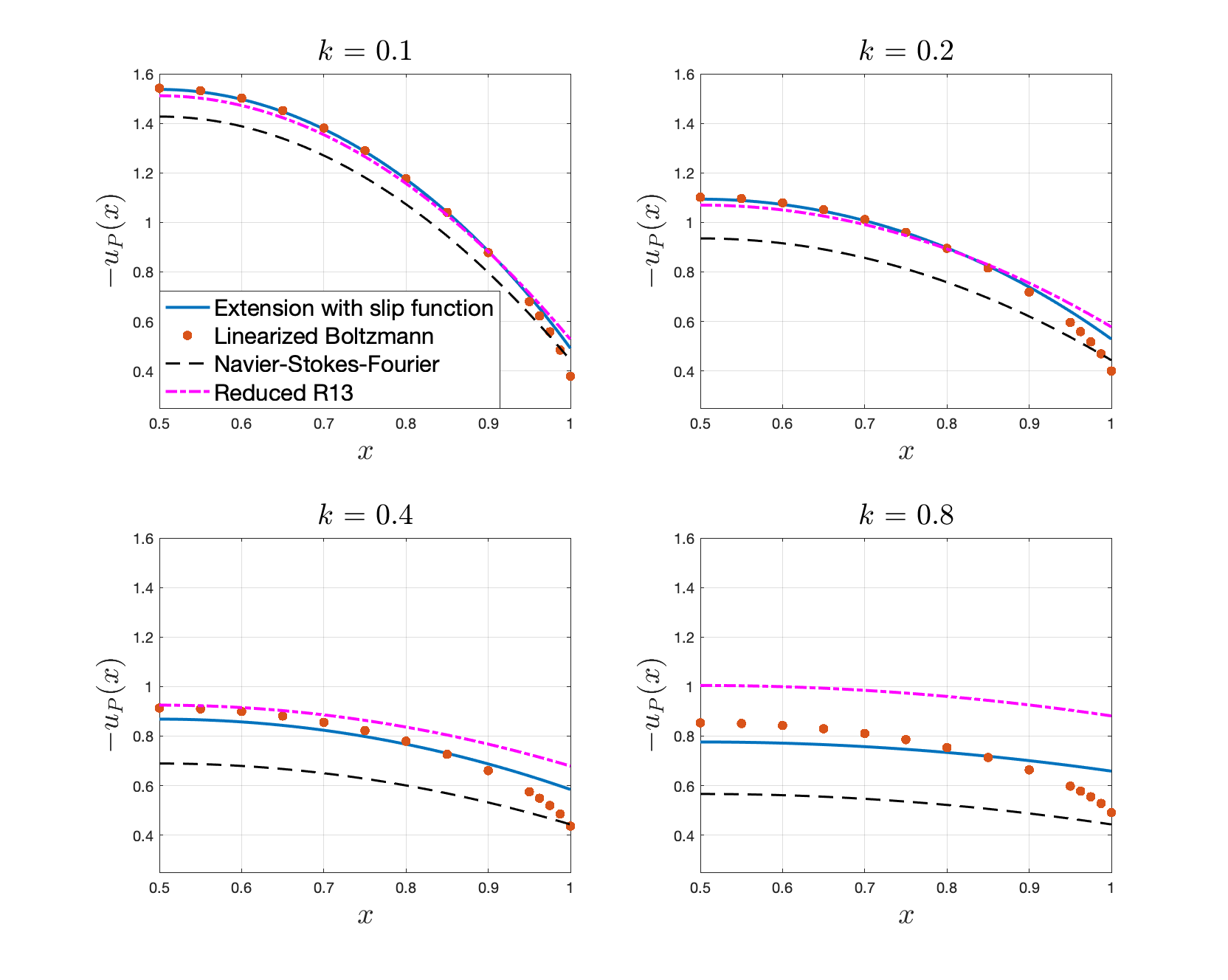}
    \caption{\small We use $\alpha = 3.543$, $h_1 = 0.18199$, $h_2=3.0858$ in Equation \eqref{eq:velSlip} to compare $\frac{2}{B\sqrt{\pi}} u\left(x;\frac{2\gamma_1}{\sqrt{\pi}}k\right)$ (blue line) with the Navier-Stokes-Fourier solution (black dash), the linearized Boltzmann solution in \cite{Ohwada1989} (red dot) and the reduced R13 solution from \cite{Struchtrup2008} for $k<1$. The range of the y-axis is fixed to $[0.25,\,1.6]$ for all plots. The added slip allows the solution of the entropy stable extension to better match the linearized Boltzmann solution (compare solution without added slip in Figure \ref{fig:soneflowmix}). In general, this solution underestimates the peak velocity and overestimates the slip at the boundary. }
    \label{fig:flowslipsmall}
\end{figure}

\begin{figure}[ht]
    \centering
\includegraphics[width=.95\textwidth]{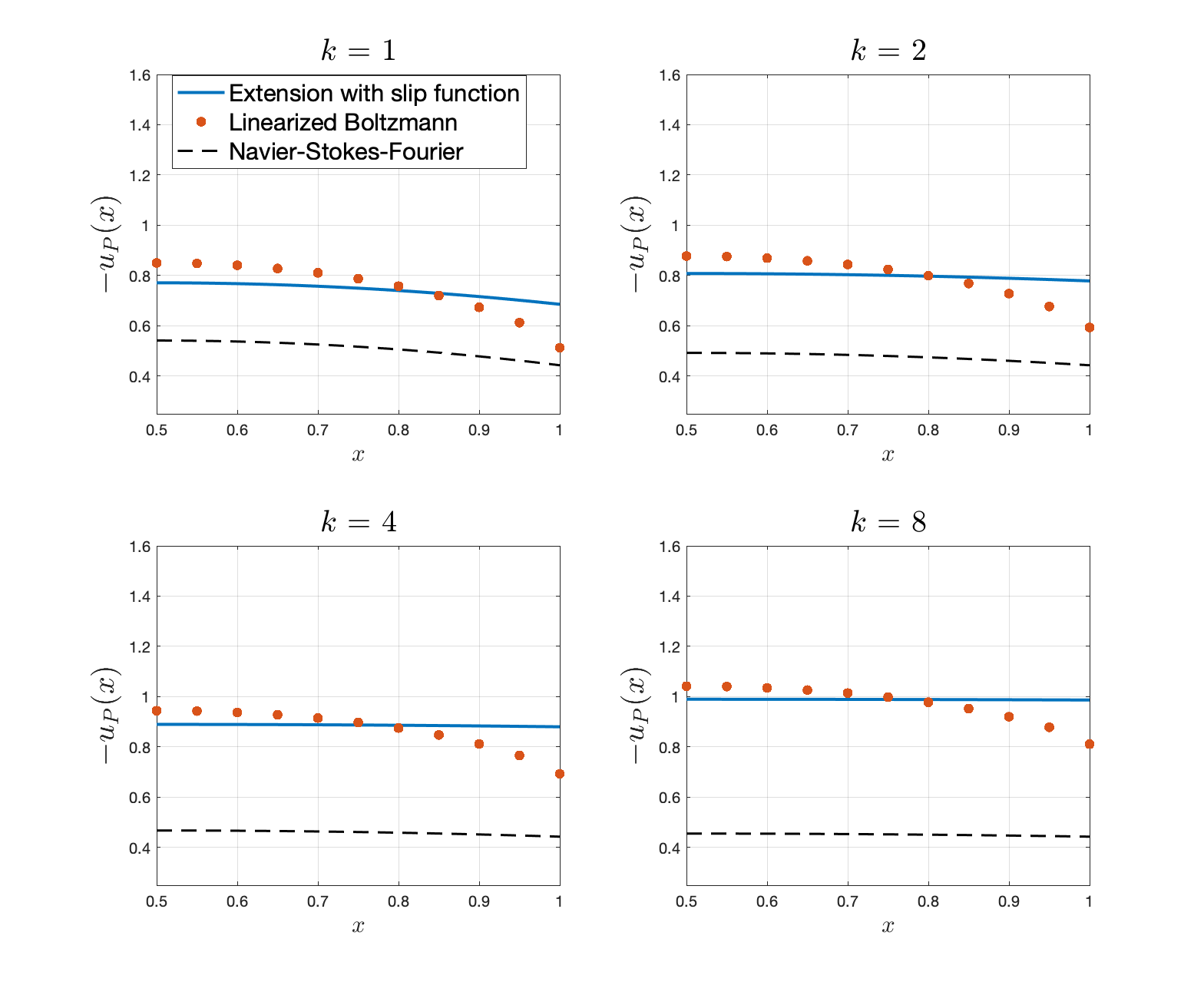}
    \caption{\small We use $\alpha = 3.543$, $h_1 = 0.18199$, $h_2=3.0858$ in Equation \eqref{eq:velSlip} to compare $\frac{2}{B\sqrt{\pi}} u\left(x;\frac{2\gamma_1}{\sqrt{\pi}}k\right)$ (blue line) with the Navier-Stokes-Fourier solution (black dash) and the linearized Boltzmann solution (red dot) for $k$ ten times larger than in Figure \ref{fig:flowslipsmall}. Range of the y-axis is fixed to $[0.25,\,1.6]$ for all plots. The mismatch between the slip from the extension and that from the linearized Boltzmann equation is much more pronounced but the non-constant slip leads to a better overall flow profile than in the version without the slip function. The velocity slip mismatch could indicate that effects due to the Knudsen layer are not sufficiently resolved by the extension.  }
    \label{fig:flowslipbig}
\end{figure}

We then use these parameters in the velocity equation \eqref{eq:velSlip} and compare the result to the velocity data in Table I and II of \cite{Ohwada1989} in Figures \ref{fig:flowslipsmall} and \ref{fig:flowslipbig}. In general, the fitted velocity underestimates the peak velocity whilst overestimating the velocity slip. However this solution is superior to Equation \eqref{eq:AltBPoiss} in matching the linearized Boltzmann solution.

\section{Conclusion}
In this manuscript, we derived a set of fourth order PDEs that serve as an alternative to the Burnett equations  through a reformulation of the process of deriving closures for the deviatoric stress and heat flux from the Boltzmann equation. This reformulation subsumes the Chapman-Enskog expansion whilst opening the door to other possibilities for closure. In particular, our closure is crafted so as to obtain entropy stability at all Knudsen numbers for both non-linear and linearized versions of the resulting conservation equations. \textcolor{black}{We conclude the derivation with a discussion of the transport coefficients and the entropy inequality for the linear and a subsystem of the nonlinear entropy stable extension of the Navier-Stokes-Fouier equations.}

The rest of the paper focuses on the linearized  Hard spheres version of the equations which we apply to the 1D stationary heat transfer problem and the Poiseuille channel. By deriving a symmetric weak form for the momentum and energy equations, we are able to deduce the left hand sides of the natural boundary conditions for these two problems. We obtain the full form of the first set of natural boundary conditions {\color{black} (i.e. those obtained when we integrate by parts the first time)} by directly substituting our closure distribution into the linearized Boltzmann diffuse reflection boundary condition. For the second set of natural boundary conditions, we assume the right hand side takes a particular form with free parameters whose best values we determine by comparing to the Boltzmann equation.   

For the stationary heat problem, we are able to obtain an almost perfect match to the linearized Boltzmann heat flux of Ohwada et al. \cite{Ohwada1989StatHeat} by fitting just two free parameters in the second set of natural boundary conditions. We then compute the temperature distribution within the channel with the obtained free parameters and find that it agrees remarkably well with the temperature distribution predicted by a Grad-Hilbert expansion from \cite{SoneBook} in the interior of the domain. The discrepancy between the two solutions at the boundary slowly increases as the Knudsen number is increased.

For the Poiseuille channel, we find that even though the fitted parameters lead to remarkable agreement of our mass flux to that of the linearized Boltzmann equation, the boundary conditions as set up do not provide any correction to the Navier-Stokes-Fourier velocity slip at the boundary. We found that introducing a slip function into the boundary conditions not only provided a correction to the velocity slip but also led to a mass flux that better matched the linearized Boltzmann mass flux.

Considering that these closures were derived as asymptotic corrections to the Navier-Stokes-Fourier equations into the early transition regime, it cannot be overemphasized how surprising it is that they produce accurate solutions far beyond the expected regime of validity. As the simple nature of the parallel plate test domain might partially explain this, it is imperative to test these equations on more complicated domains analytically and numerically. Of particular interest for future research is the behavior of these equations in dynamic problems especially in situations where the assumption that $\text{St} = O(\epsilon)$ is no longer true. 

Deriving solutions for these equations will also require a far better understanding of the boundary conditions that need to be imposed. On this front, the model problems in this paper suggest the importance of picking extra boundary conditions that interpolate the continuum and collisionless limit behaviors of the system under consideration. What's more, we believe that the parameters in the boundary condition will depend on the surface derivative of the normal vector and, as a result, the local curvature of the boundary. A systematic derivation of the boundary conditions for these equations that correspond to kinetic analogues (eg. diffuse reflection) would be {\color{black} also} needed.


\textcolor{black}{Our extensions to the Navier-Stokes-Fourier equations have been shown to be entropy stable in both linear and nonlinear settings. However, questions regarding the well-posedness of these systems remain open, particularly in light of the connections between entropy stability and well-posedness (e.g., \cite{Godunov1961,friedrichs1971,SHIZUTA1985, Kawashima2004, Jiang2011})
. Future work will also focus on extending our analysis of the nonlinear subsystem to the full system. This extension introduces significant challenges, notably in characterizing the contributions of the bilinear operator $\mathcal{X}_{\mathcal{M}}[\dx{k}\!\ln\mathcal{M}, \, g]$ (\ref{eq:DerivativeSwitchIdentity}) to the closure relations, which demands substantial computational effort. Nevertheless, these nonlinear equations offer a broader range of applicability compared to their linearized counterparts. Additionally, we aim to develop robust numerical methods for these fourth-order systems to explore the properties of solutions in more general initial-boundary-value settings.}

\newpage

\newpage

\bibliography{book}

\begin{thebibliography}{10}

\bibitem{Aoki2017}
K.~Aoki, C.~Baranger, M.~Hattori, S.~Kosuge, G.~Martalò, J.~Mathiaud, and
  L.~Mieussens.
\newblock Slip boundary conditions for the compressible {N}avier–{S}tokes
  equations.
\newblock {\em Journal of Statistical Physics}, 169, 11 2017.

\bibitem{Baidoo2025}
F.~Baidoo.
\newblock {\em The {V}ariational {M}ultiscale {M}oment {M}ethod for the
  {B}oltzmann Equation: Derivation of an Entropy Stable Extension to the
  {N}avier-{S}tokes-{F}ourier Equations}.
\newblock PhD thesis, The University of Texas at Austin, 8 2025.

\bibitem{bardos:1991vf}
C.~Bardos, F.~Golse, and D.~Levermore.
\newblock Fluid dynamic limits of kinetic equations. i. formal derivations.
\newblock {\em Journal of Statistical Physics}, 63:323--344, 4 1991.

\bibitem{Bazilevs2007}
Y.~Bazilevs, V.~Calo, J.~Cottrell, T.~Hughes, A.~Reali, and G.~Scovazzi.
\newblock Variational {M}ultiscale residual-based turbulence modeling for large
  eddy simulation of incompressible flows.
\newblock {\em Computer Methods in Applied Mechanics and Engineering},
  197:173--201, 12 2007.

\bibitem{bhatnagar:1954hc}
P.~Bhatnagar, E.~Gross, and M.~Krook.
\newblock {A model for collision processes in gases. {I.} Small amplitude
  processes in charged and neutral one-component systems.}
\newblock {\em Phys. Rev.}, 94(511---525), 1954.

\bibitem{bobylev1982chapman}
A.~V. Bobylev.
\newblock The {C}hapman-{E}nskog and {G}rad methods for solving the {B}oltzmann
  equation.
\newblock {\em Akademiia Nauk SSSR Doklady}, 262(1):71--75, 1982.

\bibitem{bobylev2006}
A.~V. Bobylev.
\newblock Instabilities in the {C}hapman-{E}nskog expansion and hyperbolic
  {B}urnett equations.
\newblock {\em Journal of statistical physics}, 124(2):371--399, 2006.

\bibitem{Bobylev2018}
A.~V. Bobylev.
\newblock Boltzmann equation and hydrodynamics beyond {N}avier-{S}tokes.
\newblock {\em Philosophical Transactions of the Royal Society A: Mathematical,
  Physical and Engineering Sciences}, 376, 4 2018.

\bibitem{Burnett1935}
D.~Burnett.
\newblock The distribution of velocities in a slightly non-uniform gas.
\newblock {\em Proceedings of the London Mathematical Society}, s2-39:385--430,
  1935.

\bibitem{Cai2024}
Z.~Cai, M.~Torrilhon, and S.~Yang.
\newblock Linear {R}egularized 13-moment equations with {O}nsager boundary
  conditions for {G}eneral {G}as {M}olecules.
\newblock {\em SIAM Journal on Applied Mathematics}, 84:215--245, 2 2024.

\bibitem{cercignani1988}
C.~Cercignani.
\newblock {\em The Boltzmann Equation and Its Applications}.
\newblock Springer New York, New York, NY, 1988.

\bibitem{Cercignani1963}
C.~Cercignani and A.~Daneri.
\newblock Flow of a rarefied gas between two parallel plates.
\newblock {\em Journal of Applied Physics}, 34:3509--3513, 12 1963.

\bibitem{cercignani1994}
C.~Cercignani, R.~Illner, and M.~Pulvirenti.
\newblock {\em The Mathematical Theory of Dilute Gases}.
\newblock Springer New York, New York, NY, 1994.

\bibitem{WangChang1952}
C.~W. Chang and G.~Uhlenbeck.
\newblock On the propagation of sound in monatomic gases.
\newblock Technical report, Engineering Research Institute, University of
  Michigan Ann Arbor, 1952.

\bibitem{chapman1990mathematical}
S.~Chapman and T.~G. Cowling.
\newblock {\em The mathematical theory of non-uniform gases: an account of the
  kinetic theory of viscosity, thermal conduction and diffusion in gases}.
\newblock Cambridge university press, 1990.

\bibitem{Comeaux1995}
K.~Comeaux, D.~Chapman, and R.~MacCormack.
\newblock An analysis of the {B}urnett equations based on the second law of
  thermodynamics.
\newblock In {\em 33rd Aerospace Sciences Meeting and Exhibit}, pages 1 -- 19.
  American Institute of Aeronautics and Astronautics, 1 1995.

\bibitem{Coron1989}
F.~Coron.
\newblock Derivation of slip boundary conditions for the {N}avier-{S}tokes
  system from the {B}oltzmann equation.
\newblock {\em Journal of Statistical Physics}, 54:829--857, 2 1989.

\bibitem{friedrichs1971}
K.~Friedrichs and P.~Lax.
\newblock {Systems of Conservation Equations with a Convex Extension}.
\newblock {\em Proceedings of the National Academy of Science}, 68:1686--1688,
  Aug. 1971.

\bibitem{gamba2018}
I.~M. Gamba and S.~Rjasanow.
\newblock Galerkin--{P}etrov approach for the {B}oltzmann equation.
\newblock {\em Journal of Computational Physics}, 366:341--365, 2018.

\bibitem{GARCIACOLIN2008}
L.~Garciacolin, R.~Velasco, and F.~Uribe.
\newblock Beyond the {N}avier–{S}tokes equations: {B}urnett hydrodynamics.
\newblock {\em Physics Reports}, 465:149--189, 8 2008.

\bibitem{Godunov1961}
S.~Godunov.
\newblock Interesting class of quasilinear systems.
\newblock {\em Dokl. Akd. Nauk SSSR}, 139:521--523, 1961.

\bibitem{golse2005}
F.~Golse.
\newblock The {B}oltzmann equation and its hydrodynamic limits.
\newblock {\em Evolutionary equations}, 2:159--301, 2005.

\bibitem{Grad1963}
H.~Grad.
\newblock Asymptotic theory of the {B}oltzmann {E}quation.
\newblock {\em Physics of Fluids}, 6, 1963.

\bibitem{Gu2007}
X.~Gu and D.~Emerson.
\newblock A computational strategy for the regularized 13 moment equations with
  enhanced wall-boundary conditions.
\newblock {\em Journal of Computational Physics}, 225, 7 2007.

\bibitem{Gu2009}
X.-J. Gu and D.~R. Emerson.
\newblock A high-order moment approach for capturing non-equilibrium phenomena
  in the transition regime.
\newblock {\em Journal of Fluid Mechanics}, 636:177--216, 10 2009.

\bibitem{Gust2009}
E.~D. Gust and L.~E. Reichl.
\newblock Molecular dynamics simulation of collision operator eigenvalues.
\newblock {\em Physical Review E}, 79:031202, 3 2009.

\bibitem{Hashiguchi2020}
K.~Hashiguchi.
\newblock {\em Mathematical fundamentals}, chapter~1, pages 1--59.
\newblock Elsevier, 2020.

\bibitem{Hu2020}
Z.~Hu, S.~Yang, and Z.~Cai.
\newblock Flows between parallel plates: Analytical solutions of regularized
  13-moment equations for inverse-power-law models.
\newblock {\em Physics of Fluids}, 32, 12 2020.

\bibitem{hughes1995}
T.~J. Hughes.
\newblock Multiscale phenomena: Green's functions, the {D}irichlet-to-{N}eumann
  formulation, subgrid scale models, bubbles and the origins of stabilized
  methods.
\newblock {\em Computer Methods in Applied Mechanics and Engineering},
  127:387--401, 11 1995.

\bibitem{Hughes1998}
T.~J. Hughes, G.~R. Feijóo, L.~Mazzei, and J.~B. Quincy.
\newblock The variational multiscale method—a paradigm for computational
  mechanics.
\newblock {\em Computer Methods in Applied Mechanics and Engineering},
  166:3--24, 11 1998.

\bibitem{hughes2007}
T.~J. Hughes and G.~Sangalli.
\newblock Variational multiscale analysis: the fine-scale {G}reen’s function,
  projection, optimization, localization, and stabilized methods.
\newblock {\em SIAM Journal on Numerical Analysis}, 45(2):539--557, 2007.

\bibitem{CartesianTensorBook}
H.~Jeffreys.
\newblock {\em Cartesian Tensors}.
\newblock Cambridge University Press, 1961.

\bibitem{Jiang2011}
N.~Jiang and C.~D. Levermore.
\newblock Weakly nonlinear-dissipative approximations of hyperbolic–parabolic
  systems with entropy.
\newblock {\em Archive for Rational Mechanics and Analysis}, 201:377--412, 8
  2011.

\bibitem{Jin2001}
S.~Jin and M.~Slemrod.
\newblock Regularization of the {B}urnett equations via relaxation.
\newblock {\em Journal of Statistical Physics}, 103:1009--1033, 2001.

\bibitem{Kauf2010}
P.~Kauf, M.~Torrilhon, and M.~Junk.
\newblock Scale-induced closure for approximations of kinetic equations.
\newblock {\em Journal of Statistical Physics}, 141:848--888, 12 2010.

\bibitem{Kawashima2004}
S.~Kawashima and W.-A. Yong.
\newblock Dissipative structure and entropy for hyperbolic systems of balance
  laws.
\newblock {\em Archive for Rational Mechanics and Analysis}, 174:345--364, 12
  2004.

\bibitem{Kearsley1975}
E.~A. Kearsley and J.~T. Fong.
\newblock Linearly independent sets of isotropic {C}artesian tensors of ranks
  up to eight.
\newblock {\em Journal of Research of the National Bureau of Standards, Section
  B: Mathematical Sciences}, 79B:49, 1 1975.

\bibitem{Knudsen1909}
M.~Knudsen.
\newblock Die {G}esetze der {M}olekularströmung und der inneren
  {R}eibungsströmung der {G}ase durch {R}öhren.
\newblock {\em Annalen der Physik}, 333:75--130, 1909.

\bibitem{Le2012}
N.~T. Le, C.~White, J.~M. Reese, and R.~S. Myong.
\newblock Langmuir– and {L}angmuir–{S}moluchowski boundary conditions for
  thermal gas flow simulations in hypersonic aerodynamics.
\newblock {\em International Journal of Heat and Mass Transfer}, 55:5032--5043,
  9 2012.

\bibitem{LevermoreUnpublished}
C.~D. Levermore.
\newblock Gas dynamics beyond {N}avier-{S}tokes.
\newblock (private communication).

\bibitem{levermore1996}
C.~D. Levermore.
\newblock {Moment closure hierarchies for kinetic theories}.
\newblock {\em Journal of Statistical Physics}, 83:1021--1065, 1996.

\bibitem{Levermore1998gaussian}
C.~D. Levermore and W.~J. Morokoff.
\newblock The {G}aussian {M}oment {C}losure for {G}as {D}ynamics.
\newblock {\em SIAM Journal on Applied Mathematics}, 59:72--96, 1 1998.

\bibitem{Maxwell1879}
J.~C. Maxwell.
\newblock Vii. on stresses in rarified gases arising from inequalities of
  temperature.
\newblock {\em Philosophical Transactions of the Royal Society of London},
  170:231--256, 12 1879.

\bibitem{Ohwada1989StatHeat}
T.~Ohwada, K.~Aoki, and Y.~Sone.
\newblock Heat transfer and temperature distribution in a rarefied gas between
  two parallel plates with different temperatures: Numerical analysis of the
  {B}oltzmann equation for a hard sphere molecule.
\newblock In E.~Muntz, D.~Weaver, and D.~Campbell, editors, {\em Rareﬁed Gas
  Dynamics: Theoretical and Computational Techniques}, pages 70--81. American
  Institute of Aeronautics and Astronautics, 1989.

\bibitem{Ohwada1989}
T.~Ohwada, Y.~Sone, and K.~Aoki.
\newblock Numerical analysis of the {P}oiseuille and thermal transpiration
  flows between two parallel plates on the basis of the {B}oltzmann equation
  for hard‐sphere molecules.
\newblock {\em Physics of Fluids A: Fluid Dynamics}, 1:2042--2049, 12 1989.

\bibitem{rana2013}
A.~Rana, M.~Torrilhon, and H.~Struchtrup.
\newblock A robust numerical method for the {R}13 equations of rarefied gas
  dynamics: Application to lid driven cavity.
\newblock {\em Journal of Computational Physics}, 236:169 -- 186, 2013.

\bibitem{saint-raymond2009}
L.~{Saint-Raymond}.
\newblock {\em {Hydrodynamic Limits of the Boltzmann Equation}}.
\newblock Number v. 1971 in {Lecture Notes in Mathematics}. Springer, 2009.

\bibitem{saint-raymond2014}
L.~{Saint-Raymond}.
\newblock A mathematical {PDE} perspective on the {C}hapman--{E}nskog
  expansion.
\newblock {\em Bulletin of the American Mathematical Society}, 51(2):247--275,
  2014.

\bibitem{SHIZUTA1985}
Y.~Shizuta and S.~Kawashima.
\newblock Systems of equations of hyperbolic-parabolic type with applications
  to the discrete {B}oltzmann equation.
\newblock {\em Hokkaido Mathematical Journal}, 14, 2 1985.

\bibitem{SoneBook}
Y.~Sone.
\newblock {\em Molecular Gas Dynamics}.
\newblock Birkhäuser Boston, 2007.

\bibitem{Struchtrup20052}
H.~Struchtrup.
\newblock Derivation of 13 moment equations for rarefied gas flow to second
  order accuracy for arbitrary interaction potentials.
\newblock {\em Multiscale Modeling \& Simulation}, 3:221--243, 1 2005.

\bibitem{Struchtrup20053}
H.~Struchtrup.
\newblock Failures of the {B}urnett and super-{B}urnett equations in steady
  state processes.
\newblock {\em Continuum Mechanics and Thermodynamics}, 17:43--50, 4 2005.

\bibitem{Struchtrup2003}
H.~Struchtrup and M.~Torrilhon.
\newblock Regularization of {G}rad’s 13 moment equations: Derivation and
  linear analysis.
\newblock {\em Physics of Fluids}, 15:2668--2680, 9 2003.

\bibitem{Struchtrup2007}
H.~Struchtrup and M.~Torrilhon.
\newblock H theorem, regularization, and boundary conditions for linearized 13
  moment equations.
\newblock {\em Physical Review Letters}, 99:014502, 7 2007.

\bibitem{Struchtrup2008}
H.~Struchtrup and M.~Torrilhon.
\newblock Higher-order effects in rarefied channel flows.
\newblock {\em Physical Review E}, 78:046301, 10 2008.

\bibitem{SmoluchowskivonSmolan1898}
M.~S. von Smolan.
\newblock Ueber wärmeleitung in verdünnten gasen.
\newblock {\em Annalen der Physik}, 300:101--130, 1 1898.

\bibitem{Weyl1946TheRepresentations}
H.~Weyl.
\newblock {\em {The {C}lassical {G}roups: Their {I}nvariants and
  {R}epresentations}}.
\newblock Princeton University Press, 1946.

\bibitem{Zhong1993}
X.~Zhong, R.~W. MacCormack, and D.~R. Chapman.
\newblock Stabilization of the {B}urnett equations and application to
  hypersonic flows.
\newblock {\em AIAA Journal}, 31:1036--1043, 6 1993.

\end{thebibliography}

\renewcommand{\thesection}{\Alph{section}}
\begin{appendices}
\section{Brief Primer on Cartesian Tensor Summation}\label{sec:Tensor}
We give below several examples showing how to translate the implied summation of repeated indices. Consider a scalar $\mathrm{G}$, vectors $\bm{\mathrm{x}}= (\mathrm{x_1}, \mathrm{x_2},\dots,\mathrm{x}_D)^{\mathsf{T}}$, $\,\bm{\mathrm{v}}=(\mathrm{v_1}, \mathrm{v_2},\dots,\mathrm{v}_D)^{\mathsf{T}}$ and matrices $\bm{M} = (M_{ij})_{1\le i,j\le D}$ and $\bm{N} = (N_{ij})_{1\le i,j\le D}$
\begin{align}\label{eq:TensorSumExample}
&\mathrm{v}_k\partial_{\mathrm{x}_k}\mathrm{G} := \sum_{k=1}^D \mathrm{v}_k \partial_{\mathrm{x}_k}\mathrm{G} = \bm{\mathrm{v}}\cdot\nabla_{\bm{\mathrm{x}}} \mathrm{G} &\partial_{\mathrm{x}_i}(\bm{M}_{ij} \mathrm{v}_j) := \sum_{i=1}^D \partial_{\mathrm{x}_i}\left(\sum_{j=1}^D{M}_{ij} \mathrm{v}_j\right) = \text{div}_{\bm{\mathrm{x}}}(\bm{M} \bm{\mathrm{v}}) \nonumber\\ 
&\     \\
&\partial_{\mathrm{x_k}}\partial_{\mathrm{x_k}}\mathrm{G}  := \sum_{k=1}^D \partial_{\mathrm{x_k}}^2\mathrm{G} = \Delta_{\bm{\mathrm{x}}} \mathrm{G} 
&M_{ij}N_{ij}:= \sum_{i=1}^D \sum_{j=1}^D M_{ij}N_{ij} = \mathrm{Tr}(\bm{M}^T\bm{N}) \nonumber
\end{align}
This convention comes into its own when dealing with higher rank tensors where explicit summation quickly become cumbersome.

{\color{black}
\section{Isotropic Tensor Integrals}\label{appendix:isotropicTensorIntegrals}
Due to Property \ref{assume:linsymm}, a tensor made up of integrals whose integrand is composed of a Maxwellian $\mathcal{M} = \mathcal{M}_{\varrho, \bm{w}, \vartheta}$, monomials of $(\bm{v}-\bm{w})$, functions of $|\bm{v} - \bm{w}|$ and $\iLM$ (or $\LM$) will be an isotropic tensor. A proof of this assertion can be found in Proposition 3.1 of \cite{Baidoo2025}. Isotropy in tensors is such a restrictive condition that an isotropic tensor of a given rank can be expected to take a particular form. For example,
\begin{enumerate}
    \item Every rank 2 isotropic tensor is a scalar multiple of Kronecker delta, i.e. $C \, \delta_{ij}$ for some scalar $C$.
    \item Every rank 4 isotropic tensor is of the form $T_{ijkl} = C_1\delta_{ij}\delta_{kl} + C_2 \delta_{ik}\delta_{jl} + C_3 \delta_{il}\delta_{jk}$ for scalars $C_1$, $C_2$ and $C_3$.
\end{enumerate}
A description of the form of rank 2 to rank 8 isotropic tensors can be found in \cite{Kearsley1975}.  More generally, in \cite{Weyl1946TheRepresentations}, Weyl shows that even rank isotropic tensor takes the form of linear combinations of products Kronecker delta tensors. 

The macroscopic closures described in this manuscript are computed from tensor integrals of the form described above. For example, the deviatoric stress for Navier-Stokes-Fourier closure involves computing the integral 
\begin{align*}
    W_{ij}^{nl} &= \bracke{(v_i - u_i)(v_j - u_j) - \frac{\delta_{ij}}{D}|\bm{v}- \bm{u}|^2,\, \mu\, \iLmu\!\sqbrac{(v_n - u_n)(v_l - u_l) - \frac{\delta_{nl}}{D}|\bm{v}- \bm{u}|^2} }\\
    &= \bracke{(v_1 - u_1)(v_3 - u_3),\, \mu\, \iLmu\!\sqbrac{(v_1 - u_1)(v_3 - u_3) } } \left(\delta_{in}\delta_{jl} + \delta_{il}\delta_{jn} - \frac{2}{D}\delta_{ij}\delta_{nl} \right)
\end{align*}
In the second line, we take advantage of the fact that $W_{ij}^{nl}$ is an isotropic tensor such that $W_{ij}^{nl} = W_{ji}^{nl}$ and $W_{kk}^{nl} = 0$. What is more, because the indices of the tensors we shall encounter are entirely determined by monomials of $(\bm{v} - \bm{w})$, all odd ranked isotropic tensor integrals we encounter shall be the zero tensor. This is can be proved by performing the change of variables $(\bm{v} - \bm{w}) \rightarrow -(\bm{v} - \bm{w})$ in the integral in order to observe that the isotropic tensor integrals described are such that 
$$ T_{i_1\,i_2\dots i_n} = (-1)^n\, T_{i_1\,i_2\dots i_n}$$

Taking advantage of isotropy also allows us to compute moments of the Maxwellian, i.e. expressions for integrals of the form 
\begin{equation}\label{intform}
\int_{\mathbb R^D} c_{i_1}\dots c_{i_m}|\bm c |^k \mathcal M_{\varrho, \bm{w}, \vartheta} \, d\bm c 
\end{equation}
where $\bm c = \bm v -\bm w$ and $m$  and $k$ are positive even integers. 

Note first that generalized spherical coordinates can be used to show that 
\begin{equation}\label{scalarform}
    \int_{\mathbb R^D} |\bm c|^{2n} \mathcal M_{\varrho,\bm{w},\vartheta} \, d\bm c = \left(\prod_{k=0}^{n-1} (D + 2k)\right)\varrho\vartheta^n 
\end{equation}
When integrals of the form \eqref{intform} are transformed to their isotropic representations, the resulting scalars take the form of the left-hand side of \eqref{scalarform}. This observation allows us to quickly compute many integrals of the form \eqref{intform}.

We present below several integrals we will use in later sections of the Appendix.
\begin{align}
    \int_{\mathbb R^D}c_i c_j\mathcal M \,d\bm c &= \frac{\delta_{ij}}{D}\int_{\mathbb R^D}|\bm c|^2\mathcal M \,d\bm w = \varrho\vartheta\delta_{ij}\label{eq:1}\\
    \int_{\mathbb R^D}c_i c_j|\bm c|^2\mathcal M \,d\bm w &= (D+2)\,\varrho\vartheta^2\delta_{ij}\label{eq:2}\\
    \int_{\mathbb R^D} c_i c_j c_k c_l \mathcal M \,d\bm c &= \varrho\vartheta^2\, (\delta_{ij}\delta_{kl} + \delta_{ik}\delta_{jl} + \delta_{il}\delta_{jk})\label{eq:4}\\
    \int_{\mathbb R^D} c_i c_j c_k c_l |\bm c|^2 \mathcal M \,d\bm c&= (D+4)\varrho\vartheta^3\, (\delta_{ij}\delta_{kl} + \delta_{ik}\delta_{jl} + \delta_{il}\delta_{jk})\label{eq:5}
\end{align}

}

    \section{Derivation of Projection Operator}
\label{appendix:projection}
{ In this section, we show that equation (\ref{eq:orthproj}) (shown below as equation (\ref{eq:proj})) defines an orthogonal projection onto the space of collision invariants $\mathscr I$ with respect to the $\mathscr L^2(\mathcal M_{\rho, \bm{u}, \theta}\, d\bm v)$ inner product.}
\begin{multline}\label{eq:proj}
\PI [g] = 
\frac{1}{\rho}
\bracke{1, \,
\mathcal{M}\,g} +
\frac{(\bm v- \bm u)}{\rho\theta}
\cdot
\bracke{(\bm v- \bm u),\,\mathcal{M}\,g } 
\\
+
\left( \frac{|\bm v- \bm u|^2}{2\theta} -\frac{D}{2} \right) 
\frac{2}{D\rho}
\bracke{ \left( \frac{|\bm v- \bm u|^2}{2\theta} -\frac{D}{2} \right),\, \mathcal{M}\,g }      
\end{multline}
The derivation amounts to switching from the basis $\{1, v_1, \dots, v_D, |v|^2/2\}$ to the $\mathscr L^2(\mathcal M d\bm v)$-orthonormal basis and using this basis for our projection. To that end, we perform the orthogonalization via Gram-Schmidt.

The first basis element $e_1 = \frac{1}{\sqrt{\rho}}$ follows immediately from the fact that \\$ \bracke{1,\, \mathcal{M}} = \rho$. To deduce $e_2$, note that $$q_{2}^1 = v_1 - \bracke{e_1,\,\mathcal{M}\,v_1} \, e_1= v_1 - \frac{1}{\rho}\bracke{ 1,\,\mathcal{M}\,v_1} = v_1 - u_1$$
 has norm 
$$||\, q_{2}^1 \, ||^2_\mathcal{M} := \bracke{ (v_1 - u_1),\,\mathcal{M}\,(v_1-u_1)} = \rho\theta $$ where we used equation (\ref{eq:1}) to obtain the final equality. 
Thus our second orthonormal basis term is $\,e_{2}^1 = (v_1 - u_1)/\sqrt{\rho\theta}\,$.  More generally if we define $\,e_{2}^i = (v_i - u_i)/\sqrt{\rho\theta}\,$, we see that the set  $\{e_{2}^i\}_{i=1}^D$ is orthonormal because, by Equation \eqref{eq:1}, 
\begin{equation*}
    \bracke{ (v_i-u_i),\, \mathcal{M}\, (v_j - u_j)} = \rho\theta \delta_{ij}\,= \,0, \quad \text{for $i\not=j$ }
\end{equation*} 
Thus we are left with just $|\bm{v}|^2$. We first orthogonalize
\begin{equation}\label{q3}
    q_3 = \frac{|\bm{v}|^2}{2} - \frac{1}{2}\bracke{ e_1,\,\mathcal{M}\,|\bm{v}|^2}\, e_1 - \frac{1}{2} \bracke{\,e_2^i,\, \mathcal{M}\,|\bm{v}|^2}  \,e_2^i = \frac{|\bm{v}-\bm{u}|^2 - D\theta}{2}
\end{equation}

The norm of $q_3$ gives us three terms to integrate 
\begin{align}\label{norm3}
    \bracke{ \left(\frac{|\bm{v}-\bm{u}|^2 - D\theta}{2} \right)^2,\,\mathcal{M}}
    &= \frac{1}{4}\bracke{{|\bm{v}-\bm{u}|^4,\,\mathcal M}}
    -
    \frac{D\theta }{2}\bracke{|\bm{v}-\bm{u}|^2,\, \mathcal M}
    +  
    \frac{D^2\theta^2}{4}\bracke{1, \,\mathcal{M}} \nonumber\\
    &= \frac{D}{2}\rho\theta^2
\end{align}
where we have made use of the identity (\ref{scalarform}) in the above.
Thus our final basis element can be written as $$ e_3 = \sqrt{\frac{2}{D\rho}}\left( \frac{|\bm{v}-\bm{u}|^2}{2\theta} - \frac{D}{2}\right)  $$
  We then perform an orthonormal projection onto $\mathscr{I}$ with our orthonormal basis $\{e_1, e_2^1, \dots, e_2^D, e_3\}$, i.e. 
  $$\PI[g] = e_1\bracke{e_1,\, \mathcal{M}\,g} + e_2^i\bracke{e_2^i,\,\mathcal{M}\,g} + e_3\bracke{e_3,\,\mathcal{M}\,g} $$
  Notice that equation (\ref{eq:proj}) isn't the only orthonormal projection onto the space of collision invariants as we could, for example, obtain other orthonormal bases by switching the order in which we carry out the Gram-Schmidt procedure. However\, the fact that (\ref{eq:proj}) also appears when computing $\mu'(\mathcal M) [\mathcal M f]$ (see section \ref{sec:BGK}) suggests that this particular projection might well be considered canonical.
 
 \section{BGK collision operator}\label{sec:BGK}
{ In this section, we will detail how the BGK operator \cite{bhatnagar:1954hc} gives rise to a linearized collision operator that satisfies the properties required for our formulation. 
The BGK operator is given by }
\begin{equation}\label{eq:BGK}
\mathcal{C}^{\mathrm{BGK}}(F) = \frac{\mu(F) - F}{\tau(F)}
\end{equation}
where $\tau(F)>0$ denotes the rate of relaxation to equilibrium that may depend on moments of $F$ but not on $\bm{v}$. For example, \cite{SoneBook} and \cite{Ohwada1989} use $\tau = A_c\, \rho(F)$ where $A_c$ is a constant.

To show that $\mathcal{C}^{\mathrm{BGK}}$ complies with Galilean invariance, note that { by a change of variables in (\ref{eq:selfeq}) it can be shown that $\rho_{\mu}(\mathcal T_{\bm U} F) = \rho_{\mu}(F) = \rho_{\mu}(\mathcal T_{\bm O}F)\,$ and $\,\theta_{\mu}(\mathcal T_{\bm U}F) = \theta_{\mu}(F) = \theta_{\mu}(\mathcal T_{\bm O}F)\,$ whilst $\,\bm u_{\mu}(\mathcal T_{\bm U}F) = \bm u_{\mu}(F) +\bm U\,$ and $\,\bm u_{\mu}(\mathcal{T}_{\bm O}F) = \mathcal{T}_{\bm O}\bm u_{\mu}(F) $. This means that $\mu(\mathcal{T}_{\bm U} F) = \mathcal{T}_{\bm U} \mu(F)$ and $\mathcal{T}_{\bm O}\mu(F) = \mu(\mathcal{T}_{\bm O} F)$}.  Therefore, we establish that
\begin{equation}
\mathcal{C}^{\mathrm{BGK}}(\mathcal{T}_{\bm U}F) 
= 
\frac{
\mu({\mathcal T_{\bm U} F} )
- 
\mathcal{T}_{\bm U}F}{\tau} 
= 
\mathcal{T}_{\bm U} \mathcal{C}^{\mathrm{BGK}}(F).
\end{equation}
and 
\begin{equation}
\mathcal{C}^{\mathrm{BGK}}(\mathcal{T}_{\bm O}F) 
= 
\frac{\mu({\mathcal T_{\bm O}F}) 
- 
\mathcal{T}_{\bm O}F}{\tau} 
= 
\mathcal{T}_{\bm O} \mathcal{C}^{\mathrm{BGK}}(F),
\end{equation}
provided that $\tau$ is invariant under co-ordinate translations and rotations, which is a reasonable imposition on the relaxation rate. 
 The collision invariance property \ref{assume:collinvar} follows immediately from (\ref{eq:selfeq}) by linearity. 

To establish the dissipation relation (\ref{eq:dissipation}) for $\mathcal C^{\textrm{BGK}}$ we note that  $\ln\left({\mu(F)}\right) \in \mathscr{I}$ and therefore by (\ref{eq:selfeq}), 
\begin{equation}
\int_{\mathbb R^D} 
\ln\left({\mu(F)}\right)
\, 
(F - \mu(F))
\,
d\bm v=0. 
\end{equation}
The dissipation inequality (\ref{eq:dissipation}) for $\mathcal C^{\textrm{BGK}}$ then follows because
\begin{equation}\label{eq:BGKdissipation}
\int_{\mathbb R^D} 
\ln\left({F}\right)
\, 
\mathcal{C}^{\mathrm{BGK}}(F)
\,
d\bm v 
= 
-\frac{1}{\tau} 
\int_{\mathbb R^D} 
\ln\left(\frac{F}{\mu(F)}\right)
\, 
(F-\mu(F)) 
\, d\bm v
\leq 0.
\end{equation}
because $\ln\big(\frac{a}{b}\big)(a - b) \ge 0$ for $a,b>0$. Moreover, because equality in (\ref{eq:BGKdissipation}) holds if and only if $F = \mu(F)$, the condition 
\[
\int_{\mathbb R^D}\ln({F})\,\mathcal{C}^{\mathrm{BGK}}(F)\,d\bm v=0
\]
implies that $F$ is a Maxwellian and as a result $\ln\left({F}\right)\in\mathscr{I}$. Thus the equivalence in (\ref{eq:equilibria}) is therefore also verified.

  The linearization of the BGK collision operator about an arbitrary Maxwellian $\mathcal{M}$  is given by
{
\begin{equation}
\left. 
\frac{d}{d\epsilon}\mathcal{C}^{\mathrm{BGK}}
(\mathcal M +\epsilon \mathcal M f))\right|_{\epsilon=0} 
= 
\frac{\frac{d}{d\epsilon}\left. \mu(\mathcal{M} + \epsilon\mathcal{M}f) \right|_{\epsilon=0}
- 
\mathcal M f}{\tau(\mathcal{M})}
\end{equation}
We note that the derivative of a Maxwellian $\mathcal{M}_{\rho, \bm{u}, \theta}$ with respect to some arbitrary argument of $(\rho, \bm{u}, \theta)$ is given by
\begin{equation}\label{eq:dmaxwellian}
\partial\mathcal{M}_{\rho,\bm u,\theta}
= \mathcal{M}_{\rho,\bm u, \theta}\left(
\frac{\partial\rho}{\rho} 
+ 
\frac{(\bm v-\bm u)\cdot 
\partial\bm u}{\theta}
+ 
\left(\frac{|\bm v - \bm u|^2}{2\theta}-\frac{D}{2}\right) \frac{
\partial\theta}{\theta}\right)
\end{equation}
Thus in order to calculate $\frac{d}{d\epsilon}\left. \mu(\mathcal{M} + \epsilon\mathcal{M}f) \right|_{\epsilon=0}$, we need expressions for $\frac{d}{d\epsilon}\left. \rho_\mu(\mathcal{M} + \epsilon\mathcal{M}f) \right|_{\epsilon=0} $, $\frac{d}{d\epsilon}\left. \bm{u}_\mu(\mathcal{M} + \epsilon\mathcal{M}f) \right|_{\epsilon=0} $ and $\frac{d}{d\epsilon}\left. \theta_\mu(\mathcal{M} + \epsilon\mathcal{M}f) \right|_{\epsilon=0}$. Working from (\ref{eq:mass}), (\ref{eq:momentum}) and (\ref{eq:intenergy}), we have that:
 \begin{align}
    \frac{d}{d\epsilon}\left. \rho_\mu(\mathcal{M} + \epsilon\mathcal{M}f) \right|_{\epsilon=0} 
     &=  
     \bracke{ f,\,\mathcal M}
     \\
     \frac{d}{d\epsilon}\left. \bm{u}_\mu(\mathcal{M} + \epsilon\mathcal{M}f) \right|_{\epsilon=0}  
     &= 
     \frac{1}{\rho_{\mu}(\mathcal M)}
     \bracke{
     (\bm v - \bm u_{\mu}(\mathcal M) ),\, \mathcal M\,f}
      \\
    \frac{d}{d\epsilon}\left. \theta_\mu(\mathcal{M} + \epsilon\mathcal{M}f) \right|_{\epsilon=0} 
     &= 
     \frac{2}{\rho_{\mu}(\mathcal M)}
     \bracke{ 
     \left(\frac{|\bm v - \bm u_{\mu}(\mathcal M)|^2}{2D} 
     -
     \frac{\theta_{\mu}(\mathcal M)}{2}\right) ,\,\mathcal M\,f}
 \end{align}
 Substituting the above into (\ref{eq:dmaxwellian}) gives (\ref{eq:proj}). Thus we have proved that
 }
\begin{equation}\label{eq:linbgk}
\mathcal{L}^{\mathrm{BGK}}_{\mathcal M}[f] 
= 
- \frac{1}{\tau(\mathcal{M})}\left(\mathrm{Id}-\Pi_{\mathcal M} \right)[f]
\end{equation}
Compliance of $\mathcal{L}^{\mathrm{BGK}}_{{\mathcal M}}[f] $ with properties (\ref{assume:selfadj})--(\ref{assume:fredholm}) follow directly from the fact that $\Pi_{\mathcal M}$ is an orthogonal projection of $\mathscr L^2(\mathcal M d\bm v)$ onto $\mathscr{I}$. In fact, it could be argued that \eqref{eq:linbgk} is the simplest possible linearized collision operator that satisfies these properties.

{\color{black}
 \section{A Useful Hierarchy of Tensors}\label{appendix:alphabet}
In order to compute the macroscopic closures that arise from the various fine-scale approximations introduced in this text, we introduce the following hierarchy of tensor functions: For a Maxwellian $\mathcal{M} = \mathcal{M}_{\varrho, \bm{w}, \vartheta}$, we have
\begin{subequations}\label{eq:Alphabet}
    \begin{align}
        A^\mathcal{M}_{ij} &= (\Id - \Pi_\mathcal{M})[v_i v_j] = (v_i - w_i)(v_j - w_j) - \frac{\delta_{ij}}{3}|\bm{v} - \bm{w}|^2 \label{eq:Atensor}
        \\
        B_i^\mathcal{M} &= (\Id -\Pi_\mathcal{M})\left[\frac{v_i|\bm{v}-\bm{w}|^2}{2} \right] = \frac{v_i - w_i}{2} (|\bm{v}-\bm{w}|^2 - 5\,\vartheta) \label{eq:Btensor}
        \\
        C^\mathcal{M}_{ijk} &= (v_k -w_k) \, \iLM\big[A^\mathcal{M}_{ij} \big]\label{eq:Ctensor}
        \\
        D^\mathcal{M}_{ij} &= (v_j - w_j) \, \iLM\big[B^\mathcal{M}_i \big] \label{eq:Dtensor}
        \\
        E^{\mathcal{M}}_{ijkl} &= (v_l - w_l) \, \iLM\sqbrac{C_{ijk}^\mathcal{M}}\label{eq:Etensor}
        \\
        F^\mathcal{M}_{ijk} &= (v_k - w_k) \, \iLM\sqbrac{D^\mathcal{M}_{ij}}\label{eq:Ftensor}
    \end{align}
\end{subequations}
The Maxwellians of interest to us in this manuscript are, of course, the global Maxwellian $ M = \mathcal{M}_{\rho_0, \bm{0}, \theta_0}$ and the self-consistent Maxwellian $ \mu = \mathcal{M}_{\rho, \bm{u}, \theta}$. However the notation introduced above allows us to discuss both cases simultaneously. 

Notice that the exact form of the tensors \eqref{eq:Ctensor}-\eqref{eq:Ftensor} depends on the linearized collision operator. For the BGK operator, $\mathcal L^{-1}_{\mathcal M}[\cdot] = -\tau (\text{Id} - \Pi_{\mathcal M})\left[\cdot\right]$, these tensors can be computed explicitly with the tools developed in the previous sections of the Appendix.
\begin{align}
 \label{eq:C_BGK} C^{\text{BGK}}_{ijk} &= -\tau\left(c_i c_j c_k - \frac{|\bm c|^2}{D}c_k \delta_{ij} \right)\\
     \label{eq:D_BGK}D^{\text{BGK}}_{ij}&= -\frac{\tau }{2}\left( |\bm c|^2 c_i c_j - (D+2)\vartheta c_i c_j \right)\\ F^{\text{BGK}}_{ijk} &= \frac{\tau^2}{2} \Bigg( |\bm c|^2 c_i c_j c_k - (D+2)\vartheta c_i c_j c_k - 2\left(1 + \frac{2}{D}\right)\left(\frac{|\bm c|^2}{2\vartheta} - \frac{D}{2} \right)\vartheta^2 c_k\delta_{ij}\Bigg)\label{eq:F_BGK}
 \end{align}
 where $\bm{c} = \bm{v} - \bm{w}$.

\section{Computing Macroscopic Closures}\label{appendix:alphabet2}
The linearized and non-linear deviatoric stress and heat flux closures are calculated from the integrals
\begin{align*}
    &\tilde{\sigma}^{(n)}_{ij} = -\abrac{A^M_{ij},\,M\, \breve{f}_{(n)}} 
    &{\sigma}^{(n)}_{ij} = -\epsilon \abrac{A^\mu_{ij},\,\mu\, {f}_{(n)}} \\
    &\tilde{q}^{(n)}_{i} = \abrac{B^M_{i},\,M\, \breve{f}_{(n)}} 
    &{q}^{(n)}_{i} = \epsilon\abrac{B^\mu_{i},\,\mu\, {f}_{(n)}}
\end{align*}
where $\hat{f}_{(n)}$ and $f_{(n)}$ are the linear and non-linear fine-scale approximations respectively. In this section, we will focus on the Navier-Stokes-Fourier closure (linear and non-linear), Burnett closure (linear only) and the (Simplified) Entropy Stable Extension (linear and non-linear).

\subsection{Navier-Stokes-Fourier}
With the linear and non-linear coarse-scale terms given by
\begin{subequations}\label{Aeq:coarse-scales}
\begin{align}
    &\dx{k}\bar{f} = \frac{\dx{k}\tilde{\rho}}{\rho_0}\, +\, {v}_n\frac{\dx{k}{\tilde{u}_n}}{\theta_0}\, +\, \left(\frac{|\bm{v}|^2}{2}-\frac{3\theta_0}{2} \right)\frac{\dx{k}\tilde{\theta}}{\theta_0^2}\\
    &\dx{k}\ln\mu = \frac{\dx{k}\rho}{\rho} + (v_n - u_n)\frac{\dx{k}u_n}{\theta} + \left(\frac{|\bm{v} - \bm{u}|^2}{2} - \frac{3\theta}{2} \right)\frac{\dx{k}\theta}{\theta^2}
\end{align}
\end{subequations}
the fine-scale closures of interest are 
\begin{subequations}
\begin{align}
  \label{Aeq:linFineScale}  \breve{f}_{(1)} &= \epsilon\, \iLm\!\sqbrac{ \vdx{k}\bar{f}\,}= \epsilon \frac{\dx{k}{\tilde{u}_n}}{\theta_0}\,\iLm[v_kv_n] + \epsilon\frac{\dx{k}\tilde{\theta}}{\theta_0^2}\, \iLm\!\sqbrac{\frac{v_k|\bm{v}|^2}{2}}\notag\\
    &= \epsilon \frac{\dx{k}{\tilde{u}_n}}{\theta_0}\,\iLm\!\sqbrac{A^M_{kn}\,}  + \epsilon\frac{\dx{k}\tilde{\theta}}{\theta_0^2}\, \iLm\!\sqbrac{B^M_k\,}\\
    \label{Aeq:nonlinFineScale}
    f_{(1)} &=  \iLmu\!\sqbrac{\vdx{k}\ln\mu\,}\notag\\
    &=  \frac{\dx{k}u_n}{\theta}\iLmu\!\sqbrac{A_{kn}^\mu\,} +  \frac{\dx{k}\theta}{\theta^2} \iLmu\!\sqbrac{B_k^\mu\,}
\end{align}
\end{subequations}
We arrive at \eqref{Aeq:linFineScale} and \eqref{Aeq:nonlinFineScale} by using collision invariance to focus only on the terms that belong in the orthogonal complement to the space of collision invariants.
Starting with the deviatoric stress,
\begin{align*}
   \tilde{\sigma}^{(1)}_{ij} &=  -\abrac{A^M_{ij},\,M\, \breve{f}_{(1)}} = - \epsilon \frac{\dx{k}{\tilde{u}_n}}{\theta_0}\abrac{ A^M_{ij},\,M\, \iLm\!\sqbrac{A^M_{kn}\,}} -\epsilon \frac{\dx{k}\tilde{\theta}}{\theta_0^2} \abrac{A^M_{ij},\,M\,\iLm\!\sqbrac{B^M_k\,} }\\
   &=  - \epsilon \frac{\dx{k}{\tilde{u}_n}}{\theta_0}\abrac{ A^M_{ij},\,M\, \iLm\!\sqbrac{A^M_{kn}\,}}\\
   \sigma^{(1)}_{ij}&=  -\epsilon\abrac{A^\mu_{ij},\,\mu\, {f}_{(1)}} = - \epsilon \frac{\dx{k}u_n}{\theta} \abrac{A^\mu_{ij},\, \mu\, \iLmu\!\sqbrac{A^\mu_{kn}\,}} - \epsilon\frac{\dx{k}\theta}{\theta^2}\abrac{A^\mu_{ij},\,\mu\,\iLmu\!\sqbrac{B_k^\mu\,}}\\
   &= - \epsilon \frac{\dx{k}u_n}{\theta} \abrac{A^\mu_{ij},\, \mu\, \iLmu\!\sqbrac{A^\mu_{kn}\,}}
\end{align*}
In both the linear and non-linear case, the last simplification comes about because the integral $\abrac{A^\mathcal{M}_{ij}, \mathcal{M}\,\iLM[B^\mathcal{M}_k]}$ yields an isotropic tensor with an odd number of indices and will thus equal zero. For the rank four isotropic tensor that remains, we can use the fact that $\mathcal{A}_{ij}^\mathcal{M}$ is symmetric and traceless to obtain
$$ \abrac{A^\mathcal{M}_{ij}, \mathcal{M}\,\iLM[A^\mathcal{M}_{kn}]} = C\, \parent{\delta_{ik}\delta_{jn} + \delta_{in}\delta_{jk} - \frac{2}{D}\delta_{ij}\delta_{kn}}$$
for some scalar $C$. It is easy to check that 
$$C= \abrac{A^\mathcal{M}_{13}, \mathcal{M}\,\iLM[A^\mathcal{M}_{13}]} = \abrac{A^\mathcal{M}_{12}, \mathcal{M}\,\iLM[A^\mathcal{M}_{12}]} = \abrac{A^\mathcal{M}_{23}, \mathcal{M}\,\iLM[A^\mathcal{M}_{23}]}$$
Thus the linear and non-linear deviatoric stresses work out to
\begin{align*}
    \tilde{\sigma}_{ij} &= \parent{-\epsilon \frac{\abrac{A^{M}_{13}, \,{M}\,\iLm[A^{M}_{13}]}}{\theta_0}} \parent{\dx{i}\tilde{u}_j + \dx{j}\tilde{u}_i - \frac{2}{D}\delta_{ij}\dx{k}\tilde{u}_k}\\
    {\sigma}_{ij} &= \parent{-\epsilon \frac{\abrac{A^{\mu}_{13}, \,{\mu}\,\iLmu[A^{\mu}_{13}]}}{\theta}} \parent{\dx{i}{u}_j + \dx{j}{u}_i - \frac{2}{D}\delta_{ij}\dx{k}{u}_k}
\end{align*}
and once we define the linear and non-linear viscosity 
\begin{align*}
    \tilde{\omega} &:= -\epsilon \frac{\abrac{A^{M}_{12}, \,{M}\,\iLm[A^{M}_{12}]}}{\theta_0}\\
    \omega &:= -\epsilon \frac{\abrac{A^{\mu}_{12}, \,{\mu}\,\iLmu[A^{\mu}_{12}]}}{\theta}
\end{align*}
we arrive at the Navier-Stokes-Fourier closure for the deviatoric stress.

The symmetric traceless isotropic tensor used to obtain the above closure will appear enough times later on to warrant a clear definition 
\begin{defn}[The Kronecker Phi tensor]
    \begin{align}\label{eq:KroneckerPhi}
        \KrPhi_{ab}^{cd} := \delta_{ac}\delta_{bd} + \delta_{ad}\delta_{bc} - \frac{2}{D}\delta_{ab}\delta_{cd}
    \end{align}
\end{defn}
\noindent Note that $\KrPhi_{ab}^{cd} = \KrPhi_{ba}^{cd} = \KrPhi_{ab}^{dc}$ and $\KrPhi_{kk}^{cd} = 0=\KrPhi_{ab}^{kk}$. What is more,
\begin{align*}
\KrPhi_{ab}^{cd}\ \dx{c}{w_d} &=  \dx{a}{w_b}+  \dx{b}{w_a}  - \frac{2}{D}\delta_{ab}\text{div}\bm{w}= 2\, \tepsilon_{ab}(\bm{w})\\
  \KrPhi_{ab}^{cd}\ \dx{c}\dx{d}g &= 2 \parent{\dx{a}\dx{b} - \frac{\delta_{ab}}{D}\Delta}[g] 
\end{align*}

The calculation for the Navier-Stokes-Fourier heat flux flows in a similar, if not simpler, manner to the deviatoric stress calculation
\begin{align*}
    \tilde{q}_i &= \abrac{B^M_{ij},\,M\, \hat{f}_{(1)}}= \epsilon \frac{\dx{k}\tilde{\theta}}{\theta_0^2} \abrac{B^M_{i},\,M\,\iLm\!\sqbrac{B^M_k\,} } = - \parent{- \epsilon \frac{\abrac{B^M_{1},\,M\,\iLm\!\sqbrac{B^M_1\,} }}{\theta_0^2} }\delta_{ik}\dx{k}\tilde{\theta}\\
    q_i &= \epsilon\abrac{B_i^\mu,\, \mu \, f_{(1)}} = \epsilon\frac{\dx{k}\theta}{\theta^2}\abrac{B_i^\mu,\,\mu\,\iLmu\!\sqbrac{B^\mu_k}}= - \parent{- \epsilon \frac{\abrac{B^\mu_{1},\,\mu\,\iLmu\!\sqbrac{B^\mu_1\,} }}{\theta^2} }\dx{i}{\theta}
\end{align*}
With the linear and non-linear heat conductivity defined as 
\begin{align*}
    \tilde{\kappa} &:= - \epsilon \frac{\abrac{B^M_{1},\,M\,\iLm\!\sqbrac{B^M_1\,} }}{\theta_0^2} \\
    \kappa &:= - \epsilon \frac{\abrac{B^\mu_{1},\,\mu\,\iLmu\!\sqbrac{B^\mu_1\,} }}{\theta^2}
\end{align*}
we arrive at Fourier's law as the closure for the heat flux.

\subsection{Burnett}

The Burnett correction terms are 
\begin{align*}
    \breve{f}_{(2^{B1})} - \breve{f}_{(1)} &= \epsilon^2 \mathcal{L}_M^{-1}(\text{St } \partial_t + v_k\partial_{x_k}) \mathcal{L}_M^{-1}\left[ v_l\partial_{x_l}\bar{f}\, \right] \\
    &= \epsilon^2\St\parent{\frac{\partial_t\dx{l}\tilde{u}_n}{\theta_0}\Lm^{-2}\!\sqbrac{A^M_{ln}} + \frac{\partial_t\dx{l}\tilde{\theta}}{\theta_0^2}\Lm^{-2}\!\sqbrac{B^M_{l}} } + \epsilon^2 \parent{ \frac{\dx{k}\dx{l}\tu_n}{\theta_0}\iLm\!\sqbrac{C^M_{lnk}} + \frac{\dx{k}\dx{l}\ttheta}{\theta_0^2}  \iLm\!\sqbrac{D_{lk}^M}       }
\end{align*}
Starting with the deviatoric stress term and ignoring all tensor integrals with an odd number of indices yields
\begin{align*}
{\tilde{\sigma}_{ij}^{(2^{B1})} - \tilde{\sigma}_{ij}^{(1)}} &=  -\epsilon^2 \left(\frac{\text{St}\,\partial_t\partial_{x_l} \tilde{u}_n }{\theta_0}\left\langle \mathcal{L}_M^{-1}[A_{ij}^M], M \mathcal{L}^{-1}_M[A_{nl}^M] \right\rangle + \frac{\partial_{x_k}\partial_{x_l} \tilde{\theta}}{\theta_0^2}\left\langle \mathcal{L}_M^{-1}[A_{ij}^M], M D_{lk}^M \right\rangle \right)\\
&= -\Xi\, \KrPhi_{ij}^{ln}\,\St\partial_t\partial_{x_l} \tilde{u}_n - {\Psi}\KrPhi_{ij}^{lk}\,\dx{k}\dx{l}\ttheta\\
&= - 2\,\Xi\ \St\partial_t\tepsilon_{ij}(\bm{\tu}) - 2\Psi \parent{\dx{i}\dx{j} - \frac{\delta_{ij}}{D}\Delta}\!\big[\,\ttheta\,\big] 
\end{align*}
where 
\begin{align*}
    &\Xi = \frac{\epsilon^2}{\theta_0}\left\langle \mathcal{L}_M^{-1}[A_{13}^M], M \mathcal{L}^{-1}_M[A_{13}^M] \right\rangle &\Psi = \frac{\epsilon^2}{\theta_0^2}\abrac{\iLm[A_{13}^M\,],\,M\,D_{13}^M} = \frac{\epsilon^2}{\theta_0^2} \abrac{C^M_{133},\, M \iLm[B_1^M\,]}
\end{align*}
Similarly for the heat flux,
\begin{align*}
    \tilde{q}_i^{(2^{B1})} - \tilde{q}_{i}^{(1)} &= \epsilon^2 \parent{ \frac{\St\partial_t \dx{l}\ttheta}{\theta_0^2} \abrac{\iLm[B^M_i], \, M \,\iLm[B^M_l] } + \frac{\dx{k}\dx{l}\tu_n}{\theta_0} \abrac{\iLm[B_i^M],\, M\, C_{lnk}^M} }\\
    &= \Upsilon \, \St\partial_t\dx{i}\ttheta \,+ \, \theta_0 {\Psi}\ \KrPhi_{ln}^{ik}\dx{k}\dx{l}\tu_n  \\
    &= \Upsilon \, \St\partial_t\dx{i}\ttheta +2 \theta_0 \Psi \dx{k}\tepsilon_{ik}(\bm{\tu})
\end{align*}
where 
$$ \Upsilon = \frac{\epsilon^2}{\theta^2} \abrac{\iLm[B^M_2], \, M \,\iLm[B^M_2] } $$

\subsection{(Simplified) Entropy Stable Extension}
From \eqref{eq:nonlinaltburnclose} and \eqref{eq:linaltburnclose}, the correction terms of interest can be written as
\begin{align*}
 \breve{f}_{(2^E)} - \breve{f}_{(1)} &=  -\epsilon^3\mathcal{L}^{-1}_M v_k \partial_{x_k}\left[ \mathcal{L}_M^{-1}\left[ v_l\, \mathcal{L}^{-1}_M\left[ \vdx{n}\dx{l}\bar{f}\ \right]\right]\right]
 \\
{f}_{(2^E)} - {f}_{(1)} &=  -\epsilon^2 \mathcal{L}_\mu^{-1}\vdx{k}\left[ \mathcal{L}_\mu^{-1}\left[  v_l \,\mathcal{L}_\mu^{-1}\big[\vdx{n}\dx{l}\!\ln\mu \big] \right]\right]   + \text{term with }\mathcal{X}_\mu 
    \end{align*}  
where the identity \eqref{eq:DerivativeSwitchIdentity} has been applied in the non-linear case. 

\subsubsection{Deviatoric Stress}
For the linear deviatoric stress,
\begin{align*}
    \tilde{\sigma}_{ij}^{(2^E)} - \tilde{\sigma}_{ij}^{(1)} &= \epsilon^3 \abrac{v_k \iLm[A_{ij}^M],\, M\, \dx{k}\!\left[ \mathcal{L}_M^{-1}\left[ v_l\, \mathcal{L}^{-1}_M\left[ \vdx{n}\dx{l}\bar{f}\ \right]\right]\right] } \\
     &= \epsilon^3 \dx{k}\!\sqbrac{\abrac{C_{ijk}^M,\, M\, \mathcal{L}_M^{-1}\left[C_{npl}^M\right] }\frac{\dx{n}\dx{l}\tu_p}{\theta_0} }\\ &:= -\dx{k}\tilde{\mathfrak{S}}_{ijk}
\end{align*}
In its most general form, the above rank 6 isotropic tensor will be 
\begin{align*}
   _MK_{ijk}^{npl}:= \abrac{C_{ijk}^M,\, M\, \mathcal{L}_M^{-1}\left[C_{npl}^M\right] } = \tilde{a}_1\,\delta_{in}\delta_{jp}\delta_{kl} + \tilde{a}_2\,\delta_{in}\delta_{jk}\delta_{pl} + \tilde{a}_3\,\delta_{in}\delta_{jl}\delta_{kp}+\dots \text{(12 other terms)}   
\end{align*}
By using $C_{ijk}^M = C_{jik}^M$, $C_{npl}^M = C_{pnl}^M$ and $C_{jjk} = C_{ppl} = 0$, this can be simplified further to 
\begin{align*}
 _MK_{ijk}^{npl} = \tilde{a}_1\, \delta_{kl}\KrPhi_{ij}^{np}+ \tilde{a}_2 \left(\delta_{ln}\KrPhi_{ij}^{pk} +\delta_{lp}\KrPhi_{ij}^{nk} -\frac{2}{D}\delta_{np}\KrPhi_{ij}^{kl} \right)+\tilde{a}_3\left(\delta_{kn}\KrPhi_{ij}^{pl}+\delta_{kp}\KrPhi_{ij}^{nl} -\frac{2}{D}\delta_{np}\KrPhi_{ij}^{kl} \right)  
\end{align*}
where $\tilde{a}_1 =\ _MK_{123}^{123}$, $\tilde{a}_2 =\ _MK_{122}^{133}$ and $\tilde{a}_3 =\ _M\!K_{123}^{132}$.
Thus,
\begin{align*}
     \tilde{\sigma}_{ij}^{(2^E)} - \tilde{\sigma}_{ij}^{(1)} &= \frac{\epsilon^3}{\theta_0} \dx{k}\!\Big[ K_{ijk}^{npl}\dx{l}\dx{n}\tu_{p}  \big]\\
     &= \dx{k}\! \Big[\ 2\tilde{a}_1 \dx{k}\tepsilon_{ij}(\bm{\tu})+ \tilde{a}_3 \left({\color{violet}\dx{k}\dx{i}\tu_j} + {\color{teal}\dx{k}\dx{j}\tu_i} - {\color{blue} \frac{2}{D}\delta_{ij}\dx{k}\divut} \right) \\&\hspace{1.5cm}+ \tilde{a}_3 \left({\color{violet} \dx{i}\dx{j}\tu_k} + {\color{teal}\dx{i}\dx{j}\tu_k} - {\color{blue}\frac{2}{D}\delta_{ij}\Delta\tu_k} \right)\\&\hspace{1.5cm} -\frac{2}{D}\tilde{a}_3 \left({\color{teal}\delta_{ik}\dx{j}\divut} + {\color{violet}\delta_{jk}\dx{i}\divut} - {\color{blue}\frac{2}{D}\delta_{ij}\dx{k}\divut} \right) \\&\hspace{1.5cm} + \tilde{a}_2 \delta_{ik}\left(\Delta\tu_{j} +\dx{j}\divut - \frac{2}{D}\dx{j}\divut  \right)  \\
     &\hspace{1.5cm}+ \tilde{a}_2 \delta_{jk}\left(\Delta\tu_{i} +\dx{i}\divut - \frac{2}{D}\dx{i}\divut  \right) \\
     &\hspace{1.5cm}- \frac{2}{D}\tilde{a}_2 \delta_{ij}\left(\Delta\tu_{k} +\dx{k}\divut - \frac{2}{D}\dx{k}\divut  \right) 
     &\Big]
\end{align*}
Noting that $2\dx{a}\tepsilon_{ab}(\bm{\tu}) = \Delta\tu_{b} + \dx{b}\divut - \frac{2}{D}\dx{b}\divut$ and grouping the terms of the same color, the above simplifies to 
\begin{align*}
      \dx{k}\tilde{\mathfrak{S}}_{ijk} &=  2\dx{k}\!\Bigg[ \tilde{\alpha}_1 \dx{k}\tepsilon_{ij}(\bm{\tu}) + \tilde{\alpha}_3 \left(\dx{i}\tepsilon_{jk}(\bm{\tu})+  \dx{j}\tepsilon_{ik}(\bm{\tu})  -\frac{2}{D}\delta_{ij}\dx{l}\tepsilon_{lk}(\bm{\tu}) \right)\\ &\hspace{3.5cm}
      +\tilde{\alpha}_2 \left(\delta_{ik}\dx{l}\tepsilon_{lj}(\bm{\tu}) +\delta_{jk}\dx{l}\tepsilon_{li}(\bm{\tu}) - \frac{2}{D}\delta_{ij}\dx{l}\tepsilon_{lk}(\bm{\tu})\right)
      \Bigg]
\end{align*}
where
\begin{align}\label{eq:linESEAlphaDef}
   &\tilde{\alpha}_1 = - \frac{\epsilon^3 \tilde{a}_1}{\theta_0}        & \tilde{\alpha}_2 = -\frac{\epsilon^3\tilde{a}_2}{\theta_0} &&\tilde{\alpha}_3 = - \frac{\epsilon^3 \tilde{a}_3}{\theta_0}  
\end{align}
In the next section, we shall see that $\tilde{a}_1 = \tilde{a}_3$ and thus the final expression is actually 
\begin{align}\label{eq:linESESpressFinal}
   \tilde{\mathfrak{S}}_{ijk} &=  2 \tilde{\alpha}_1 \left(\dx{k}\tepsilon_{ij}(\bm{\tu}) +  \dx{i}\tepsilon_{jk}(\bm{\tu})+  \dx{j}\tepsilon_{ik}(\bm{\tu})  -\frac{2}{D}\delta_{ij}\dx{l}\tepsilon_{lk}(\bm{\tu}) \right)\nonumber\\ &\hspace{1.5cm}
      +2\,\tilde{\alpha}_2 \left(\delta_{ik}\dx{l}\tepsilon_{lj}(\bm{\tu}) +\delta_{jk}\dx{l}\tepsilon_{li}(\bm{\tu}) - \frac{2}{D}\delta_{ij}\dx{l}\tepsilon_{lk}(\bm{\tu})\right)
\end{align}

For the non-linear deviatoric stress,
\begin{align*}
    {\sigma}_{ij}^{(2^E)} - {\sigma}_{ij}^{(1)} &= \epsilon^3 \abrac{v_k \iLmu[A_{ij}^\mu], \mu\, \dx{k}\!\left[ \mathcal{L}_\mu^{-1}\left[  v_l \,\mathcal{L}_\mu^{-1}\big[\vdx{n}\dx{l}\!\ln\mu \big] \right]\right]  } + \mathcal{X}_\mu \text{ term}\\
    &= \epsilon^3 \dx{k}\!\abrac{v_k \iLmu[A_{ij}^\mu], \mu\,  \mathcal{L}_\mu^{-1}\!\left[  v_l \,\mathcal{L}_\mu^{-1}\big[\vdx{n}\dx{l}\!\ln\mu \big] \right]  } \\
    &\hspace{0.5cm} - \epsilon^3 \abrac{\frac{1}{\mu} \vdx{k}\!\sqbrac{\mu \iLmu[A_{ij}^\mu]},\,\mu\, \mathcal{L}_\mu^{-1}\!\left[  v_l \,\mathcal{L}_\mu^{-1}\big[\vdx{n}\dx{l}\!\ln\mu \big] \right]    } + \mathcal{X}_\mu \text{ term}\\
    &= \epsilon^3 \dx{k}\!\abrac{v_k \iLmu[A_{ij}^\mu], \mu\,  \mathcal{L}_\mu^{-1}\!\left[  v_l \,\mathcal{L}_\mu^{-1}\big[\vdx{n}\dx{l}\!\ln\mu \big] \right]  } \\
    &\hspace{0.5cm} - \epsilon^3 \abrac{ v_k \iLmu[\dx{k}A_{ij}^\mu],\,\mu\, \mathcal{L}_\mu^{-1}\!\left[  v_l \,\mathcal{L}_\mu^{-1}\big[\vdx{n}\dx{l}\!\ln\mu \big] \right]    } + \mathcal{X}_\mu \text{ terms}\\
\end{align*}
Because $\dx{k}A_{ij}^\mu$ is a collision invariant, the second term will disappear. This means that, as with the linear case, we will have 
\begin{align*}
    -\dx{k}\mathfrak{S}_{ijk} &:=  \epsilon^3 \dx{k}\!\abrac{v_k \iLmu[A_{ij}^\mu], \mu\,  \mathcal{L}_\mu^{-1}\!\left[  v_l \,\mathcal{L}_\mu^{-1}\big[\vdx{n}\dx{l}\!\ln\mu \big] \right]  }\\
    &= \epsilon^3 \dx{k}\!\!\sqbrac{ \frac{\dx{n}\dx{l}u_p}{\theta}\abrac{v_k \iLmu[A_{ij}^\mu], \mu\,  \mathcal{L}_\mu^{-1}\!\left[  v_l \,\mathcal{L}_\mu^{-1}\big[A_{np}^\mu] \right]  } + \frac{\dx{n}\dx{l}\theta}{\theta^2}\abrac{v_k \iLmu[A_{ij}^\mu], \mu\,  \mathcal{L}_\mu^{-1}\!\left[  v_l \,\mathcal{L}_\mu^{-1}\big[B_{n}^\mu] \right]  } }
\end{align*}
Making the substitution $v_r = (v_r - u_r) + u_r$ and keeping only the terms whose integrals have an even number of indices then gives
\begin{align*}
  \dx{k}\mathfrak{S}_{ijk} &= -\epsilon^3 \dx{k}\!\!\sqbrac{\frac{\dx{n}\dx{l}u_p}{\theta}\abrac{C_{ijk}^\mu, \mu\,  \mathcal{L}_\mu^{-1}\!\left[ C_{npl}^\mu \right]  }  + u_k \frac{\bm{u}\!\cdot\!\nabla[\dx{n}u_p]}{\theta} \abrac{ \iLmu[A_{ij}^\mu], \mu\,   \,\mathcal{L}_\mu^{-2}\big[A_{np}^\mu]   }    }  \\
  &\hspace{1cm} -\epsilon^3 \dx{k}\!\!\sqbrac{\frac{\bm{u}\!\cdot\!\nabla[\dx{n}\theta]}{\theta^2}\abrac{C_{ijk}^{\mu},\,\mu\, \Lmu^{-2}[B_n^\mu]}    + u_k\frac{\dx{n}\dx{l}\theta}{\theta^2}\abrac{\iLmu[A_{ij}^\mu],\,\mu\, \iLmu[D_{nl}^\mu]}} 
\end{align*}
Based off the work carried out earlier,
\begin{align*}
_\mu {K}_{ijk}^{npl}&:= \abrac{C^\mu_{ijk}, \mu \iLmu\big[C_{npl}^\mu \big]}\\
&=\ _\mu{K}_{123}^{123}\, \left(\delta_{kl}\KrPhi_{ij}^{np}+\delta_{kn}\KrPhi_{ij}^{pl}+\delta_{kp}\KrPhi_{ij}^{nl} -\frac{2}{D}\delta_{np}\KrPhi_{ij}^{kl} \right)+\ _\mu {K}_{122}^{133} \left(\delta_{ln}\KrPhi_{ij}^{pk} +\delta_{lp}\KrPhi_{ij}^{nk} -\frac{2}{D}\delta_{np}\KrPhi_{ij}^{kl} \right)
\\
\mathcal{A}_{ij}^{np}&:= \abrac{\iLmu[A^\mu_{ij}], \mu \Lmu^{-2}[A_{np}^\mu]}  = \mathcal{A}_{13}^{13}\ \KrPhi_{ij}^{np}
\\
\mathcal{P}_{ijk}^{n}&:= \abrac{C_{ijk}^\mu,\mu\,\Lmu^{-2}[B_n^\mu]} = \mathcal{P}_{131}^1\ \KrPhi_{ij}^{kn}
\\
\mathcal{D}_{ij}^{nl}&:= \abrac{\iLmu[A_{ij}^\mu], \mu\, \iLmu[D_{nl}^\mu]} = \mathcal{D}_{13}^{13}\ \KrPhi_{ij}^{nl}
\end{align*}
Thus
\begin{align*}
\mathfrak{S}_{ijk} &=  2 {\alpha}_1 \left(\dx{k}\tepsilon_{ij}(\bm{u}) +  \dx{i}\tepsilon_{jk}(\bm{u})+  \dx{j}\tepsilon_{ik}(\bm{u})  -\frac{2}{D}\delta_{ij}\dx{l}\tepsilon_{lk}(\bm{\tu}) \right)\nonumber\\ \nonumber&\hspace{1.5cm}
      +2\,{\alpha}_2 \left(\delta_{ik}\dx{l}\tepsilon_{lj}(\bm{u}) +\delta_{jk}\dx{l}\tepsilon_{li}(\bm{u}) - \frac{2}{D}\delta_{ij}\dx{l}\tepsilon_{lk}(\bm{u})\right)
      \\\nonumber&\hspace{1cm}
      +2 \mathcal{A}\, u_k\, \uD[\tepsilon_{ij}(\bm{u})]+ 2\mathcal{P}\, \uD\!\left[\frac{\delta_{ik}\partial_{x_j}\theta +\delta_{jk}\partial_{x_i}\theta}{2} - \frac{1}{D}\delta_{ij}\partial_{x_k}\theta  \right] 
      \\&\hspace{1.5cm}
      + 2\mathcal{D}\, u_k\, \left(\partial_{x_i}\partial_{x_j} - \frac{\delta_{ij}}{D}\Delta\right)[\theta]
\end{align*}
where 
\begin{align*}
    &\alpha_1 = -\frac{\epsilon^3\ _\mu{K}_{123}^{123}}{\theta} &\alpha_2 = -\frac{\epsilon^3\ _\mu\mathcal{K}_{122}^{133}}{\theta} &&\mathcal{A} = -\frac{\epsilon^3\mathcal{A}_{13}^{13}}{\theta}
    \\
    &\mathcal{P} = -\frac{\epsilon^3\mathcal{P}_{313}^1}{\theta^2} &\mathcal{D} = - \frac{\epsilon^2\mathcal{D}_{13}^{13}}{\theta^2}
\end{align*}

\subsubsection{Heat flux}
For the linear case,
\begin{align*}
    \tilde{q}_{i}^{(2^E)} - \tilde{q}_{i}^{(1)} &= -\epsilon^3 \abrac{v_k \iLm[B_{i}^M],\, M\, \dx{k}\!\left[ \mathcal{L}_M^{-1}\left[ v_l\, \mathcal{L}^{-1}_M\left[ \vdx{n}\dx{l}\bar{f}\ \right]\right]\right] }\\
    &= -\epsilon^3 \dx{k}\!\sqbrac{\frac{\dx{l}\dx{n}\theta}{\theta^2}\ {\abrac{D_{ik}^M,\, M\, D_{nl}^M }} }, & \ {\color{teal}_MS_{ik}^{nl}:= \abrac{D_{ik}^M,\, M\, D_{nl}^M } }\\
    &= -\epsilon^3 \dx{k}\!\sqbrac{\frac{\dx{l}\dx{n}\theta}{\theta^2} (\ _MS_{13}^{13} \delta_{in}\delta_{kl} + \ _MS_{13}^{31} \delta_{il}\delta_{kn} + \ _MS_{11}^{33} \delta_{ik}\delta_{nl})   }\\
    &= \dx{k}\sqbrac{\ \delta_{ik}\,\tilde{\Lambda}_0\,\Delta\theta +\tilde{\Lambda}_1\, \dx{i}\dx{k}\theta\ }\\
    &:= \dx{k}\tilde{\mathfrak{H}}_{ik}
\end{align*}
where 
\begin{align*}
    &\tilde{\Lambda}_0 = - \epsilon^3\, \frac{_MS_{11}^{33}}{\theta^2} &\tilde{\Lambda}_1= -\epsilon^3\, \frac{\ _MS_{13}^{13} +\ _MS_{13}^{31}}{\theta^2} 
\end{align*}

The non-linear case gives
\begin{align*}
     {q}_{i}^{(2^E)} - {q}_{i}^{(1)} &= - \epsilon^3 \dx{k}\!\abrac{v_k \iLmu[B_{i}^\mu], \mu\,  \mathcal{L}_\mu^{-1}\!\left[  v_l \,\mathcal{L}_\mu^{-1}\big[\vdx{n}\dx{l}\!\ln\mu \big] \right]  } \\
    &\hspace{0.5cm} + \epsilon^3 \abrac{ v_k \iLmu[\dx{k}B_{i}^\mu],\,\mu\, \mathcal{L}_\mu^{-1}\!\left[  v_l \,\mathcal{L}_\mu^{-1}\big[\vdx{n}\dx{l}\!\ln\mu \big] \right]    }\ + \mathcal{X}_\mu \text{ terms}
\end{align*}
Using $\iLmu[\dx{k}B_{i}^\mu] = -\dx{k}\!u_p\,\iLmu[A^\mu_{ip}]$, we then get
\begin{align*}
     {q}_{i}^{(2^E)} - {q}_{i}^{(1)} &= - \epsilon^3 \dx{k}\!\abrac{v_k \iLmu[B_{i}^\mu], \mu\,  \mathcal{L}_\mu^{-1}\!\left[  v_l \,\mathcal{L}_\mu^{-1}\big[\vdx{n}\dx{l}\!\ln\mu \big] \right]  }  + \dx{k}\!u_p\, \mathfrak{S}_{ipk}\ + \mathcal{X}_\mu\text{ terms}\\
     &:= \dx{k}\mathfrak{H}_{ik} + \dx{k}\!u_p\, \mathfrak{S}_{ipk}\ + \mathcal{X}_\mu\text{ terms}
\end{align*}

Making the substitution $v_r = (v_r - u_r) + u_r$ and keeping only the terms whose integrals have an even number of indices then yields
\begin{align*}
    \mathfrak{H}_{ik} &= -\epsilon^3 \abrac{D_{ik}^\mu, \mu \Lmu^{-2}[A_{lp}^\mu]}\frac{\uD[\dx{l}u_p]}{\theta}
    -\epsilon^3\abrac{\iLmu[B_{i}^\mu], \mu \iLmu[C_{npl}^\mu]}\,u_k\,\frac{\dx{l}\dx{n}u_p}{\theta}
     \\&\hspace{1cm}
     -\epsilon^3\abrac{D_{ik}^\mu, \mu\iLmu[D_{nl}^\mu]}\frac{\dx{l}\dx{n}\theta}{\theta^2} -\epsilon^3\abrac{\iLmu[B_i^\mu],\mu \Lmu^{-2}[B_n^\mu]}\,u_k\,\frac{\uD[\dx{n}\theta]}{\theta^2}\\
     &= 2\mathcal{D}\,\theta\, \uD[\tepsilonu{ik}] + 2\mathcal{P}\, \theta \, u_k \dx{l}\tepsilonu{li} + \mathcal{B}\,u_k\uD[\dx{i}\theta]\notag
    \\ &\hspace{1cm}
    + \Lambda_0\, \Delta\theta \delta_{ik} + \Lambda_1 \dx{i}\dx{k}\theta
\end{align*}
where 
\begin{align*}
    &\Lambda_0 = -\epsilon^3\,\frac{ _\mu S_{33}^{11}}{\theta^2}  &\Lambda_1 = -\epsilon^3\, \frac{ _\mu S_{13}^{13} +\ _\mu S_{13}^{31}}{\theta^2} &&\mathcal{B} = -\frac{\epsilon^3}{\theta^2}\abrac{\iLmu[B_3^\mu],\mu\Lmu^{-2}[B_3^\mu]}
\end{align*}

\section{Hard Spheres Fluid Parameters}\label{appendix:HardSphereCalc}
For a Maxwellian $\mathcal{M} = \mathcal{M}_{\varrho, \bm{w},\vartheta}$, recall that the linearized Hard Spheres collision operator is given by
 $$\mathcal{L}_{\mathcal{M}}[g]= \frac{\epsilon}{2\pi\rho_0\ell\sqrt{2}}\int_{\mathbb{R}^3\times\mathbb{S}^2} \mathcal{M}(\bm{v}_*)\Big(g(\bm{v}{'})+ g(\bm v{'}_*) -g(\bm v_*) -g(\bm v)  \Big)\big|\bm{\hat\eta}\cdot(\bm{v}_*-\bm{v})\big|\ d\bm{\hat\eta}\, d\bm v_* $$
Introducing the non-dimensionalized velocity $\bm{c}= \frac{\bm{v} - \bm{w}}{\sqrt{2\vartheta}}$ and the Gaussian $\gauss(\bm{c}) = e^{-|\bm{c}|^2}$, the linearized collision operator can be written as
\begin{align}\label{eq:ColNonDim}
    \LM[g](\bm{v}) &= \frac{\varrho\,\mathfrak{b}_{\mathcal{M}}}{\rho_0\pi^{\frac{3}{2}}}\int_{\mathbb{R}^3\times\mathbb{S}^2}\! \gauss_*\, \big((g\circ\bm{v})'+(g\circ\bm{v})'_* - (g\circ\bm{v}) - (g\circ\bm{v})_*\big) \widetilde{\mathfrak{B}}(\odir, \bm{c}, \bm{c}_*) d\odir d\bm{c}_* \notag
    \\
    &=: \frac{\varrho\,\mathfrak{b}_{\M}}{\rho_0\pi^{\frac{3}{2}}}\, \LG[g\circ\bm{v}](\bm{c})
\end{align}
where $\gauss_* = \gauss(\bm{c}_*)$,  $(g\circ\bm{v})'= g(\bm{v}(\bm{c'}))$, $(g\circ\bm{v})_* = g(\bm{v}(\bm{c}_*))$, etc. The collision kernel has been decomposed into a non-dimensional collision kernel $\widetilde{\mathfrak{B}}$ and a dimensional scaling factor ${\mathfrak{b}_\M}/{\rho_0}$
 \begin{align*}
        &\mathfrak{b}_\M = \frac{\epsilon\,\sqrt{\vartheta}\,}{2\pi \ell}  &\widetilde{\mathfrak{B}} = |\odir\cdot(\bm{c}_* - \bm{c})| 
    \end{align*}
For a function $g$ in the orthogonal complement $\mathscr{I}^{\perp_\M}$, we can use the identity $\LM[\iLM[g]]=g$ to show that the inverse linearized operator will be
\begin{align}\label{eq:InvColNonDim}
    \iLM[g](\bm{v}) = \frac{\rho_0 \pi^{\frac{3}{2}}}{\varrho\, \mathfrak{b}_\M} \ \iLG[g\circ\bm{v}](\bm{c})
\end{align}
The calculations that will follow also make use of a non-dimensionalized hierarchy of tensors:
\begin{align*}
    &A^\gauss_{ij}(\bm{c}) = c_i c_j -\frac{|\bm{c}|^2}{3}\delta_{ij} = \frac{1}{2\vartheta}A_{ij}^\M(\bm{v}(\bm{c}))
    &B^\gauss_{i}(\bm{c}) = c_i\left(|\bm{c}|^2 - \frac{5}{2} \right) = \frac{1}{\sqrt{2\vartheta^3}}\, B^\M_i(\bm{v}(\bm{c}))\\
    &C^\gauss_{ijk}(\bm{c}) = c_k \iLG[A_{ij}^\gauss] = \frac{ \mathfrak{b}_\M}{(2\pi\vartheta)^{\frac{3}{2}}}\,\frac{\varrho}{\rho_0}\,C^\M_{ijk}(\bm{v}(\bm{c}))
    &D^\gauss_{ij}(\bm{c}) = c_j \iLG[B_i^\gauss] = \frac{\mathfrak{b}_\M}{2\vartheta^2\pi^{\frac{3}{2}}}\,\frac{\varrho}{\rho_0}\,D^\M_{ij}(\bm{v}(\bm{c}))
\end{align*}
and an $\gauss$-weighted inner product
\begin{align*}
    \Ebrac{g, \, h} := \int_{\mathbb{R}^3} g(\bm{c})\, h(\bm{c})\, \gauss(\bm{c})\, d\bm{c}
\end{align*}
For suitable functions $a(\bm{v})$ and $b(\bm{v})$, 
\begin{align*}
    \abrac{a,\,\M \, b} = \frac{\varrho}{\pi^{\frac{3}{2}}}\Ebrac{\,(a\circ\bm{v}),\, (b\circ \bm{v})\,}
\end{align*}

With these, we can calculate the Hard Spheres fluid parameters for both the linear and non-linear cases simply by working with $\iLG$ and appropriately rescaling the result. The fluid parameters are calculated with the aid of Burnett functions \cite{Burnett1935, Gust2009, WangChang1952, gamba2018}
\begin{equation}\label{eq:BurnettFxn}
    \chi_{n,l,m}(\bm{c}) = \sqrt{\frac{2(n!)}{\Gamma\!\left(n + l + \frac{3}{2} \right)}}\,\ \text{L}^{l+\frac{1}{2}}_n(|\bm{c}|^2) \,|\bm{c}|^l\,\,Y^m_l(\hat{\bm{c}})
\end{equation}
where $\text{L}_n^{\alpha}$ is an associated Laguerre polynomial, $\bm{\hat{c}} = \frac{\bm{c}}{|\bm{c}|}$ and $Y_l^m$ is a spherical harmonic. The Burnett functions are orthonormal in $\mathscr{L}^2(\gauss\, d\bm{c})$:
$$ \Ebrac{ \chi_{n,l,m}^*,\ \chi_{p,q,r}} = \delta_{np}\delta_{lq}\delta_{mr}$$
where $ \chi_{n,l,m}^*$ is the complex conjugate of $ \chi_{n,l,m}$. A function $h \in \mathscr{L}^2(\gauss d\bm{c})$ can be decomposed in terms of Burnett functions
$$h(\bm{c}) = \sum_{n=0}^\infty\sum_{l=0}^\infty\sum_{m=-l}^l h_{n,l,m}\,\chi_{n,l,m}(\bm{c}), \qquad\, h_{n,l,m}= \Ebrac{\chi_{n,l,m}^*\,, \ h}$$ 
In particular with the coordinate system $\bm{\hat{c}} = (\sin\Theta\,\cos\varphi,\,\sin\Theta\, \sin\varphi,\,\cos\Theta )^{\mathtt{T}}$, the basis of the space of collision invariants decompose as
\begin{align*}
    &1 = \pi^{\frac{3}{4}}\chi_{0,0,0}(\bm{c})&\quad c_3 = \frac{\pi^{\frac{3}{4}}}{\sqrt{2}}\chi_{0,1,0}(\bm{c})\\ &c_2 = \frac{\mathbf{i}\pi^{\frac{3}{4}}}{2}\big( \chi_{0,1,1}(\bm{c}) + \chi_{0,1,-1}(\bm{c}) \big)  &c_1 = \frac{\pi^{\frac{3}{4}}}{2}\big( \chi_{0,1,-1}(\bm{c}) - \chi_{0,1,1}(\bm{c}) \big)\\
    &|\bm{c}|^2 = \pi^{\frac{3}{4}} \left(\frac{3}{2}\chi_{0,0,0}(\bm{c}) - \sqrt{\frac{3}{2}}\chi_{1,0,0}(\bm{c}) \right)
\end{align*}

Burnett functions are important to us because the action of a linearized collision operator on them can be characterized in terms of symmetric collision matrices
\begin{align*}
    &\Ebrac{\chi_{nlm}^*,\  \LG[\chi_{pqr}]\,} =  \big\{\mathbf{L}^l\big\}_{np}\, \delta_{l q}\,\delta_{mr}
    \\
    &\Ebrac{\chi_{nlm}^*,\ \iLG[\chi_{pqr}]\,} =    \big\{\mathbf{L}^l\big\}^{-1}_{np}\, \delta_{l q}\,\delta_{mr}
\end{align*}
From \cite{Gust2009}, we have that the entries for Hard Spheres collision matrix are  
\begin{equation*}
    \{\mathbf{L}^l\}_{np}\! = \pi^2 \sqrt{\frac{2(n!\,p!)}{ \Gamma(n+l+\frac{3}{2})\Gamma(p+l+\frac{3}{2})}}\,\sum_{j=0}^{\text{min}\{n,\,p\}}\sum_{k=0}^l\frac{l!\,\Gamma(-\frac{1}{2} + n+p+l - 2j - k)}{(n-j)!(p-j)!(l-k)!\, 2^{n+p+l-2j-k} }\,B^k_j
\end{equation*}
where $B_j^k = \frac{(j+k+1)!}{j!k!} + \delta_{j0}\delta_{k0} - 2^{1-k}\binom{2j+k+1}{k}$. From practical calculations, we compute a $200\times200$ truncation of $ \{\mathbf{L}^l\}$ for $l\in\{0,1,2,3,4 \}$ and compute the pseudo-inverse of these matrices in order to obtain the entries of $ \{\mathbf{L}^l\}^{-1}$.

\subsubsection{Burnett function decompositions}
Using the coordinate system $\bm{\hat{c}}= (\sin\Theta\,\cos\varphi,\,\sin\Theta\, \sin\varphi,\,\cos\Theta)^{\mathtt{T}} $, the following Burnett function representations can be made
\begin{align*}
    &A_{13}^\gauss(\bm{c}) = \frac{\pi^{\frac{3}{4}}}{2}\ \frac{\chi_{0,2,-1}(\bm{c}) - \chi_{0,2,1}(\bm{c})}{\sqrt{2}} &{A}_{12}^{\gauss}(\bm{c}) = \frac{\mathbf{i}\,\pi^{\frac{3}{4}}}{2} \,\frac{\chi_{0,2,-2}(\bm{c}) - \chi_{0,2,2}(\bm{c})}{\sqrt{2}}
    \\
     &{B}_{3}^\gauss(\bm{c}) = -\frac{\pi^{\frac{3}{4}}\sqrt{{5}}}{2} \,\chi_{1,1,0}(\bm{c}) 
 & {B}_1^\gauss(\bm{c}) = -\,{\frac{\pi^{\frac{3}{4}}\sqrt{5}}{2}}\,\frac{\chi_{1,1,-1}(\bm{c}) - \chi_{1,1,1}(\bm{c}) }{\sqrt{2}}
\end{align*}
Applying the operator $\iLG$ to the above gives
\begin{align*}
    &\iLG[A_{13}^\gauss](\bm{c}) = \frac{\pi^{\frac{3}{4}}}{2}\ \sum_{n=0}^{\infty}\{\mathbf{L}^2\}_{0n}^{-1} \frac{\chi_{n,2,-1}(\bm{c}) - \chi_{n,2,1}(\bm{c})}{\sqrt{2}} 
    &\iLG[{A}_{12}^{\gauss}](\bm{c}) = \frac{\mathbf{i}\,\pi^{\frac{3}{4}}}{2} \,\sum_{n=0}^\infty \{ \mathbf{L}^2\}^{-1}_{0n}\,\frac{\chi_{n,2,-2}(\bm{c}) - \chi_{n,2,2}(\bm{c})}{\sqrt{2}}
    \\
    &\iLG[{B}_{3}^\gauss](\bm{c}) = -\frac{\pi^{\frac{3}{4}}\sqrt{{5}}}{2} \,\sum_{n=0}^\infty \{\mathbf{L}^1 \}_{1n}^{-1}\ \chi_{n,1,0}(\bm{c}) 
    & \iLG[{B}_1^\gauss](\bm{c}) = -\,{\frac{\pi^{\frac{3}{4}}\sqrt{5}}{2}}\, \sum_{n=0}^\infty \{\mathbf{L}^1 \}_{1n}^{-1} \frac{\chi_{n,1,-1}(\bm{c}) - \chi_{n,1,1}(\bm{c}) }{\sqrt{2}}
\end{align*}
The viscosity and heat conductivity can be calculated immediately
\begin{align*}
    & -\frac{\epsilon}{\vartheta}\abrac{A_{13}^\M,\,\M\iLM[A_{13}^\M]} = -\frac{4\epsilon\rho_0 \vartheta}{\mathfrak{b}_\M} \Ebrac{A_{13}^\gauss,\, \iLG[A_{13}^\gauss]} = - \frac{\epsilon\,\pi^{\frac{3}{2}} }{\mathfrak{b}_\M} \{\mathbf{L}^2 \}^{-1}_{00}\ \rho_0 \vartheta
    \\
    & -\frac{\epsilon}{\vartheta^2}\abrac{B_{3}^\M,\,\M\iLM[B_{3}^\M]} = -\frac{2\epsilon\, \rho_0\vartheta}{\mathfrak{b}_\M}\Ebrac{B_3^\gauss,\,\iLG[B_3^\gauss]} =\frac{5}{2} \left(-\frac{\epsilon\,\pi^{\frac{3}{2}}}{\mathfrak{b}_\M}\{\mathbf{L}^1 \}_{11}^{-1} \right)\rho_0\vartheta
\end{align*}
For brevity, we shall only describe the process of computing $\alpha_1$ and $\tilde{\alpha}_1$. The computation of the other ESE fluid parameters follows a similar process. We start by calculating $C_{123}^\gauss$. By deducing identities similar to that in \cite[page 20]{WangChang1952}, we can show that
\begin{align*}
     &{C}_{123}^\gauss(\bm{c}) = {C}_{132}^\gauss(\bm{c}) = \frac{\mathbf{i}\pi^{\frac{3}{4}}}{2} \sum_{n=0}^{\infty}\{\mathbf{L}^2 \}_{0n}^{-1} \sum_{n_1 = 0}^{\infty} \mathfrak{e}_{n,n_1} \frac{\chi_{n_1,3,-2}(\bm{c}) - \chi_{n_1,3,2}(\bm{c})}{\sqrt{2}} 
\end{align*}
where
\begin{align*}
    \mathfrak{e}_{n,n_1} &= \delta_{n,n_1}\sqrt{\frac{7+2n}{14}} - \delta_{n-1, n_1}\sqrt{\frac{n}{7}}
\end{align*}
Applying the non-dimensional linearized operator then gives
\begin{align*}
 &\iLG[{C}^\gauss_{123}] = \iLG[{C}^\gauss_{132}]= \frac{\mathbf{i}\pi^{\frac{3}{4}}}{2} \sum_{n=0}^{\infty}\{\mathbf{L}^2 \}_{0n}^{-1} \sum_{n_1 = 0}^{\infty} \mathfrak{e}_{n,n_1}\sum_{n_2=0}^{\infty} \{\mathbf{L}^3\}^{-1}_{n_1,n_2} \frac{\chi_{n_2,3,-2}(\bm{c}) - \chi_{n_2,3,2}(\bm{c})}{\sqrt{2}} 
 \end{align*}
Thus,
\begin{align*}
    \Ebrac{\big(C^\gauss_{123}\big)^*,\ \iLG[{C}^\gauss_{123}] } = \frac{\pi^\frac{3}{2}}{4} \sum_{n_1 = 0}^\infty \parent{ \sum_{n=0}^\infty \{\mathbf{L}^2 \}_{0,n}^{-1} \mathfrak{e}_{n, n_1}  } \sum_{n_2 = 0}^\infty \{\mathbf{L}^3 \}_{n_1,n_2}^{-1} \parent{ \sum_{n_3=0}^\infty \{\mathbf{L}^2 \}_{0,n_3}^{-1} \mathfrak{e}_{n_3, n_2}  }
\end{align*}
Although the sums shown above are infinite, in practice using the first twenty terms for each sum gives a result accurate to seven significant figures.
Calculating $\alpha_1$ (resp. $\tilde{\alpha}_1$) is a matter of picking $\mathcal{M} = \mu$ (resp. $\mathcal{M}= M$) so that
\begin{align*}
    \alpha_1 &= -\epsilon^3\, \frac{_\mu K_{123}^{123}}{\theta} \\
    &= - \epsilon^3\frac{8\pi^3}{\mathfrak{b}_\mu^3}\,\frac{ \rho_0^3 \theta^2}{\rho^2} \Ebrac{\big(C^\gauss_{123}\big)^*,\ \iLG[{C}^\gauss_{123}] }
\end{align*}
Note that because $C^\gauss_{123} = C^\gauss_{132}$, we will have that $\alpha_1 = \alpha_3$ and $\tilde{\alpha}_1 = \tilde{\alpha}_3$.

Calculations for the BGK fluid parameters can be calculated in the same way as the Hard Spheres case by noting that the corresponding entries to the collision matrices are
\begin{align*}
    &\{ \mathbf{L}^l\}_{np} = \{ \mathbf{L}^l\}^{-1}_{np} = \begin{cases}
     0\,, & (n, l) \in \{(1, 0),\, (0,0),\,(0, 1) \}
     \\
         -\delta_{np}\,, & \text{otherwise}
     \end{cases}
\end{align*}

}

\end{appendices}
\begin{appendices}

\end{appendices}
\end{document}